\RequirePackage{mathtools}
\documentclass[epj]{svjour}
\def\makeheadbox{}
\usepackage{booktabs}
\usepackage{dblfloatfix}

\usepackage{tikz}
\usepackage{amssymb,latexsym}
\usepackage[shortskips]{methtools}
\usepackage{dsfont}
\usepackage{slashed}
\usepackage{mathrsfs}
\usepackage{widetable}
\usepackage{dcolumn}
\usepackage{ecolumn}
\usepackage{multirow}
\usepackage[numbers,sort&compress]{natbib}
\usepackage[squaren,Gray]{SIunits}

\usepackage{hyperref}

\newcommand{\mcemd}{\multicolumn{1}{c}{---}}
\newcommand*{\timex}{\multicolumn{1}{c}{$\times$}}%
\newcommand{\overbar}[1]{\mkern 1.5mu\overline{\mkern-1.5mu#1\mkern-1.5mu}\mkern 1.5mu}
\newcommand{\MSbar}{\overbar{\text{MS}}}
\newcommand*{\BnoL}{B\mathord{\smash{\neq}}\Lambda}
\newcommand{\SU}[1]{\ensuremath{\text{SU}(#1)}}
\renewcommand*{\cdot}{\mathbin{\mbox{\textperiodcentered}}}
\definecolor{matplotgreen}{rgb}{0,0.5,0}%
\newcommand{\trm}{\textcolor{matplotgreen}{trM}}
\newcommand{\msc}{\textcolor{red}{msc}}
\newcommand{\sym}{\textcolor{blue}{sym}}
\newcommand*{\VmA}{V\mathord{-}A}%
\newcommand*{\?}{\tmspace{-}{1.5mu}{.08333em}}%
\newcommand*{\B}{{\!B}}%
\newlength{\Wfockuud}
\newcommand*{\fockpalette}[2]{\settowidth{\Wfockuud}{$#1u^\gooduparrow u^\gooduparrow d^\gooduparrow$}\mathmakebox[\Wfockuud]{#2}}
\newcommand*{\fock}[1]{\boxed{\mathpalette{\fockpalette}{#1}}}
\newcommand*{\arrowpalette}[2]{\raisebox{\depth}{\resizebox*{!}{\heightof{$#1X$}}{$#1#2$}}}
\newcommand*{\gooduparrow}{{\mathpalette{\arrowpalette}{\uparrow}}}
\newcommand*{\gooddownarrow}{{\mathpalette{\arrowpalette}{\downarrow}}}
\newcommand*{\goodupdownarrow}{{\raisebox{.25ex}{$\scriptstyle\gooduparrow$}\mkern-2mu\raisebox{-.25ex}{$\scriptstyle\gooddownarrow$}}}
\newcommand*{\goodopen}{{\mathpalette{\arrowpalette}{(}}}
\newcommand*{\goodclose}{{\mathpalette{\arrowpalette}{)}}}
\makeatletter
\newif\iftag@here
\newcommand*{\taghere}[1][0pt]
{\ifmeasuring@\else
  \global\tag@heretrue
  \tikz[remember picture,overlay]{\coordinate (taghere) at (0pt,#1);}%
\fi}
\def\place@tag{%
    \iftagsleft@
      \kern-\tagshift@
      \iftag@here
        \global\tag@herefalse
        \tikz[remember picture,overlay]%
          {\path (taghere) -| node[anchor=base]{\rlap{\boxz@}} (0pt,0pt);}%
      \else
        \if1\shift@tag\row@\relax
            \rlap{\vbox{%
                \normalbaselines
                \boxz@
                \vbox to\lineht@{}%
                \raise@tag
            }}%
        \else
            \rlap{\boxz@}%
        \fi
        \kern\displaywidth@
      \fi
    \else
      \kern-\tagshift@
      \iftag@here
        \global\tag@herefalse
        \tikz[remember picture,overlay]%
          {\path  (taghere) -|  node[anchor=base]{\llap{\boxz@}} (0pt,0pt);}%
      \else
        \if1\shift@tag\row@\relax
            \llap{\vtop{%
                \raise@tag
                \normalbaselines
                \setbox\@ne\null
                \dp\@ne\lineht@
                \box\@ne
                \boxz@
            }}%
        \else \llap{\boxz@}%
        \fi
      \fi
    \fi
}
\makeatother

\begin{document}%
%
%
\title{%
Light-cone distribution amplitudes of octet baryons\\ from lattice QCD
}
%
%
%
\author{%
\protect\rule[0sp]{0sp}{1.8cm}\includegraphics{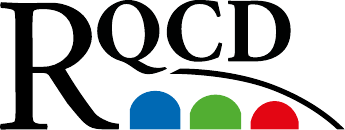}\\[\baselineskip]
Gunnar~S.~Bali		\inst{1,2}	\and
Vladimir~M.~Braun	\inst{1}	\and
Simon~B\"urger		\inst{1}	\and
Sara~Collins		\inst{1}	\and
Meinulf~G\"ockeler	\inst{1}	\and
Michael~Gruber		\inst{1}	\and
Fabian~Hutzler		\inst{1}	\and
Piotr~Korcyl		\inst{3}	\and
Andreas~Sch\"afer	\inst{1}	\and
Wolfgang~S\"oldner	\inst{1}	\and
Andr\'e~Sternbeck	\inst{4}	\and
Philipp~Wein		\inst{1}
}%
\authorrunning{RQCD Collaboration}
%
%
\institute{%
Institut f{\"u}r Theoretische Physik, Universit{\"a}t Regensburg, Universit{\"a}tsstra{\ss}e 31, 93040 Regensburg, Germany
\and
Department of Theoretical Physics, Tata Institute of Fundamental Research, Homi Bhabha Road, Mumbai 400005, India
\and
Marian Smoluchowski Institute of Physics, Jagiellonian University, ul.\ \L ojasiewicza 11, 30-348 Krak\'ow, Poland
\and
Theoretisch-Physikalisches Institut, Friedrich-Schiller-Universit{\"a}t Jena, 07743 Jena, Germany
}%
%
%
\abstract{%
We present lattice QCD results for the wave function normalization constants and the first moments of the distribution amplitudes for the lowest-lying baryon octet. The analysis is based on a large number of $N_f=2+1$ ensembles comprising multiple trajectories in the quark mass plane including physical pion (and kaon) masses, large volumes, and, most importantly, five different lattice spacings down to $a=\unit{0.039}{\femto\meter}$. This allows us to perform a controlled extrapolation to the continuum and infinite volume limits by a simultaneous fit to all available data. We demonstrate that the formerly observed violation of flavor symmetry breaking constraints can, indeed, be attributed to discretization effects that vanish in the continuum limit.
}%
\date{}%
%
%
\maketitle
%
%
\section{Introduction}
Hard exclusive reactions involving large momentum transfer between the initial and final state hadron are most sensitive to the leading Fock states with a small number of partons and to the distribution of the longitudinal momentum amongst these constituents, which is encoded in light-cone distribution amplitudes~(DAs)~\cite{Lepage:1979zb,Lepage:1980fj,Efremov:1979qk}. This opens up a complementary view on hadron structure compared to the usual parton distributions and form factors, which do not provide information on the individual Fock states.\par%
Pinning down DAs solely from experimental measurements is a formidable task, since they appear in convolutions with a hard scattering kernel and, at available momentum transfers, are affected by soft corrections. These can be taken into account using light-cone sum rules~\mbox{\cite{Braun:1988qv,Balitsky:1989ry,Chernyak:1990ag}}. Furthermore, exclusive channels face the generic difficulty that the final state phase space is very small and that reactions with a small number of final state hadrons are power-suppressed at high momentum transfer. Hence, complementary input and guidance from theoretical tools that can be applied in the nonperturbative regime and allow us to narrow down the set of possible models and pa\-ram\-e\-triza\-tions is of vital importance.\par%
This article is the capstone of a long-term effort to determine baryon distribution amplitudes from lattice QCD using dynamical clover fermions. This project began more than a de\-cade ago with the $N_f=2$ studies of nucleon DAs~\cite{Gockeler:2008xv,Braun:2008ur} and subsequently also explored the negative parity partner of the nucleon~\cite{Braun:2009jy,Schiel:2011av,Braun:2012zza,Braun:2014wpa}. The new $N_f=2+1$ ensembles generated more recently within the CLS (Coordinated Lattice Simulations) effort~\cite{Bruno:2014jqa}\footnote{\url{https://wiki-zeuthen.desy.de/CLS/}} include dynamical strange quarks and have recently allowed for a first determination of hyperon distribution amplitudes~\cite{Bali:2015ykx} on four ensembles at a single lattice spacing. The current work extends this analysis to $40$~ensembles covering the whole CLS landscape, which contains multiple trajectories in the quark mass plane (including physical pion and kaon masses), large volumes, and five different lattice spacings down to $a=\unit{0.039}{\femto\meter}$. Using this wealth of data we determine, for the first time, the moments of nucleon and hyperon DAs in the physical continuum limit by performing a controlled extrapolation to the physical meson masses, the infinite volume, and, most importantly, to zero lattice spacing. For moments of pseudoscalar meson DAs we have recently presented a similar determination in Ref.~\cite{Bali:2019dqc}.\par%
Relying on the groundwork of Refs.~\cite{Bali:2015ykx,Wein:2016ozo,Gruber:2017ozo,Hutzler:2018ozo}, we will give a concise discussion of the theoretical formalism in Section~\ref{sect_DAs}, followed, in Section~\ref{sect_lattice}, by a brief presentation of the lattice framework. Section~\ref{sect_strategy_and_results} contains a detailed description of our analysis strategy as well as a discussion of the final results. We summarize our findings in Section~\ref{sect_summary}.\par%
\section{Three-quark distribution amplitudes\label{sect_DAs}}
\subsection{Leading-twist distribution amplitudes\label{sect_leadingtwist}}
Baryon DAs~\cite{Lepage:1980fj,Efremov:1979qk,Chernyak:1983ej} are defined as matrix elements of renormalized three-quark operators (we use the scheme proposed in Ref.~\cite{Kraenkl:2011qb}) at light-like separations:%
\begin{align}\label{eq_BDA}%
 \MoveEqLeft[1] \langle 0 | \bigl[f_\alpha(a_1 n) g_\beta(a_2 n) h_\gamma(a_3 n)\bigr]^{\MSbar} | B_{p,\lambda} \rangle \notag \\
&= \frac14 \int \! [dx] \; e^{-ip \cdot n \sum_i a_i x_i} \Bigl( v^B_{\alpha\beta;\gamma} V^B(x_1,x_2,x_3) \\&\quad+ a^B_{\alpha\beta;\gamma} A^B(x_1,x_2,x_3) + t^B_{\alpha\beta;\gamma} T^B(x_1,x_2,x_3)+\ldots \Bigr) \,. \notag
\end{align}%
On the l.h.s.\ the Wilson lines as well as the color antisymmetrization are not written out explicitly but implied. $| B_{p,\lambda} \rangle$ is the baryon state with momentum $p$ and helicity $\lambda$, while $\alpha,\beta,\gamma$ are Dirac indices, $n$ is a light-cone vector ($n^2=0$), the $a_i$ are real numbers, and $f,g,h$ are quark fields of the given flavor, chosen to match the valence quark content of the baryon $B$:%
\def\myrule{\rule[-2pt]{0sp}{12pt}}
\begin{align}
\begin{tabular}[b]{lccc}
 \myrule$\hphantom{N}\mathrel{B}$ & $f$ & $g$ & $h$ \\ \midrule 
 \myrule$\hphantom{N}\mathllap{N} \equiv p$ & $u$ & $u$ & $d$ \\
 \myrule$\hphantom{N}\mathllap{\Sigma} \equiv \Sigma^-$ & $d$ & $d$ & $s$ \\
 \myrule$\hphantom{N}\mathllap{\Xi} \equiv \Xi^0$ & $s$ & $s$ & $u$ \\
 \myrule$\hphantom{N}\mathrel{\Lambda}$ & $u$ & $d$ & $s$
\end{tabular} \,.
\end{align}
Here we assume isospin symmetry and select one representative for each isospin multiplet.\par%
The general Lorentz decomposition on the r.h.s.\ of Eq.~\eqref{eq_BDA} consists of $24$~terms~\cite{Braun:2000kw}. In this decomposition the three leading-twist DAs, $V^B$, $A^B$, and $T^B$, appear in conjunction with the structures
\begin{align}%
 v^B_{\alpha\beta;\gamma} &= (\tilde{\vphantom{n}\smash{\slashed{n}}}C)_{\alpha\beta}(\gamma_5 u^{B,+}_{p,\lambda})_\gamma\,, \notag\\
 a^B_{\alpha\beta;\gamma} &= (\tilde{\vphantom{n}\smash{\slashed{n}}}\gamma_5C)_{\alpha\beta} (u^{B,+}_{p,\lambda})_\gamma\,, \notag\\
 t^B_{\alpha\beta;\gamma} &= (i\sigma_{\perp \tilde n} C)_{\alpha\beta}(\gamma^\perp\gamma_5 u^{B,+}_{p,\lambda} )_\gamma\,,
\end{align}
with the charge conjugation matrix $C$ and the notation%
\begin{flalign}%
 \sigma_{\perp \tilde n}\otimes\gamma^\perp &= \sigma^{\mu\rho} \tilde n_\rho g^\perp_{\mu\nu} \otimes \gamma^\nu\,, &
 \!\!g_{\mu\nu}^\perp &= g_{\mu\nu} - \frac{\tilde n_\mu n_\nu+ \tilde n_\nu n_\mu}{\tilde n\cdot n} \,, \notag \\*
 u^{B,+}_{p,\lambda} &= \frac12\frac{\tilde{\vphantom{n}\smash{\slashed{n}}}\slashed{n}}{\tilde n\cdot n} u^B_{p,\lambda} \,, &
\tilde n_\mu &= p_\mu - \frac12 \frac{m^2_B}{p\cdot n} n_\mu \,,
\end{flalign}%
where $u^B_{\smash[t]{p,\lambda}}$ is the Dirac spinor with on-shell momentum~$p$ and helicity~$\lambda$. Correct translational behavior is ensured by the exponential factor in combination with the integration measure for the longitudinal momentum fractions:%
\begin{align}
 \int \! [dx] &= \int_0^1 \!\! dx_1 \int_0^1 \!\! dx_2 \int_0^1 \!\! dx_3 \; \delta (1-x_1-x_2-x_3)\,.
\end{align}\par%
To fully exploit the benefits of \SU3 flavor symmetry it proves convenient to define the following set of DAs:%
\begin{align}
 \Phi_{\pm}^{B\neq\Lambda}(x_{123})&=\tfrac{1}{2} \Bigl([\VmA]^B(x_{123}) \pm [\VmA]^B(x_{321})\Bigr) \,, \notag
\\*
 \Pi^{B\neq\Lambda}(x_{123})&= T^B (x_{132}) \,, \notag
\\*
 \Phi_{+}^{\Lambda}(x_{123})&=\sqrt{\tfrac{1}{6}} \Bigl([\VmA]^\Lambda(x_{123}) + [\VmA]^\Lambda(x_{321}) \Bigr) \,, \notag
\\*
 \Phi_{-}^{\Lambda}(x_{123})&=-\sqrt{\tfrac{3}{2}} \Bigl([\VmA]^\Lambda(x_{123}) - [\VmA]^\Lambda(x_{321}) \Bigr) \,, \notag
\\*
 \Pi^{\Lambda}(x_{123})&=\sqrt{6} \; T^\Lambda (x_{132}) \,, \label{eq_convenient_DAs}
\end{align}%
where for brevity $(x_{ijk})\equiv (x_i,x_j,x_k)$. In the limit of \SU3 flavor symmetry (subsequently indicated by a $\star$), where $m_u=m_d=m_s$, the following relations hold:\footnote{Our phase conventions for the baryon states and the corresponding flavor wave functions are detailed in Appendix~A of Ref.~\cite{Bali:2015ykx}.}%
\begin{align}%
 \Phi_{+}^\star &\equiv \Phi_{+}^{N\star} = \Phi_{+}^{\Sigma\star} = \Phi_{+}^{\Xi\star} = \Phi_{+}^{\Lambda\star} = \Pi^{N\star} = \Pi^{\Sigma\star} = \Pi^{\Xi\star} \,, \notag
\\*
 \Phi_{-}^\star &\equiv \Phi_{-}^{N\star} = \Phi_{-}^{\Sigma\star} = \Phi_{-}^{\Xi\star} = \Phi_{-}^{\Lambda\star} = \Pi^{\Lambda\star}\,.\label{eq_SU3_leadingtwist}
\end{align}%
Therefore, the amplitudes $\Pi^B$ (or $T^B$) only need to be considered when \SU3 flavor symmetry is broken. In the case of \SU2 isospin symmetry, which is exact in our $N_f=2+1$ simulation ($m_u=m_d\equiv m_\ell$) and is only broken very mildly in the real world, the nucleon DA $\Pi^N$ is equal to $\Phi_+^N$ in the whole $m_\ell$-$m_s$-plane.\par%
DAs can be expanded in terms of orthogonal polynomials $\mathcal{P}_{nk}$ in such a way that the coefficients have autonomous scale dependence at one loop (conformal partial wave expansion). Taking into account the corresponding symmetry of the DAs defined in Eqs.~\eqref{eq_convenient_DAs}, this expansion reads%
\begin{align}%
 \Phi_{+}^{B} &= 120 x_1 x_2 x_3 \bigl( \varphi^B_{00} \mathcal P_{00} + \varphi^B_{11} \mathcal P_{11} + \dots \bigr) \, , \notag \\
 \Phi_{-}^{B} &= 120 x_1 x_2 x_3 \bigl( \varphi^B_{10} \mathcal P_{10} + \dots \bigr) \,, \notag \\
 \Pi^{B\neq\Lambda} &= 120 x_1 x_2 x_3 \bigl( \pi^B_{00} \mathcal P_{00} + \pi^B_{11} \mathcal P_{11} + \dots \bigr) \, , \notag \\
 \Pi^{\Lambda} &= 120 x_1 x_2 x_3 \bigl( \pi^{\Lambda}_{10} \mathcal P_{10} + \dots \bigr) \,.\label{eq_moments}
\end{align}%
In this way all nonperturbative information is encoded in the set of scale-dependent coefficients $\varphi^B_{nk}$, $\pi^B_{nk}$ (also called shape parameters), which can be related to matrix elements of local operators that are calculable on the lattice. All $\mathcal{P}_{nk}$ have definite symmetry (being symmetric or antisymmetric) under the exchange of $x_1$ and $x_3$~\cite{Anikin:2013aka} and in each DA only polynomials of one type, either symmetric or antisymmetric, appear. The first few polynomials are (see, e.g., Ref.~\cite{Braun:2008ia})%
\begin{align}%
\mathcal{P}_{00} &= 1 \,, \notag \\
\mathcal{P}_{10} &= 21(x_1-x_3)\,, \notag \\
\mathcal{P}_{11} &= 7(x_1-2x_2 + x_3)\,.
\end{align}%
The leading contributions in Eqs.~\eqref{eq_moments} are $120 x_1 x_2 x_3 \varphi^B_{00}$ as well as $120 x_1 x_2 x_3 \pi^{B\neq\Lambda}_{00}$ and are usually referred to as the asymptotic DAs. The corresponding normalization coefficients $\varphi^B_{00}$ and $\pi^{B\neq\Lambda}_{00}$ can be thought of as the wave functions at the origin. 
\subsection{Wave function normalization constants\label{sect_normalization}}
The leading-twist normalization constants $f^B$ and $f_T^{\BnoL}$ can be defined conveniently via matrix elements of local currents (all quark fields are at the origin),%
\begin{flalign}%
 \langle0|\bigl(f^{\gooduparrow T}C\slashed{n}g^\gooddownarrow\bigr) \slashed{n}h^\gooduparrow|(\BnoL)_{p,\lambda}\rangle &= -\tfrac12f^Bp\cdot n\mkern1mu\slashed{n}u^{B\gooduparrow}_{p,\lambda} \,, \notag \\
 \langle0|\bigl(u^{\gooduparrow T}C\slashed{n}d^\gooddownarrow\bigr) \slashed{n}s^\gooduparrow|\Lambda_{p,\lambda}\rangle &= -\tfrac12\smash{\sqrt{\rule[-2pt]{0sp}{10pt}\smash{\tfrac32}}}f^\Lambda p\cdot n\mkern1mu\slashed{n}u^{\Lambda\gooduparrow}_{p,\lambda} \,, \notag \\
 \langle0|\bigl(f^{\gooduparrow T}C\gamma^\mu\slashed{n}g^\gooduparrow\bigr)\gamma_\mu\slashed{n}h^\gooddownarrow|(\BnoL)_{p,\lambda}\rangle &= 2f_T^Bp\cdot n\mkern1mu\slashed{n}u^{B\gooduparrow}_{p,\lambda} \,, \taghere
\end{flalign}%
using chiral quark fields $q^\goodupdownarrow=\frac12(\mathds 1 \pm \gamma_5) q$ and baryon spinors $\smash{u^{B\goodupdownarrow}_{p,\lambda}} = \frac12(\mathds 1 \pm \gamma_5) u^{B}_{p,\lambda}$. These normalization constants are equivalent to the zeroth moments of the leading-twist DAs defined in the previous section:%
\begin{align}%
  f^B &= \varphi^B_{00}\,, & f^{B\neq\Lambda}_T &= \pi^B_{00}\,.
\end{align}%
Due to isospin symmetry these two couplings coincide for the nucleon, $f^N_T = f^N$. For the $\Lambda$ baryon the zeroth moment of $T^\Lambda$ vanishes by construction so that only one leading-twist normalization constant $f^\Lambda$ exists.\par%
The $21$ DAs of higher twist (indicated in Eq.~\eqref{eq_BDA} by the ellipsis on the r.h.s.) only involve two new normalization constants ($\lambda_1^B$ and $\lambda_2^B$) for the isospin-nonsinglet baryons ($N$, $\Sigma$, $\Xi$) and three ($\lambda_1^\Lambda$, $\lambda_T^\Lambda$, and $\lambda_2^\Lambda$) for the $\Lambda$~baryon. These can be defined as matrix elements of local three-quark twist-four operators without derivatives:%
\begin{align}%
 \langle 0 | \bigl(f^{\gooduparrow T} C \gamma^\mu g^\gooddownarrow\bigr) \gamma_\mu h^\gooduparrow | (\BnoL) _{p,\lambda} \rangle
&=
- \tfrac 12 \lambda_1^B m_B u^{B\gooddownarrow}_{p,\lambda} \,, \notag
\\
\langle 0 | \bigl(f^{\gooduparrow T} C \sigma^{\mu\nu} g^\gooduparrow\bigr) \sigma_{\mu\nu} h^\gooduparrow | (\BnoL) _{p,\lambda} \rangle
&= \lambda_2^B m_B u^{B\gooduparrow}_{p,\lambda} \,, \notag
\\
 \langle 0 | \bigl(u^{\gooduparrow T} C \gamma^\mu d^\gooddownarrow\bigr) \gamma_\mu s^\gooduparrow | \Lambda _{p,\lambda} \rangle
&=
 \tfrac{1}{2\sqrt{6}} \lambda_1^\Lambda m_\Lambda u^{\Lambda\gooddownarrow}_{p,\lambda} \,, \notag
\\
 \langle 0 | \bigl(u^{\gooduparrow T} C d^\gooduparrow\bigr) s^\gooddownarrow | \Lambda _{p,\lambda} \rangle
&=
\tfrac{1}{2\sqrt{6}} \lambda_T^\Lambda m_\Lambda u^{\Lambda\gooddownarrow}_{p,\lambda} \,, \notag
\\
 \langle 0 | \bigl(u^{\gooduparrow T} C d^\gooduparrow\bigr) s^\gooduparrow | \Lambda _{p,\lambda} \rangle
&= \tfrac{-1}{4\sqrt{6}} \lambda_2^\Lambda m_\Lambda u^{\Lambda\gooduparrow}_{p,\lambda} \,.
\end{align}%
The definitions are chosen such that in the flavor symmetric case%
\begin{align}%
\lambda_1^\star &\equiv \lambda_1^{N\star} = \lambda_1^{\Sigma\star} = \lambda_1^{\Xi\star} = \lambda_1^{\Lambda\star} = \lambda_T^{\Lambda\star} \,, \notag \\*
\lambda_2^\star &\equiv \lambda_2^{N\star} = \lambda_2^{\Sigma\star} = \lambda_2^{\Xi\star} = \lambda_2^{\Lambda\star} \,.\label{eq_SU3_highertwist}
\end{align}%
For details see Refs.~\cite{Braun:2000kw,Wein:2015oqa}. These twist-four couplings are also interesting in a broader context, e.g., in studies of baryon decays in generic GUT models~\cite{Claudson:1981gh} or as input parameters for QCD sum rule calculations.\par%
\section{Lattice framework\label{sect_lattice}}
\begin{figure*}[t]%
\centering%
\includegraphics[width=.9\textwidth]{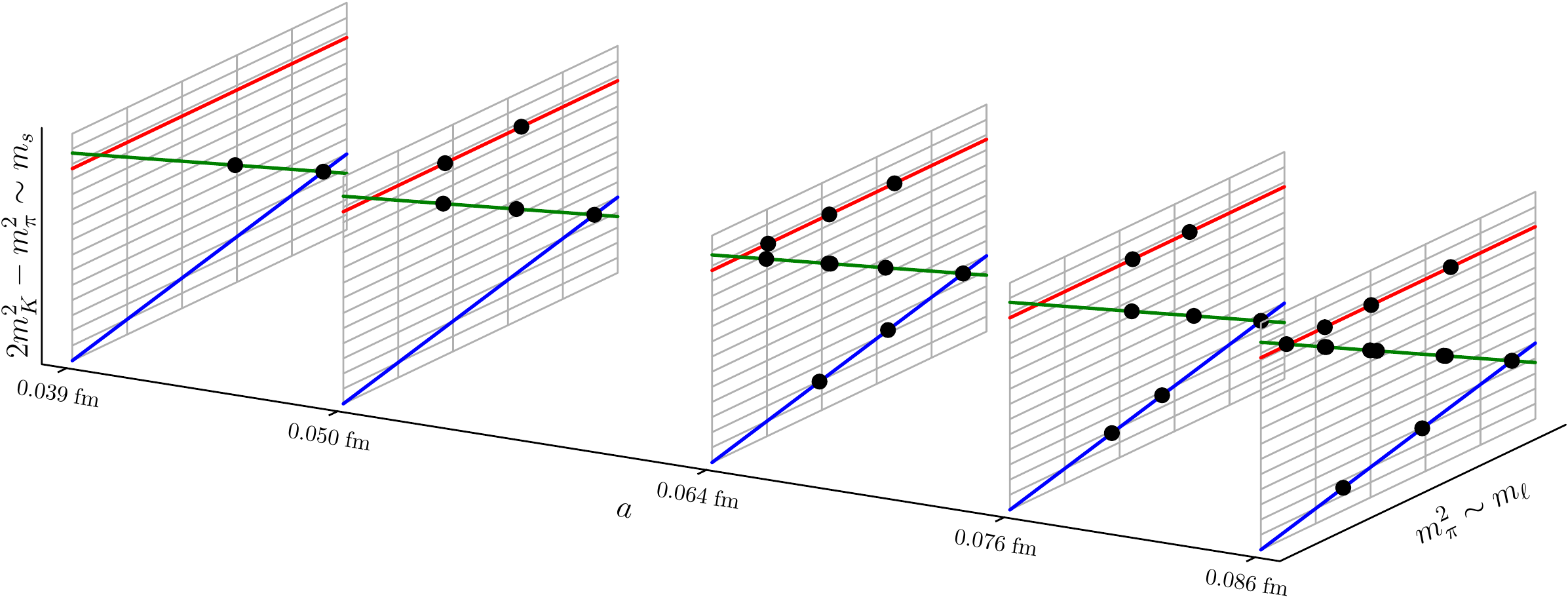}%
\caption{\label{figure_ensembles}Schematic showing the analyzed CLS ensembles in the space spanned by the lattice spacing and masses. On the flavor symmetric plane (blue), where $m_\ell=m_s$, flavor multiplets of hadrons have degenerate masses (e.g., $m_K^2=m_\pi^2$ and $m_N=m_\Sigma=m_\Xi=m_\Lambda$). The (green) lines of physical average quadratic meson mass ($2 m_K^2+m_\pi^2=\text{phys.}$) correspond to an approximately physical mean quark mass ($2m_\ell+m_s\approx\text{phys.}$). The red lines are defined by $2 m_K^2-m_\pi^2=\text{phys.}$ and indicate an almost physical strange quark mass ($m_s\approx\text{phys.}$). Physical masses are reached at the intersections of green and red lines.}%
\end{figure*}%
\begin{figure}[t]%
\centering%
\includegraphics[width=\columnwidth]{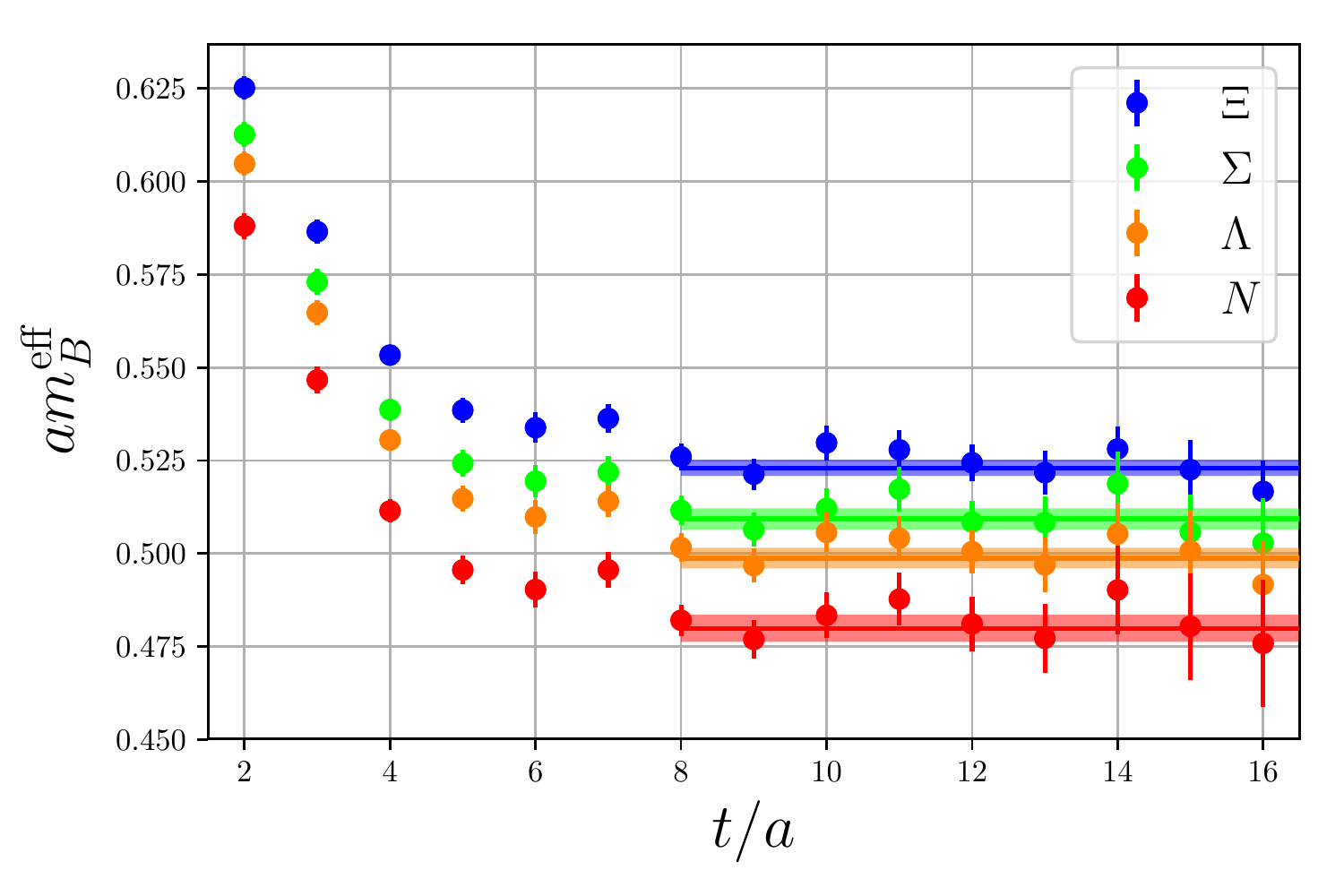}%
\caption{\label{figure_meff}The effective baryon masses obtained from averaged, source- and sink-smeared correlation functions calculated on the H102 ensemble with vanishing three-momentum. In this case the plateaus start at $t=8a$, where excited states are sufficiently suppressed. For each baryon the horizontal line represents the result of a fit to correlators in the range $8a \leq t \leq 20a$.}%
\end{figure}%
\begin{figure}[t]%
\centering%
\includegraphics[width=\columnwidth]{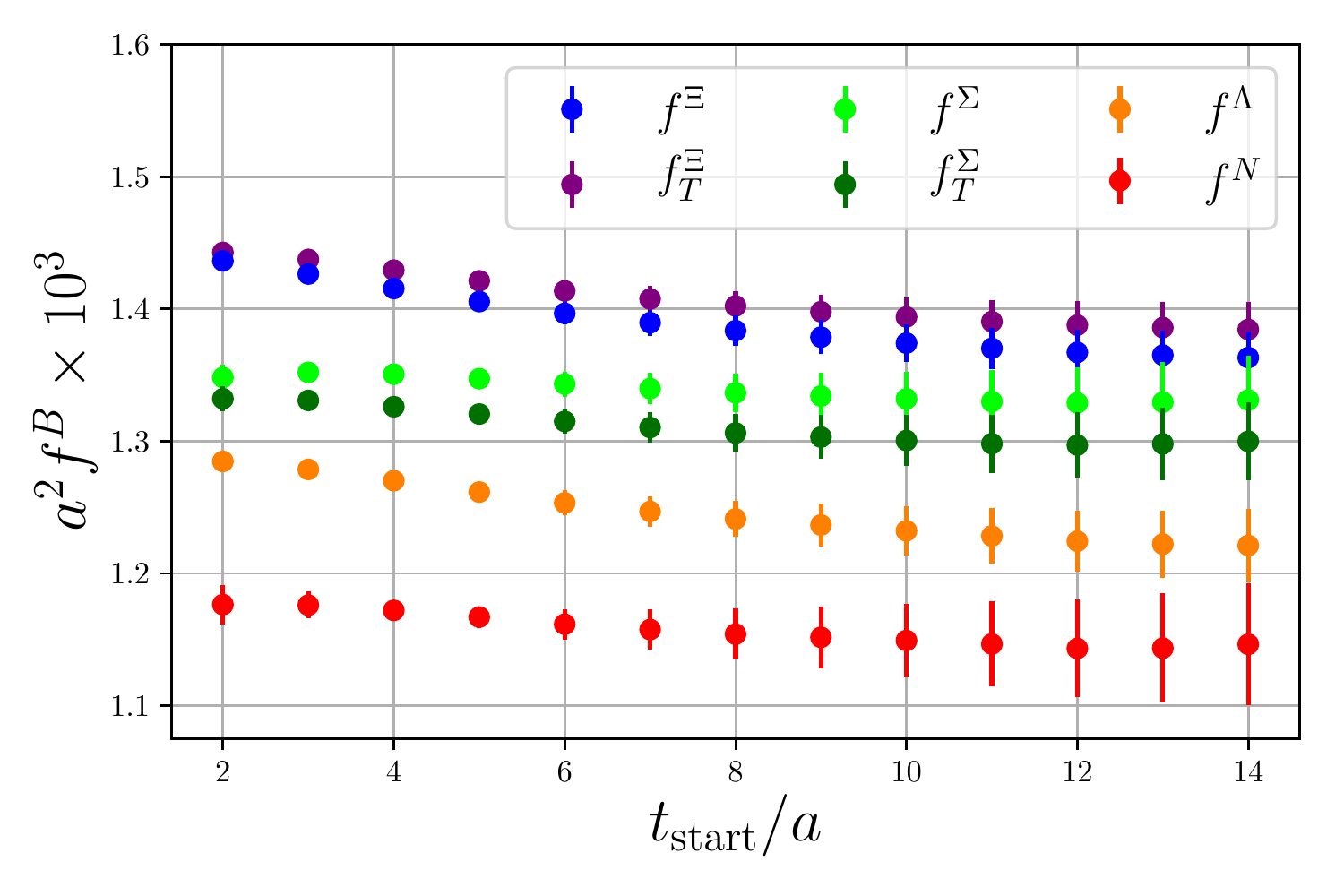}%
\caption{\label{figure_start}Typical plot (from the H105 ensemble) used for the determination of the fit ranges by varying the value of the minimal source-sink distance $t_{\text{start}}$. It shows the unrenormalized leading-twist normalization constants obtained from one-exponential fits to smeared-smeared and smeared-point correlation functions, cf.\ Ref.~\cite{Bali:2015ykx}. A conservative choice is $t_{\text{start}}=10a$, where the results have fully saturated for all leading-twist couplings. A variation of the maximal source-sink separation within reasonable bounds did not have any significant impact on the result. In this example it is always set to $t_{\text{end}}=20a$.}%
\end{figure}%
\subsection{Ensemble details\label{sect_ens}}
In this work we determine all, i.e., leading- and higher-twist, normalization constants as well as the first moments of the leading-twist distribution amplitudes for the lowest-lying spin $1/2$ positive parity baryon octet using lattice QCD. To this end, we evaluate the very same correlation functions as laid out in detail in our previous study~\cite{Bali:2015ykx} and, therefore, refrain from repeating them here. We use a large set of lattice ensembles generated within the CLS effort.\footnote{Some of the $m_\ell=m_s$ ensembles with (anti-)periodic boundary conditions in time have been generated by RQCD using the BQCD code~\cite{Nakamura:2010qh}.} These $N_f=2+1$ simulations employ the nonperturbatively order $a$ improved Wilson (clover) quark action and the tree-level Symanzik improved gauge action. We achieve an efficient and stable hybrid Monte Carlo sampling by applying twisted-mass determinant reweighting~\cite{Luscher:2012av}, which avoids near-zero modes of the Wilson Dirac operator. To enhance the ground state overlap the source interpolators are Wupper\-tal-smeared~\cite{Gusken:1989qx}, employing spatially APE-smeared~\cite{Falcioni:1984ei} gauge links.\par%
A special feature of CLS configurations is the use of open boundary conditions in the time direction~\cite{Luscher:2012av,Luscher:2011kk}. This allows for simulations on very fine lattices (in our case down to $a=\unit{0.039}{\femto\meter}$) without topological freezing. While employing open boundary conditions is crucial for fine lattice spacings, we use a mixture of lattices with open and periodic boundary conditions for the coarser spacings. In total we have five different lattice spacings ranging from~$a=\unit{0.039}{\femto\meter}$ to~$a=\unit{0.086}{\femto\meter}$, see Table~\ref{table_spacings} in Appendix~\ref{sect_plots}.\par%
A full list of the ensembles used in this work is given in Table~\ref{table_ensembles}. As schematically represented in Figure~\ref{figure_ensembles}, the available ensembles have been generated along three different trajectories covering the whole quark mass plane.\footnote{In practice the ensembles do not always lie exactly on top of the targeted green and red trajectories shown in Figure~\ref{figure_ensembles}. This does not pose a problem, and, actually can even help to further stabilize our fits of the pion and kaon mass dependence. For plotting purposes it is necessary to correct for any deviation (including also the almost negligible finite volume corrections) by projecting all ensembles onto the respective trajectory.} The ensembles cover a wide range of volumes with $2.9\leq m_\pi L\leq6.5$, where the bulk has $m_\pi L>4$, allowing us to investigate and control finite volume effects.\par%
The combination of multiple quark mass trajectories with a wide range of lattice spacings and volumes enables us to simultaneously extrapolate to physical masses, to infinite volume, and to the continuum by means of a global fit to all $40$~ensembles. In particular, this allows us to resolve mass-dependent discretization effects that we have conjectured in Ref.~\cite{Bali:2015ykx} as an explanation for the observed violation of expected \SU3 symmetry breaking patterns (derived in Ref.~\cite{Wein:2015oqa}) at a finite lattice spacing. The wealth of available CLS data combined with the extrapolation strategy explained in detail in Section~\ref{sect_global_analysis} allows us to take the continuum limit in a controlled fashion for the first time.\par%
For each gauge configuration we have carried out all measurements for $3$~different source positions. We then average over appropriate two-point functions and momenta as detailed in Ref.~\cite{Bali:2015ykx}. For the statistical analysis we generate $100$~bootstrap samples per ensemble using a bin size of $8$~configurations to eliminate autocorrelations. In order to extract the required ground state matrix elements from the correlation functions (i.e., excluding contributions from excited states) the choice for the lower bound of the fit range is crucial. Figure~\ref{figure_meff} demonstrates that, with increasing source-sink distance, the excited states decay and clear effective mass plateaus emerge. To determine the optimal minimal source-sink distance $t_{\text{start}}$ we perform multiple fits with varying fit ranges for all observables. As an example, Figure~\ref{figure_start} shows the fitted leading-twist coupling constants as a function of $t_{\text{start}}$. The latter is then chosen in such a way that fits with even larger starting times no longer show any systematic trend in the fit results.\par%
\subsection{Renormalization}%
\begin{table}[b]%
\centering%
\caption{Fit choices regarding the determination of the renormalization and mixing factors.\label{table_fits}}%
\begin{widetable}{\columnwidth}{cD{.}{.}{2.0}ccD{.}{.}{1.2}c}%
\toprule
Fit & \multicolumn{1}{c}{$\mu_1^2\,[\giga\electronvolt\squared]$} & $n_{\mathrm{loops}}$ & $n_{\mathrm{disc}}$ & \multicolumn{1}{c}{$\lambda^2_{\mathrm{scale}}$} & $\Lambda^{(3)}_{\MSbar}\,[\mega\electronvolt]$\\
\midrule
1 & 4  & 1 & 3 & 1.0  & 341\\
2 & 10 & 1 & 3 & 1.0  & 341\\
3 & 4  & 0 & 3 & 1.0  & 341\\
4 & 4  & 1 & 2 & 1.0  & 341\\
5 & 4  & 1 & 3 & 1.03 & 341\\
6 & 4  & 1 & 3 & 1.0  & 353\\
\bottomrule
\end{widetable}%
\end{table}%
The preferred renormalization scheme in phenomenological applications is based on dimensional regularization where, for baryons, there are subtleties due to contributions of evanescent operators that have to be taken into account, see Refs.~\cite{Kraenkl:2011qb,Gracey:2012gx}. For simplicity, we will refer to the prescription suggested in Ref.~\cite{Kraenkl:2011qb} as the $\MSbar$~scheme. On the lattice, we first calculate the renormalization factors nonperturbatively~\cite{Martinelli:1994ty} within the RI$^\prime$\nobreakdash-SMOM scheme~\cite{Sturm:2009kb} adapted to three-quark operators in Refs.~\cite{Gruber:2017ozo,Bali:2015ykx,Gockeler:2008we,Kaltenbrunner:2008zz}. These are then converted to the $\MSbar$~scheme using one-loop (continuum) perturbation theory. The conversion factors can be found in Ref.~\cite{Gruber:2017ozo}.\par%
The general method employed to obtain the renormalization factors has been presented in detail in our previous article~\cite{Bali:2015ykx}. For the coarser lattice spacings ($\beta=3.4$, $3.46$, $3.55$) we have ensembles with $m_\ell=m_s$ and \discretionary{(anti-)}{periodic}{(anti-)periodic} boundary conditions in time at our disposal so that we can proceed in the same way as in Ref.~\cite{Bali:2015ykx}, starting from four-point functions (in Landau gauge) of the schematic form%
\begin{align}%
\frac{a^{16}}{V} \mathop{\smash[b]{\smashoperator{\sum_{w,x,y,z}}}} e^{i(p\cdot x+q\cdot y+r\cdot z-(p+q+r)\cdot w)} \langle\mathcal{O}(w)\bar{f}(x)\bar{g}(y)\bar{h}(z)\rangle\,,
\end{align}%
where $\mathcal{O}$ represents the three-quark operators under study and $V$~denotes the four-dimensional volume of the lattice. However, on the finer lattices ($\beta=3.7$ and~$3.85$) we have to work with open boundary conditions in time. In this case the computation of the required four-point functions must be modified. We confine the momentum sources for the three external quark fields to a subvolume, keeping a distance of $4a$ from the boundaries in the time direction to stabilize the inversion of the fermion matrix. Furthermore, we restrict the (final) average over the position~$w$ of the inserted operator to points with a minimal distance of $32a$ from the temporal boundaries. This restriction ensures that the $w_4$ dependence of the remaining sums in the four-point function is negligible. As statistical errors in the renormalization are generally very small, the ensuing slight increase in the statistical fluctuations can easily be accepted. Comparisons between results obtained on ensembles that differ only in the boundary conditions (and the time extent) show differences that are of the size of the statistical errors.\par%
Another issue related to the ensembles at $\beta=3.7$ and~$3.85$ concerns the required chiral extrapolation. This difficulty becomes apparent upon a look at Table~\ref{table_ensembles_renorm}, where we have compiled the ensembles with $m_\ell=m_s$ used for the calculation of the renormalization factors. While lattices with four different values of $m_\ell=m_s$ are available for $\beta=3.4$, $3.46$, and $3.55$, only two different mass values have been used in the flavor-symmetric simulations at $\beta=3.7$ and~$3.85$. However, on the three coarser lattices the mass dependence of the amputated four-point functions is rather mild. Therefore, we are confident that the chiral extrapolations for $\beta=3.7$ and~$3.85$ do not significantly affect the reliability of our results.\par%
\begin{figure}[t]%
\centering%
\includegraphics[width=0.95\columnwidth]{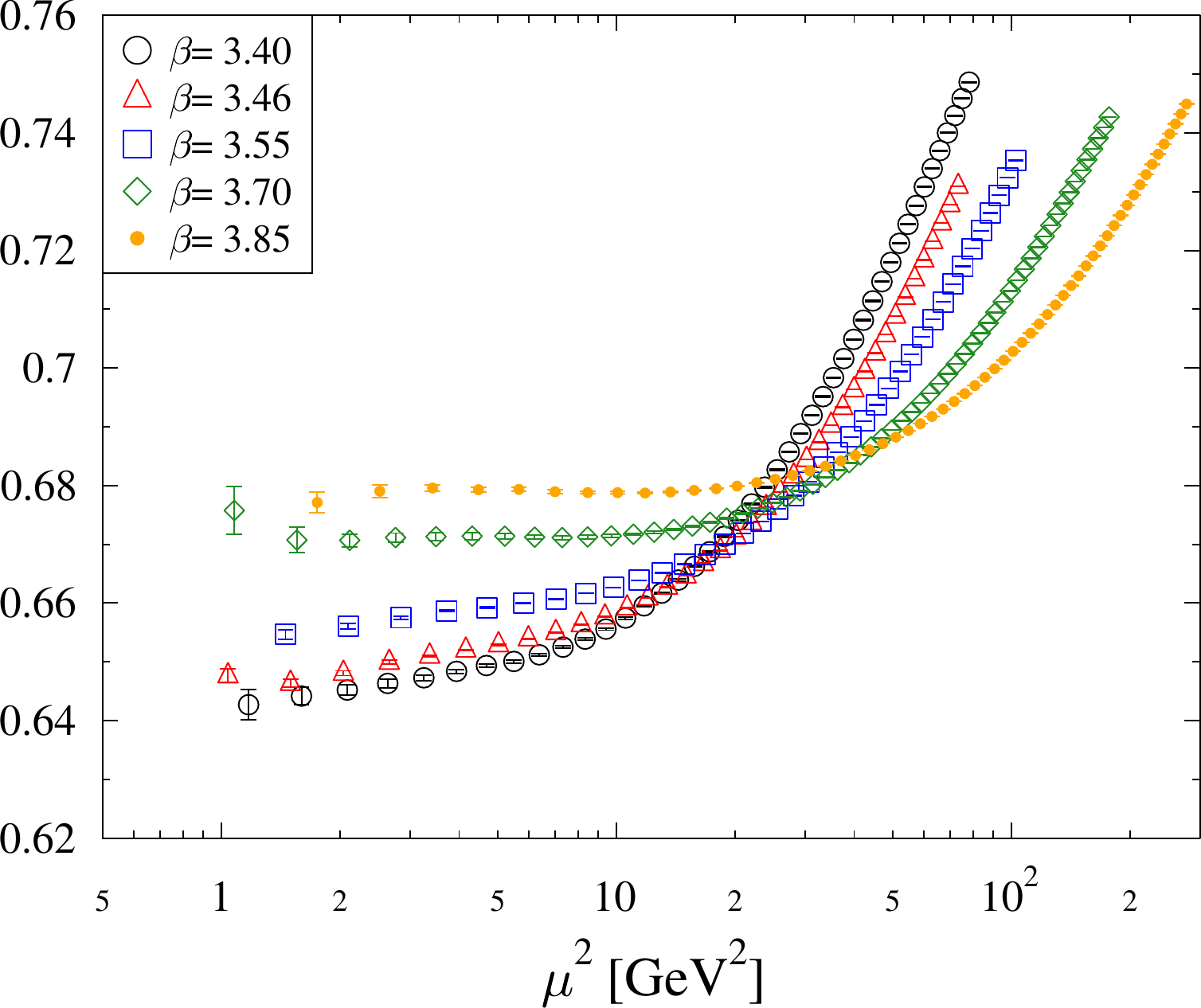}%
\caption{\label{figure_renorm} Renormalization factor $Z^{\mathscr{S}\lambda}$ in the notation of Ref.~\cite{Bali:2015ykx} rescaled to the target scale $\unit{2}{\giga\electronvolt}$.}%
\end{figure}%
The further analysis proceeds in the same way for all five values of $\beta$. A more detailed discussion and justification of our method will be given in a dedicated, forthcoming publication. As an example we show in Figure~\ref{figure_renorm} our data for one of the renormalization factors ($Z^{\mathscr{S}\lambda}$ in the notation of Ref.~\cite{Bali:2015ykx}). Three-loop renormalization group running has been used to translate the original data determined at the scale $\mu$ to the target scale $\unit{2}{\giga\electronvolt}$ so that one would expect a plateau up to higher order perturbative (and non-perturbative) contributions and power-law lattice corrections at large values of $\mu a$. The latter are clearly visible and are accounted for in the fit procedure that we outline below.\par%
In order to estimate the systematic uncertainties of our renormalization and mixing coefficients we proceed similarly to Ref.~\cite{Braun:2016wnx} and perform a number of fits of the renormalization scale dependence, varying one element of the analysis at a time: the initial scale~$\mu_1$ of the fit range, the order in perturbation theory used for the calculation of the conversion factors ($n_{\mathrm {loops}}$), and the number~$n_{\mathrm {disc}}$ of terms in the parametrization $A_1a^2\mu^2+\dots+A_{n_{\mathrm{disc}}}(a^2\mu^2)^{n_{\mathrm{disc}}}$ of the lattice artifacts. Furthermore, in order to take into account the uncertainties in the determination of the lattice spacings, the central values of $1/a^2$ following from~\mbox{Table~\ref{table_spacings}} are multiplied by a factor $\lambda^2_{\mathrm{scale}}=1.03$.\footnote{This value contains the scale uncertainty of~$8t_0^*=\mu_{\mathrm{ref}}^{*\,-2}$ given in Ref.~\cite{Bruno:2017gxd} and the largest error of our determination of~$t_0^*/a^2$, added in quadrature.} Finally, also $\Lambda^{\goodopen3\goodclose}_{\scriptscriptstyle\MSbar}=\unit{341(12)}{\mega\electronvolt}$ \cite{Bruno:2017gxd} is varied within its uncertainty. Thus we end up with six types of fits; the different settings are compiled in Table~\ref{table_fits}. In each set we take the results of fit~$1$ as our central values. Defining $\delta_i$, with $i=2,\dots,6$, as the difference between the number based on fit~$i$ and the result based on fit~$1$, we estimate the systematic uncertainties due to the renormalization procedure as $\sqrt{\delta_2^2+(0.5\cdot\delta_3)^2+\delta_4^2+\delta_5^2+\delta_6^2}$. Here we have multiplied~$\delta_3$ by~$1/2$, because going from one loop to two or more loops in the perturbative conversion factors is expected to lead to a smaller change than going from tree level (zero loops) to one-loop accuracy. Since~$\delta_3$ yields by far the largest contribution to the systematic uncertainty of the renormalization factors, replacing~$1/2$ by another factor would change the renormalization errors given in Table~\ref{tab_results} almost proportionally.\par%
\section{Extrapolation strategy and results\label{sect_strategy_and_results}}
\subsection{Global analysis}\label{sect_global_analysis}
To parametrize the quark mass plane we define the convenient linear combinations ($B_0$~is the quark condensate parameter)%
\begin{align}%
\delta m^2 &= m_K^2-m_\pi^2 \approx B_0(m_s-m_\ell)\,,\notag\\
\bar{m}^2  &= (2m_K^2+m_\pi^2)/3 \approx 2B_0(m_s+2m_\ell)/3\,,
\end{align}%
such that $\delta m=0$ corresponds to degenerate light and strange quark masses, i.e., the blue line in Figure~\ref{figure_ensembles}, while the green line of physical average masses is defined by $\bar{m}=\text{phys.}\approx\unit{412}{\mega\electronvolt}$. Along the line of an approximately physical strange quark mass, i.e., the red line in Figure~\ref{figure_ensembles}, the average mass assumes larger values; all ensembles used in this study have $\bar{m}<\unit{500}{\mega\electronvolt}$.\par%
Since the operators we use are not $\mathcal{O}(a)$ improved, the leading discretization effects are linear in the lattice spacing. These are taken into account in our combined extrapolation formula,%
\begin{align}%
\phi_\text{lat} &= \bigl(1+c_\phi^0a+\bar{c}_\phi\bar{m}^2a+\delta c_\phi\delta m^2a\bigr)\phi\,,\notag\\
\phi &= g_\phi(m_\pi,m_K,L)\bigl(\phi^0+\skew{5}{\bar}{\phi}\,\bar{m}^2+\delta\phi\,\delta m^2\bigr)\,, \label{eq_master_fit_formula}
\end{align}%
where $\phi$ is a placeholder for any of the wave function normalization constants defined in Section~\ref{sect_normalization} or for any DA moment defined in Eqs.~\eqref{eq_moments}.\par%
The prefactor $g_\phi$ contains the nonanalytic contributions calculated within covariant baryon chiral perturbation theory (BChPT) at leading one-loop order~\cite{Wein:2015oqa}, supplemented with the corresponding leading hadronic finite volume effects ($L$ is the spatial lattice extent). They are defined such that, in the infinite volume, their chiral limit approaches unity,
\begin{align}%
\lim_{m\to0} g_\phi(m,m,\infty) = 1\,,
\end{align}%
so that $\phi^0$ is indeed the value of the quantity $\phi$ in the chiral limit. The full expressions for these prefactors are lengthy and are given in Appendix~\ref{sect_beauty}.\par%
In order to be consistent with \SU3 flavor symmetry (cf.~Eqs.~\eqref{eq_SU3_leadingtwist} and~\eqref{eq_SU3_highertwist}), the parameters $c_\phi^0$, $\bar c_\phi$, $\phi^0$, and $\smash{\skew{5}{\bar}{\phi}}$ as well as $g_\phi(m,m,L)$ must have the same value for all~$\phi$ within one of the sets%
\begin{figure}[t]%
\includegraphics[width=\columnwidth]{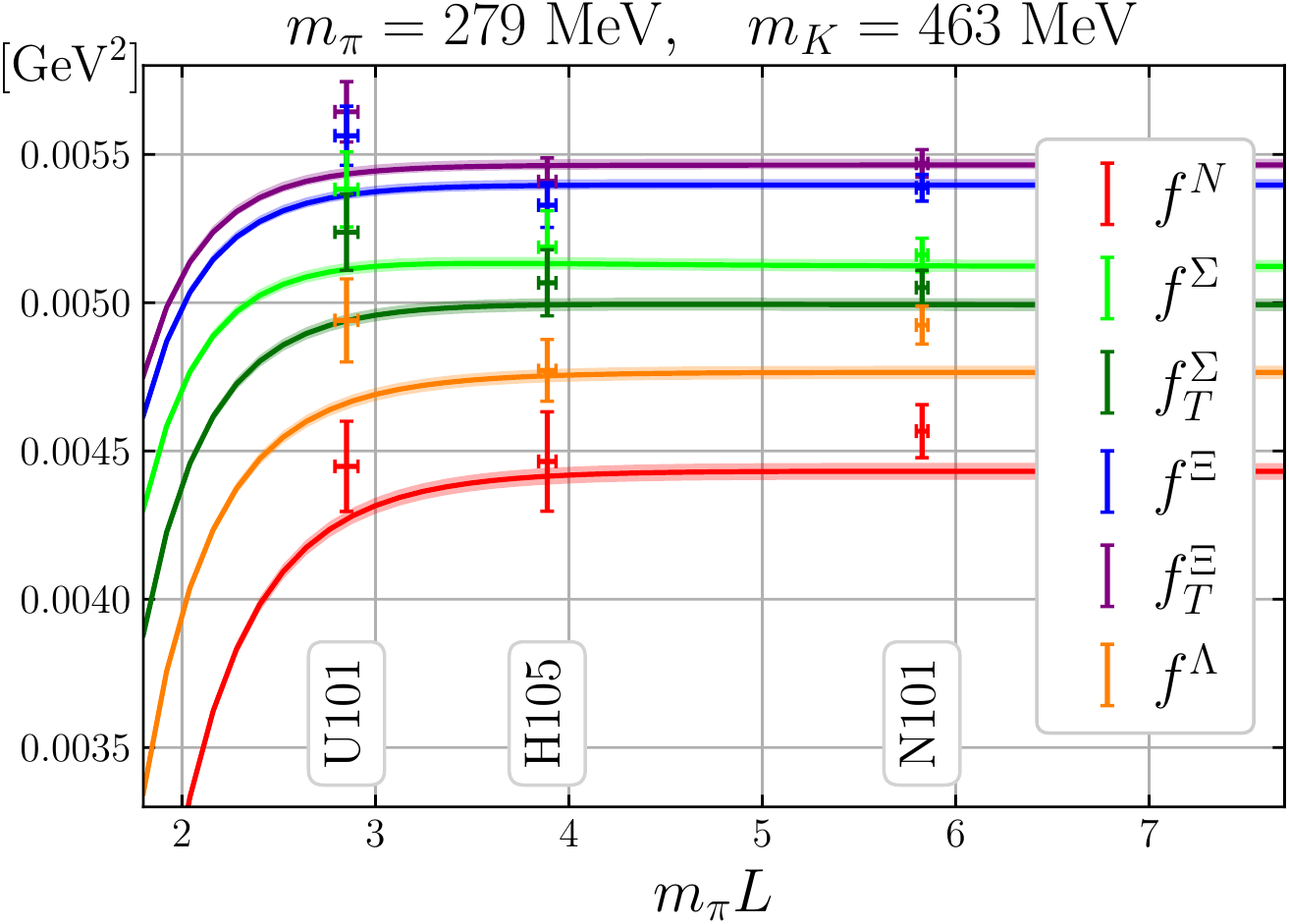}
\caption{Using the example of the leading-twist wave function normalization constants, this plot illustrates the leading hadronic volume effects that are taken into account in our fit function. The used ensembles (U101, H105, and N101) are all at the same point in the quark mass plane and at the same lattice spacing ($a=\unit{0.086}{\femto\meter}$).\label{fig_extrapolate_mpiL}}%
\end{figure}%
\begin{figure*}[t]%
\centering%
\includegraphics[width=.985\textwidth]{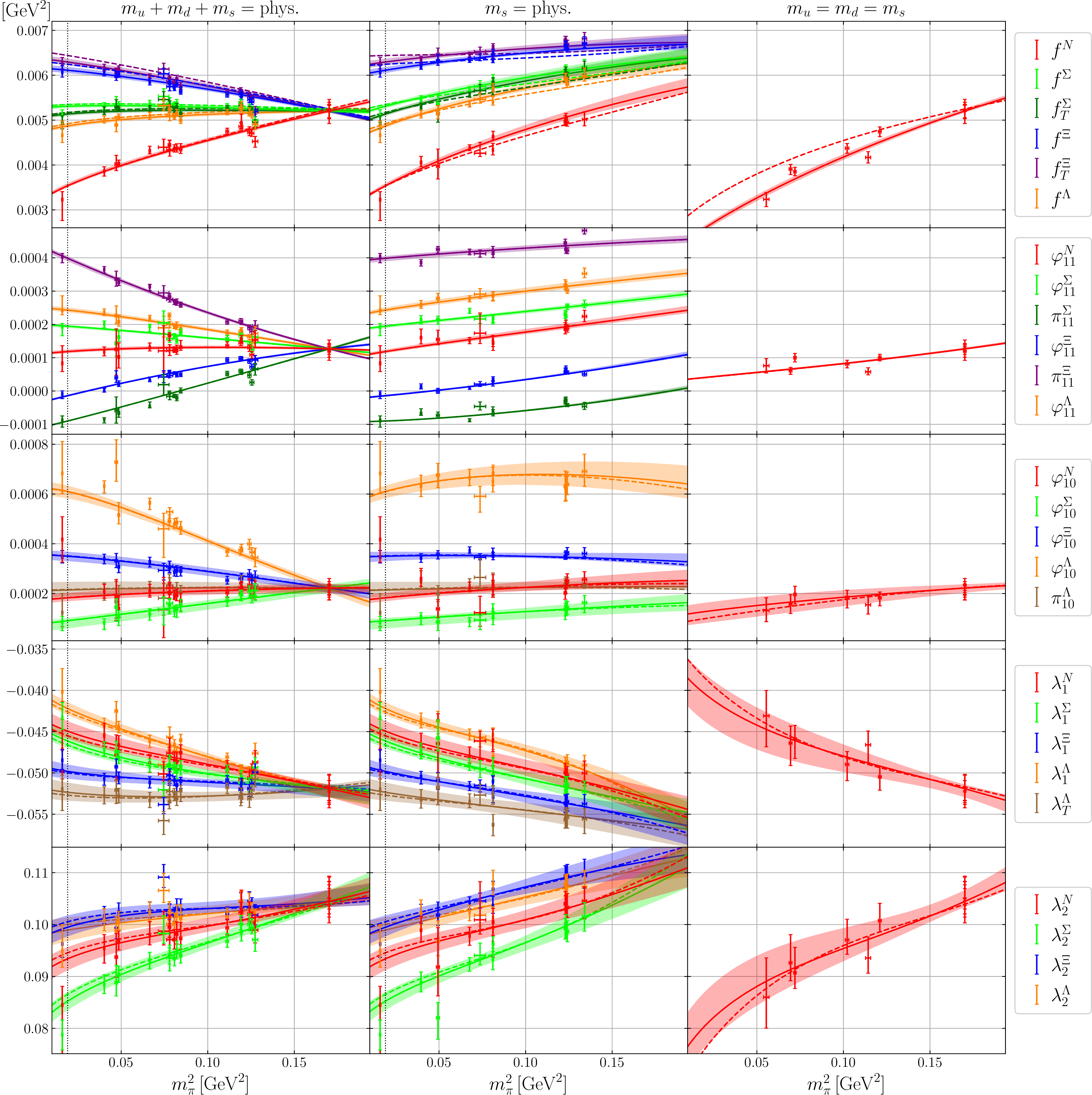}
\caption{The normalization constants and first moments along the three quark mass trajectories shown in Figure~\ref{figure_ensembles} after taking the continuum and the infinite volume limits, plotted as a function of $m_\pi^2$. The plots for individual lattice spacings can be found in Appendix~\ref{sect_plots}. The dotted vertical lines mark the point where the pion and kaon masses assume their physical values. The solid curves and shaded statistical error bands represent our main result. The points shown have been obtained by translating the data along these main fits. The dashed curves correspond to alternative fits (including terms of higher order in $\bar m$), used to estimate the parametrization dependence as described in the text.\label{fig_extrapolate_mpi}}%
\end{figure*}%
\begin{align}%
f &= \bigl\{f^N,f^\Sigma,f_T^\Sigma,f^\Xi,f_T^\Xi,f^\Lambda\bigr\}\,,\notag\\
\varphi_{11} &= \bigl\{\varphi_{11}^N,\varphi_{11}^\Sigma,\pi_{11}^\Sigma,\varphi_{11}^\Xi,\pi_{11}^\Xi,\varphi_{11}^\Lambda\bigr\}\,,\notag\\
\varphi_{10} &= \bigl\{\varphi_{10}^N,\varphi_{10}^\Sigma,\varphi_{10}^\Xi,\varphi_{10}^\Lambda,\pi_{10}^\Lambda\bigr\}\,,\notag\\
\lambda_1 &= \bigl\{\lambda_1^N,\lambda_1^\Sigma,\lambda_1^\Xi,\lambda_1^\Lambda,\lambda_T^\Lambda\bigr\}\,,\notag\\
\lambda_2 &= \bigl\{\lambda_2^N,\lambda_2^\Sigma,\lambda_2^\Xi,\lambda_2^\Lambda\bigr\}\,.
\end{align}%
The parameters $\delta \phi$ are subject to additional constraints on flavor symmetry breaking, derived in Ref.~\cite{Wein:2015oqa}. For the leading-twist wave function normalization constants these read%
\begin{align}%
0 &= \delta\?f^N+\tfrac13\bigl(\delta\?f^\Sigma+2\delta\?f_T^\Sigma\bigr)+\tfrac13\bigl(\delta\?f^\Xi+2\delta\?f_T^\Xi\bigr)\,,\notag\\
0 &= \tfrac13\bigl(\delta\?f^\Sigma+2\delta\?f_T^\Sigma\bigr)+\delta\?f^\Lambda\,,\notag\\
0 &= \delta\?f^\Sigma-\delta\?f_T^\Sigma+\delta\?f^\Xi-\delta\?f_T^\Xi\,.
\end{align}%
For the chiral-odd higher-twist couplings one obtains%
\begin{align}%
0 &= \delta\lambda_1^N+\delta\lambda_1^\Sigma+\delta\lambda_1^\Xi\,,\notag\\
0 &= \delta\lambda_1^\Sigma+\tfrac13\bigl(\delta\lambda_1^\Lambda+2\delta\lambda_T^\Lambda\bigr)\,,
\end{align}%
while in the chiral-even case%
\begin{align}%
0 &= \delta\lambda_2^N+\delta\lambda_2^\Sigma+\delta\lambda_2^\Xi\,,\notag\\
0 &= \delta\lambda_2^\Sigma+\delta\lambda_2^\Lambda\,.
\end{align}%
The constraints for the first moments of the leading-twist DAs are obtained from the equations above by replacing $(f,f_T)\mapsto(\varphi_{11},\pi_{11})$ and $(\lambda_1,\lambda_T)\mapsto(\varphi_{10},\pi_{10})$. There are no analogous constraints for the parameters $\delta c_\phi$, since our fermion formulation explicitly breaks chiral symmetry.\par%
The leading hadronic finite volume effects can be derived from the one-loop BChPT result of Refs.~\cite{Wein:2015oqa,Wein:2016ozo}. They do not entail any additional fit parameters and are contained in the prefactors $g_\phi$ appearing in Eq.~\eqref{eq_master_fit_formula}. In order to study these volume effects we have on purpose also analyzed some ensembles with small $m_\pi L<4$. We find that the volume effects are very small even for lattices with $m_\pi L\approx3$ and are entirely negligible for the vast majority of our ensembles, where $m_\pi L>4$. The effect is illustrated in Figure~\ref{fig_extrapolate_mpiL}, using the leading-twist normalization constants as an example.\par%
\def\figheight{.255\textheight}%
\begin{figure*}[t]%
\begin{minipage}[t]{\columnwidth}%
\raggedleft
\includegraphics[height=\figheight]{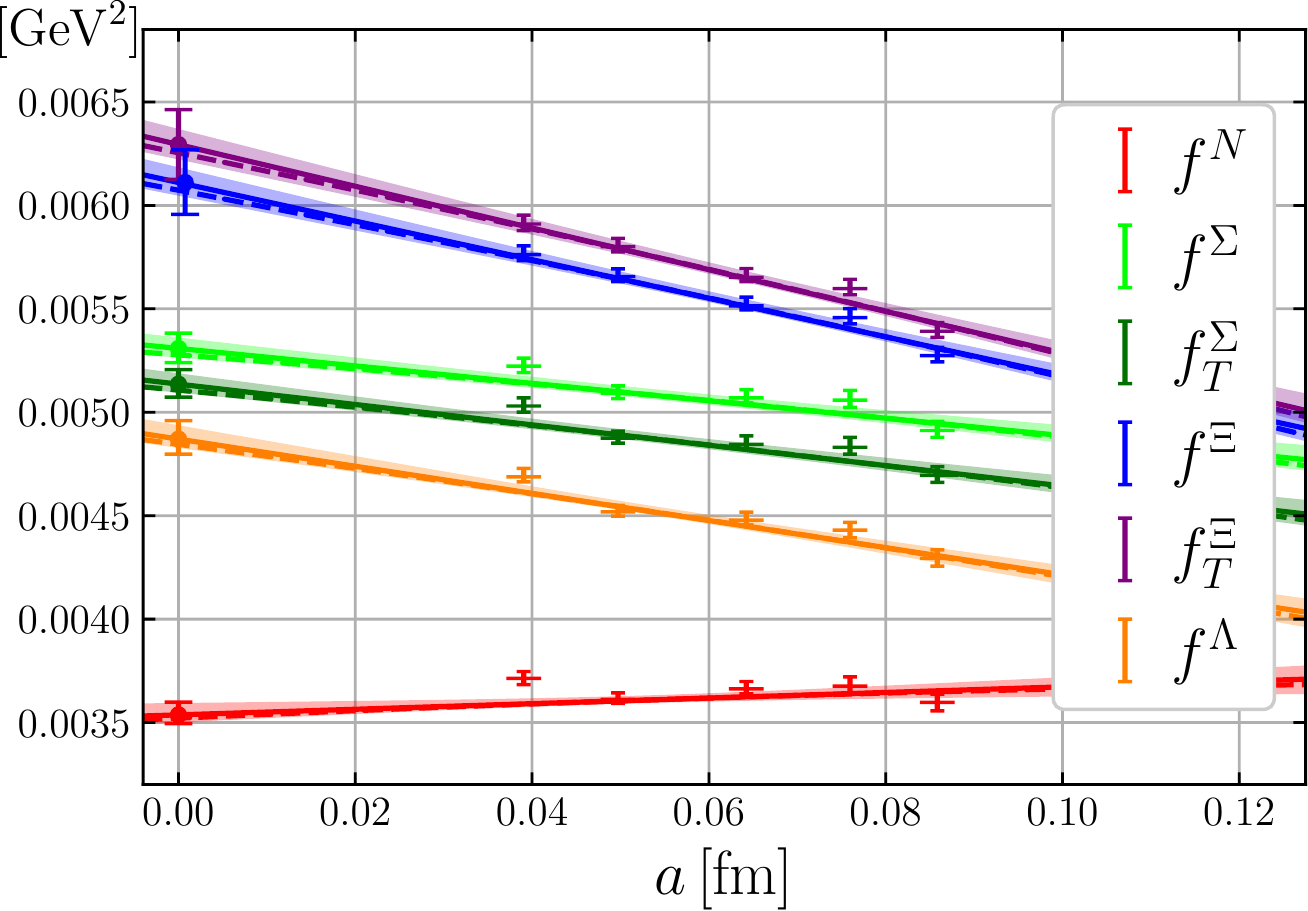}\\[.5\baselineskip]%
\includegraphics[height=\figheight]{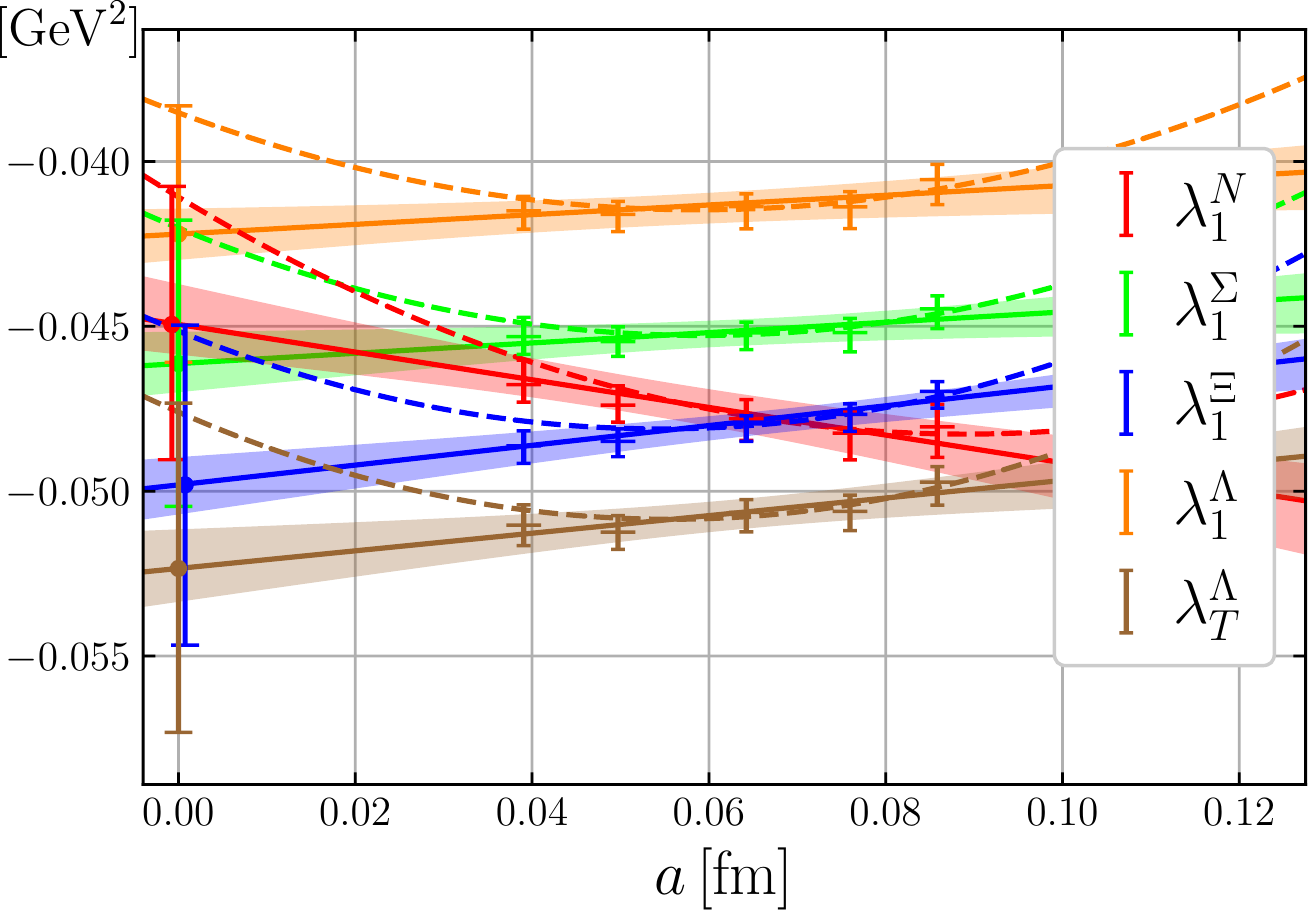}\\[.5\baselineskip]%
\includegraphics[height=\figheight]{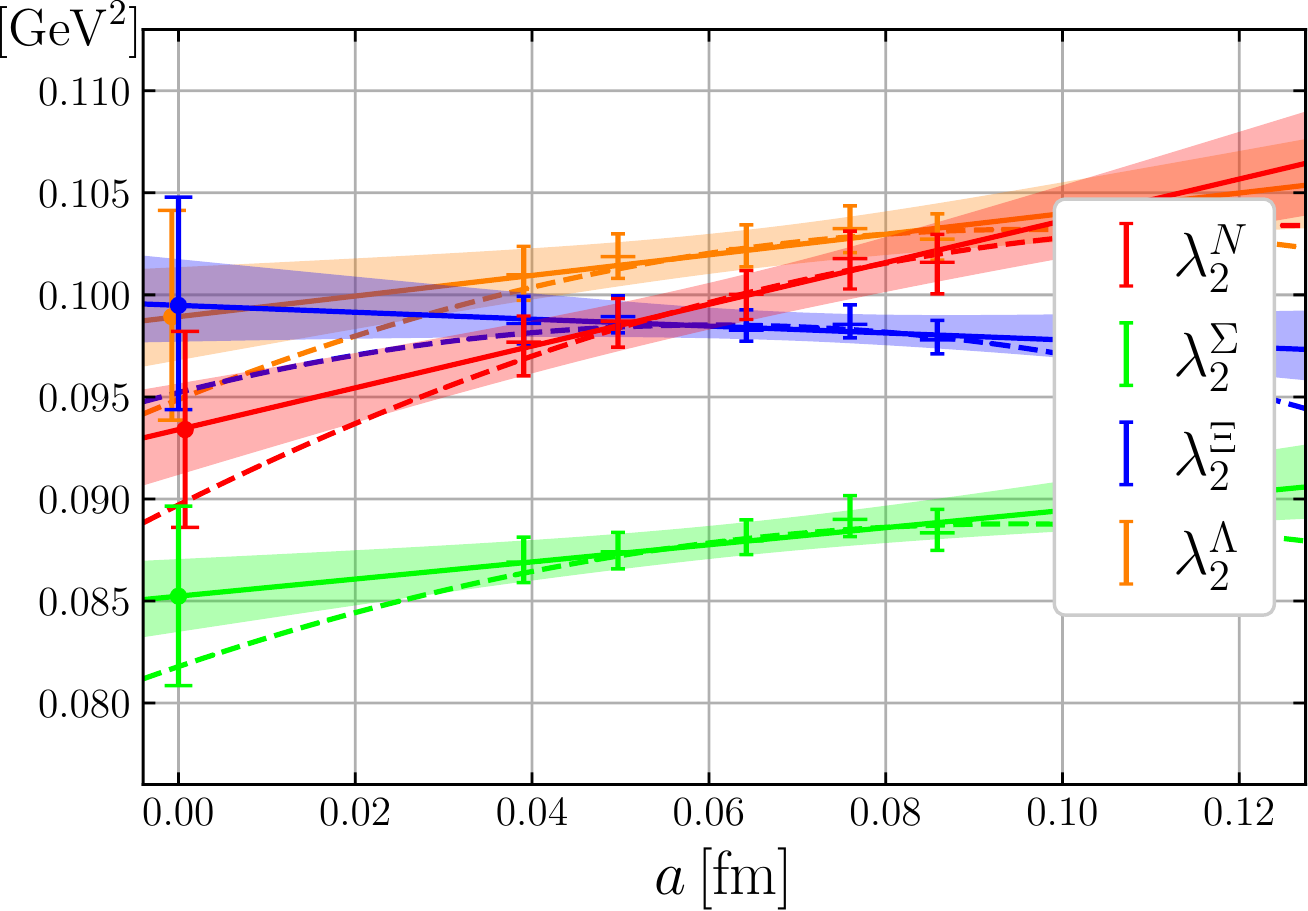}%
\end{minipage}%
\hfill%
\begin{minipage}[t]{\columnwidth}%
\raggedleft
\includegraphics[height=\figheight]{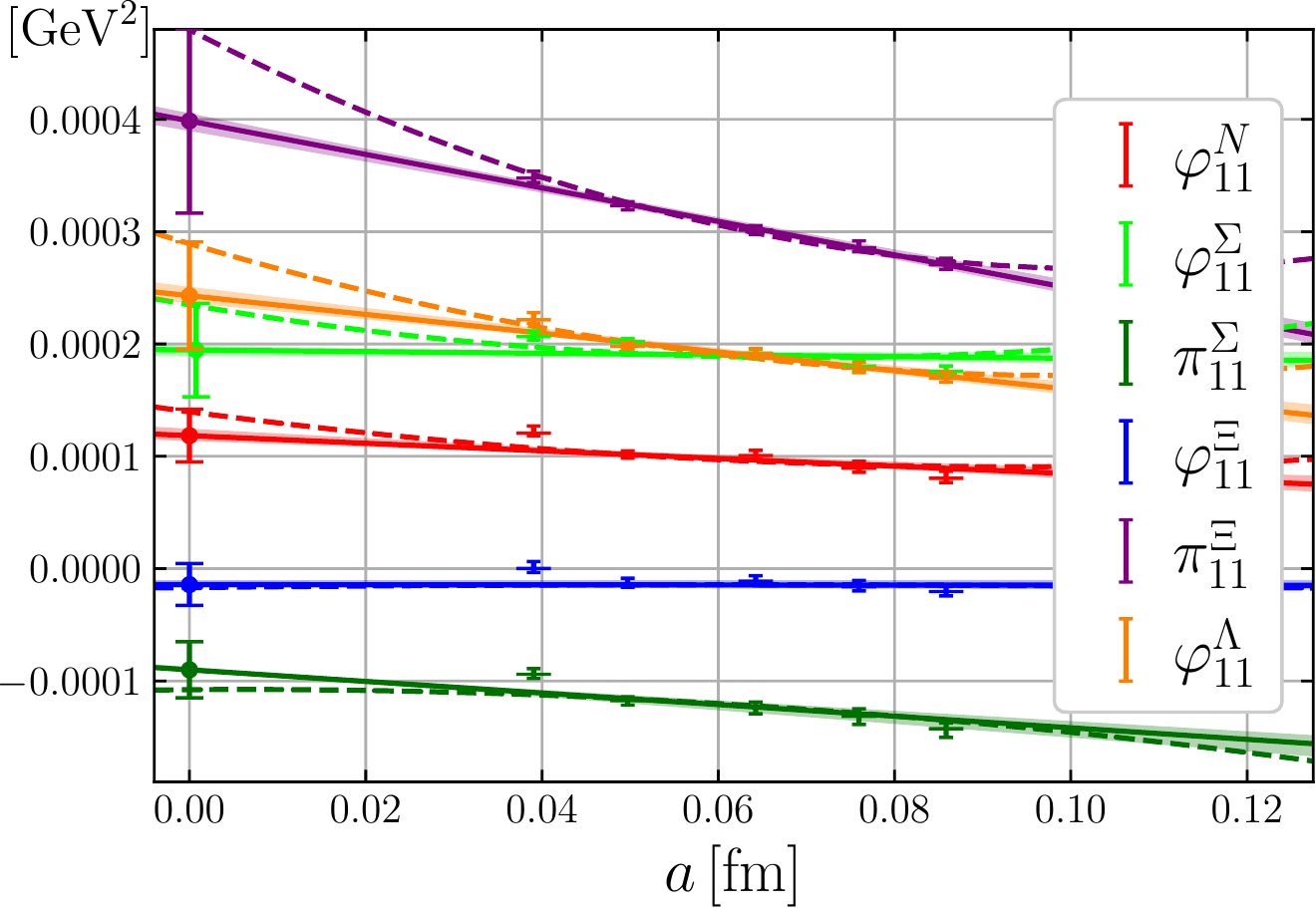}\\[.5\baselineskip]%
\includegraphics[height=\figheight]{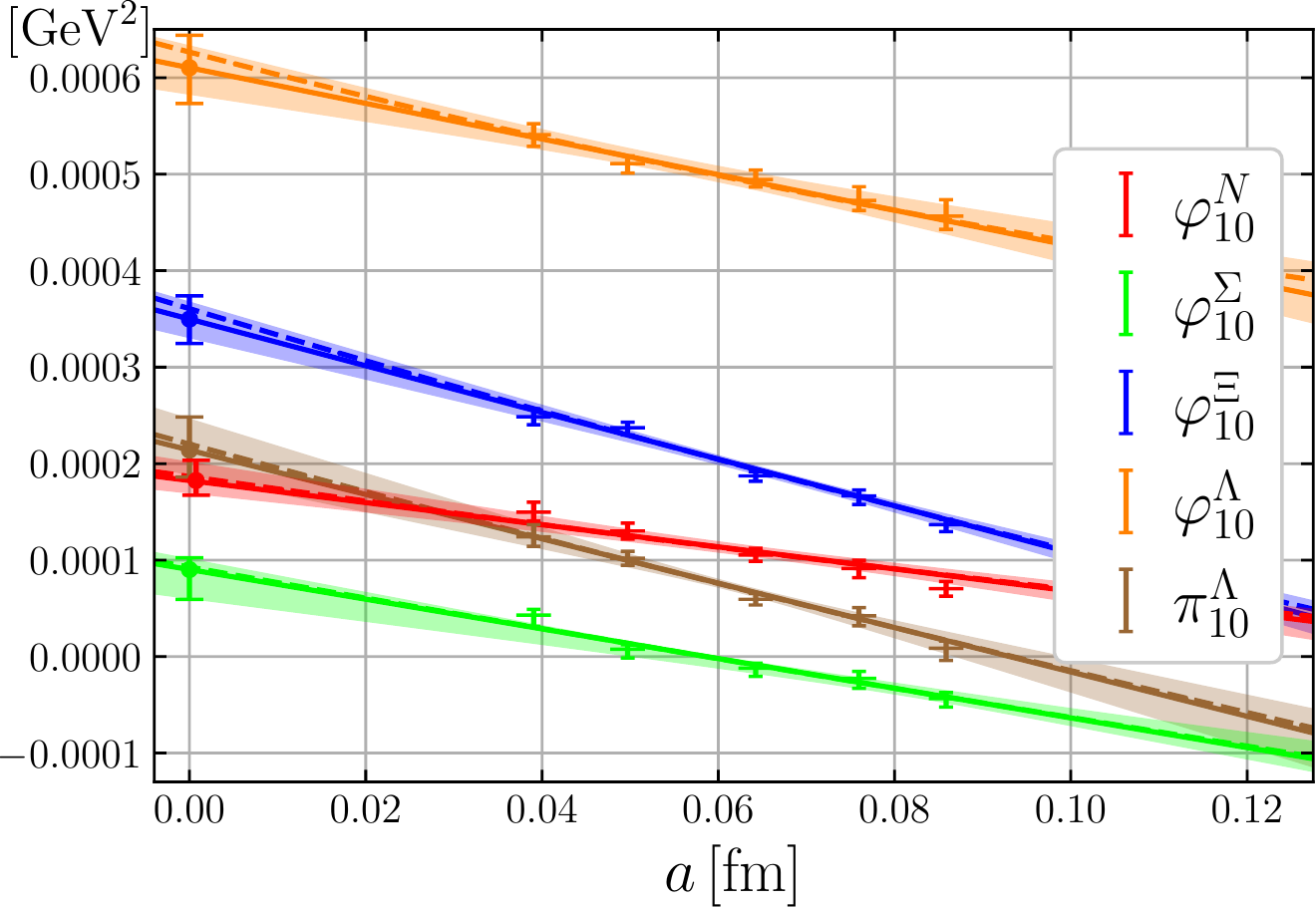}\\[.5\baselineskip]%
\caption{\label{fig_extrapolate_a}Dependence of the normalization constants and first moments on the lattice spacing~$a$, plotted for physical masses and infinite volume. The plots for the individual trajectories can be found in Appendix~\ref{sect_plots}. The solid lines and shaded statistical error bands represent our main result. The points shown have been obtained by translating all data along these main fits (keeping the lattice spacing fixed) and, then, averaging measurements with the same~$a$. From coarsest to finest lattice spacing, this corresponds to averaging the data of~$15$, $7$, $11$, $5$, or just~$2$ independent ensembles, cf. Table~\ref{table_ensembles}. The dashed curves correspond to the mean value of an alternative fit (including terms of higher order in~$a$), used to estimate the parametrization dependence as described in the text. For reference we also plot the final results as points at $a=0$ with all errors added in quadrature.}%
\end{minipage}%
\end{figure*}%
\def\myrule{\rule[-2pt]{0sp}{12pt}}%
\begin{table*}[t]%
\centering%
\caption{Results for the couplings and shape parameters. All values are given in units of $\unit{10^{-3}}{\squaren{\giga\electronvolt}}$ in the $\MSbar$ scheme at a scale $\mu=\unit{2}{\giga\electronvolt}$ with three active quark flavors. The super- and subscripts denote the statistical error after extrapolation to the physical point. The values in parentheses give estimates for the systematic error due to renormalization~($r$), continuum extrapolation~($a$), and chiral extrapolation~($m$). Due to the scale setting uncertainty~\cite{Bruno:2017gxd} all results carry an additional error of~$3\%$ (not displayed). This does not affect dimensionless quantities calculated from ratios of the couplings and shape parameters, such as those displayed in Figure~\ref{fig_barycentric} and Table~\ref{tab_not_momentum_fractions}.\label{tab_results}}%
\begin{widetable}{\textwidth}{lE{.}{.}{-}{2.7}{{}^{+0.0}_{-0.0}(.)_r(.)_a(.)_m}E{.}{.}{-}{2.7}{{}^{+0.0}_{-0.0}(.)_r(.)_a(.)_m}E{.}{.}{-}{2.7}{{}^{+0.0}_{-0.0}(.)_r(.)_a(.)_m}E{.}{.}{-}{2.7}{{}^{+0.0}_{-0.0}(.)_r(.)_a(.)_m}}%
\toprule
\myrule$B$ & \multicolumn{1}{c}{$N$} & \multicolumn{1}{c}{$\Sigma$} & \multicolumn{1}{c}{$\Xi$} & \multicolumn{1}{c}{$\Lambda$}\\
\midrule
\myrule$f^B$            & 3.54^{+6}_{-4}(1)_r(2)_a(0)_m            & 5.31^{+5}_{-4}(1)_r(3)_a(4)_m            & 6.11^{+7}_{-6}(2)_r(4)_a(13)_m           & 4.87^{+7}_{-4}(2)_r(3)_a(5)_m           \\
\myrule$f_T^B$          & 3.54^{+6}_{-4}(1)_r(2)_a(0)_m            & 5.14^{+5}_{-4}(1)_r(3)_a(3)_m            & 6.29^{+8}_{-7}(1)_r(4)_a(15)_m           & \mcemd                                  \\
\myrule$\varphi_{11}^B$ & 0.118^{+6}_{-5}(8)_r(21)_a(0)_m          & 0.195^{+4}_{-6}(10)_r(40)_a(0)_m         & -0.014^{+4}_{-3}(18)_r(3)_a(0)_m         & 0.243^{+8}_{-7}(9)_r(46)_a(0)_m         \\
\myrule$\pi_{11}^B$     & 0.118^{+6}_{-5}(8)_r(21)_a(0)_m          & -0.090^{+3}_{-2}(17)_r(18)_a(0)_m        & 0.399^{+7}_{-9}(9)_r(81)_a(0)_m          & \mcemd                                  \\
\myrule$\varphi_{10}^B$ & 0.182^{+20}_{-14}(6)_r(4)_a(1)_m         & 0.090^{+11}_{-31}(3)_r(3)_a(1)_m         & 0.350^{+18}_{-20}(11)_r(11)_a(1)_m       & 0.610^{+23}_{-28}(18)_r(16)_a(2)_m      \\
\myrule$\pi_{10}^B$     & \mcemd                                   & \mcemd                                   & \mcemd                                   & 0.214^{+33}_{-26}(7)_r(6)_a(2)_m        \\
\midrule
\myrule$\lambda_1^B$    & -44.9^{+1.2}_{-0.9}(0.9)_r(3.9)_a(0.6)_m & -46.1^{+1.0}_{-0.9}(0.9)_r(4.1)_a(0.5)_m & -49.8^{+0.8}_{-0.9}(1.0)_r(4.7)_a(0.2)_m & -42.2^{+0.8}_{-0.8}(0.9)_r(3.7)_a(0.4)_m\\
\myrule$\lambda_T^B$    & \mcemd                                   & \mcemd                                   & \mcemd                                   & -52.3^{+1.2}_{-1.0}(1.1)_r(4.7)_a(0.3)_m\\
\myrule$\lambda_2^B$    & 93.4^{+2.3}_{-2.2}(1.7)_r(3.7)_a(1.2)_m  & 85.2^{+1.8}_{-1.7}(1.6)_r(3.4)_a(1.3)_m  & 99.5^{+2.3}_{-1.8}(1.9)_r(4.3)_a(1.1)_m  & 98.9^{+2.4}_{-2.1}(1.9)_r(4.1)_a(1.1)_m \\
\bottomrule
\end{widetable}%
\end{table*}%
In Figure~\ref{fig_extrapolate_mpi} we display the mass dependence of the normalization constants and the leading-twist shape parameters along the three trajectories shown in Figure~\ref{figure_ensembles}. The results are plotted as a function of $m_\pi^2$ in the continuum limit and for infinite volume, where, for illustrative purposes, the points have been obtained by translating the data along the fitted function. A full set of plots showing the data points at individual lattice spacings can be found in Appendix~\ref{sect_plots}. One should note that all quantities plotted within one row are obtained using a combined simultaneous fit that enforces the flavor and flavor-breaking constraints discussed above. In particular for the leading-twist parameters this leads to a very good description of the data and supports the conjecture made in Ref.~\cite{Bali:2015ykx}, that the violation of flavor-breaking constraints found therein (using only one lattice spacing) was indeed due to mass-dependent discretization effects.\par%
In contrast, a fit to the higher-twist normalization constants using the parametrization~\eqref{eq_master_fit_formula} does not lead to a satisfactory description of the data (even when relaxing the \SU3 breaking constraints). Only after supplementing the chiral expansion in the continuum by hand with higher-order terms ($\bar m^4$, $\bar m^2 \delta m^2$, and $\delta m^4$), the fit describes the data well. This might indicate that the BChPT series for the higher-twist normalizations converges less rapidly than for the leading-twist quantities. To investigate the parametrization dependence, we have performed additional fits including even higher order terms in the average quark masses ($\propto\bar m^4$ for the leading-twist quantities and $\propto\bar m^6$ in the case of the higher-twist couplings). Their mean values are plotted as dashed curves in Figure~\ref{fig_extrapolate_mpi}, and we take the difference between the two extrapolations at the physical point as an estimate for the systematic error due to the chiral extrapolation.\par%
The approach to the continuum limit is depicted in Figure~\ref{fig_extrapolate_a}. All plots show the result at physical pion and kaon masses as a function of the lattice spacing. The points are obtained by averaging all ensembles at a given lattice spacing, after translating them to the physical point along the fitted curve. Note that this procedure is only applied to the points in these plots for illustrative purposes, while the bands are obtained from the actual fit performed using the original data points, cf.\ Appendix~\ref{sect_plots}. It is somewhat amusing that for most observables the nucleon (often used as a benchmark for the whole octet) shows a different behavior than the hyperons. While the leading discretization effects in our extrapolation formula have to be linear in~$a$, higher order effects could also contribute. To study these, we perform another fit (indicated by the dashed curves in Figure~\ref{fig_extrapolate_a}) where we include an additional (mass-independent) term~$\propto a^2$ in the parametrization. The difference with respect to the main fit at the physical point is taken as an estimate for the systematic uncertainty of the extrapolation to the continuum limit.\par%
\subsection{Results}%
\begin{table*}[t]%
\centering%
\caption{\label{tab_cozparison}Comparison of our final results with older $N_f=2+1$ results at a finite lattice spacing~\cite{Bali:2015ykx} ($a=\unit{0.086}{\femto\meter}$), the $N_f=2$ lattice study for the nucleon~\cite{Braun:2014wpa}, and the Chernyak--Ogloblin--Zhitnitsky (COZ) model~\cite{Chernyak:1987nu}. All values are given in units of~$\unit{10^{-3}}{\squaren{\giga\electronvolt}}$. All quantities have been converted to the conventions employed in this work and rescaled to $\mu=\unit{2}{\giga\electronvolt}$, cf.\ Ref.~\cite{Bali:2015ykx}. Note that the quantity called~$f^T_\Lambda$ in Ref.~\cite{Chernyak:1987nu} is proportional to the first moment~$\pi_{10}^\Lambda$ in our nomenclature. In Ref.~\cite{Braun:2014wpa} the normalization constants of the nucleon have been extrapolated to the continuum limit, while the first moments correspond to $a=\unit{0.06\text{--}0.08}{\femto\meter}$. As discussed in detail in the main text, only the error estimates of the present $N_f=2+1$ $(a\rightarrow0)$ result contain all relevant systematic uncertainties, cf.\ Table~\ref{tab_results}.}%
\begin{widetable}{\textwidth}{lclE{.}{.}{}{1.2}{{}^{+00}_{-00}}E{.}{.}{}{1.2}{{}^{+00}_{-00}}E{.}{.}{-}{1.3}{{}^{+00}_{-00}}E{.}{.}{-}{1.3}{{}^{+00}_{-00}}E{.}{.}{-}{1.3}{{}^{+00}_{-00}}E{.}{.}{}{1.3}{{}^{+00}_{-00}}E{.}{.}{-}{2.1}{{}^{+0.0}_{-0.0}}E{.}{.}{-}{2.1}{{}^{+0.0}_{-0.0}}E{.}{.}{}{3.1}{{}^{+0.0}_{-0.0}}}%
\toprule
$B$ & \clap{work} & method & \multicolumn{1}{c}{$f^B$} & \multicolumn{1}{c}{$f_T^B$} & \multicolumn{1}{c}{$\varphi_{11}^B$} & \multicolumn{1}{c}{$\pi_{11}^B$} & \multicolumn{1}{c}{$\varphi_{10}^B$} & \multicolumn{1}{c}{$\pi_{10}^B$} & \multicolumn{1}{c}{$\lambda_1^B$} & \multicolumn{1}{c}{$\lambda_T^B$} & \multicolumn{1}{c}{$\lambda_2^B$}\\
\midrule
\multirow{4}{*}{$N$}
&\text{this}                   & $N_f=2+1$ $(a\rightarrow0)$ & 3.54^{+6}_{-4}    & 3.54^{+6}_{-4}   & 0.118^{+24}_{-23} & 0.118^{+24}_{-23} & 0.182^{+21}_{-15} &\mcemd &-44.9^{+4.2}_{-4.1} &\mcemd& 93.4^{+4.8}_{-4.8}\\
&\text{\cite{Bali:2015ykx}}    & $N_f=2+1$ $(a\neq0)$        & 3.60^{+6}_{-6}    & 3.60^{+6}_{-6}   & 0.080^{+20}_{-20} & 0.080^{+20}_{-20} & 0.060^{+20}_{-20} &\mcemd &-49.0^{+2.0}_{-2.0} &\mcemd& 98.0^{+5.0}_{-5.0}\\
&\text{\cite{Braun:2014wpa}}   & $N_f=2$                     & 2.84^{+33}_{-33}  & 2.84^{+33}_{-33} & 0.085^{+21}_{-21} & 0.085^{+21}_{-21} & 0.082^{+29}_{-29} &\mcemd &-41.3^{+2.0}_{-2.0} &\mcemd& 81.9^{+4.0}_{-4.0}\\
&\text{\cite{Chernyak:1987nu}} & COZ                         & 4.55 & 4.55 & 0.885 & 0.885 & 0.748  &\mcemd &\timex&\mcemd&\timex\\
\midrule
\multirow{3}{*}{$\Sigma$}
&\text{this}                   & $N_f=2+1$ $(a\rightarrow0)$ & 5.31^{+7}_{-7} & 5.14^{+7}_{-6} & 0.195^{+41}_{-42} &-0.090^{+25}_{-25} & 0.090^{+12}_{-31} &\mcemd &-46.1^{+4.4}_{-4.3}   &\mcemd& 85.2^{+4.4}_{-4.4}\\
&\text{\cite{Bali:2015ykx}}    & $N_f=2+1$ $(a\neq0)$        & 5.07^{+5}_{-5} & 4.88^{+5}_{-5} & 0.170^{+20}_{-20} &-0.100^{+30}_{-30} &-0.069^{+10}_{-10} &\mcemd &-45.4^{+2.1}_{-2.1}   &\mcemd& 86.0^{+4.0}_{-4.0}\\
&\text{\cite{Chernyak:1987nu}} & COZ                         & 4.65 & 4.46 & 1.11  & 0.511 & 0.523 &\mcemd &\timex&\mcemd&\timex\\
\midrule
\multirow{3}{*}{$\Xi$}
&\text{this}                   & $N_f=2+1$ $(a\rightarrow0)$ & 6.11^{+16}_{-15} & 6.29^{+17}_{-17} &-0.014^{+19}_{-19} & 0.399^{+82}_{-82} & 0.350^{+24}_{-25} &\mcemd &-49.8^{+4.8}_{-4.9} &\mcemd& 99.5^{+5.3}_{-5.1}\\
&\text{\cite{Bali:2015ykx}}    & $N_f=2+1$ $(a\neq0)$        & 5.38^{+5}_{-5}   & 5.47^{+5}_{-5}   & 0.010^{+20}_{-20} & 0.300^{+10}_{-10} & 0.140^{+10}_{-10} &\mcemd &-47.6^{+2.3}_{-2.3} &\mcemd& 96.0^{+5.0}_{-5.0}\\
&\text{\cite{Chernyak:1987nu}} & COZ                         & 4.83 & 4.92 & 0.685 & 1.10  & 0.883 &\mcemd &\timex&\mcemd&\timex\\
\midrule
\multirow{3}{*}{$\Lambda$}
&\text{this}                   & $N_f=2+1$ $(a\rightarrow0)$ & 4.87^{+9}_{-7}  &\mcemd& 0.243^{+48}_{-48} &\mcemd & 0.610^{+34}_{-37} & 0.214^{+34}_{-28} &-42.2^{+3.9}_{-3.9} &-52.3^{+5.0}_{-5.0} & 98.9^{+5.2}_{-5.1}\\
&\text{\cite{Bali:2015ykx}}    & $N_f=2+1$ $(a\neq0)$        & 4.38^{+6}_{-6}  &\mcemd& 0.180^{+10}_{-10} &\mcemd & 0.480^{+40}_{-40} & 0.010^{+16}_{-16} &-39.0^{+2.0}_{-2.0} &-51.0^{+2.0}_{-2.0} &101.0^{+5.0}_{-5.0}\\
&\text{\cite{Chernyak:1987nu}} & COZ                         & 4.69 &\mcemd& 1.05  &\mcemd & 1.39  & 1.32  &\timex&\timex&\timex\\
\bottomrule
\end{widetable}%
\end{table*}%
While we find that discretization effects are important for the normalization constants (up to $\sim 20\%$), they can have a game-changing impact on the moments. From the right two panels of Figure~\ref{fig_extrapolate_a} it can be seen that between $a=\unit{0.086}{\femto\meter}$ (our coarsest lattice spacing) and $a=0$ there can be huge variations of the moments that can even affect the sign, e.g., of~$\varphi_{10}^\Sigma$. In particular we find a nonzero value for $\pi_{10}^\Lambda$ in the continuum, while its value crosses zero at our coarsest lattice spacing. Hence, finding the distribution amplitude $T^\Lambda$ to be strongly suppressed in our exploratory study~\cite{Bali:2015ykx} was, in hindsight, just a coincidence (the study was conducted using only ensembles with $a=\unit{0.086}{\femto\meter}$). In addition, there can also be sign changes in the chiral extrapolation (cf.~$\varphi_{11}^\Xi$ and~$\pi_{11}^\Sigma$ in the second row of Figure~\ref{fig_extrapolate_mpi}). This, together with the fact that quark mass and discretization effects are entangled, highlights the importance of performing a simultaneous extrapolation to obtain meaningful results.\par%
In the $N_f=2$ lattice study~\cite{Braun:2014wpa} a first continuum extrapolation had been carried out for the normalization constants of the nucleon using three lattice spacings in the narrow range $a\approx\unit{0.06-0.08}{\femto\metre}$. This resulted in a decrease of~$f^N$ by~$\approx 30\%$ and a somewhat smaller decrease for~$\lvert\lambda_{1,2}^N\rvert$ going from the largest spacing to the continuum limit. Using the more refined analysis method described in Section~\ref{sect_global_analysis} and employing a much larger dataset (with a different lattice action, see Section~\ref{sect_ens}) we observe only changes of~$\lesssim5\%$ for the nucleon normalization constants, while for the hyperons effects of~$\sim20\%$ are not uncommon, cf.\ Figure~\ref{fig_extrapolate_a}.\par%
In Table~\ref{tab_results} we summarize the results for the normalization constants and the first moments at the physical point obtained from the simultaneous treatment of finite volume, quark mass, and discretization effects described above, including estimates for all relevant uncertainties. In Table~\ref{tab_cozparison} we compare our values with our earlier $N_f=2+1$ results at a finite lattice spacing~\cite{Bali:2015ykx}, with the \mbox{$N_f=2$} \mbox{lattice} study for the nucleon~\cite{Braun:2014wpa}, and with the values used in the Chernyak--Ogloblin--Zhitnitsky (COZ) model~\cite{Chernyak:1987nu}. The errors given in Table~\ref{tab_cozparison} have been obtained by adding up all provided errors in quadrature. The errors given in Ref.~\cite{Bali:2015ykx} do not include estimates for the parametrization dependence of the chiral extrapolation and for discretization effects. Considering that our present results are continuum extrapolated while those in~\cite{Bali:2015ykx} were not, one would not expect an agreement within errors. The error estimates given for the $N_f=2$ study~\cite{Braun:2014wpa} contain the statistical uncertainty, and, in the case of the normalization constants, a rough estimate of the systematic uncertainty due to the continuum extrapolation. For the COZ model~\cite{Chernyak:1987nu} no error estimate is provided.\par%
For the first order shape parameters of the leading-twist DA of the nucleon, $\varphi_{11}^N=\pi_{11}^N$ and  $\varphi_{10}^N$, our results are larger than those from our earlier lattice studies that did not include a controlled continuum extrapolation~\cite{Braun:2014wpa,Bali:2015ykx}.\footnote{Our $\varphi^N_{nk}$ correspond to the products $f_N \varphi^N_{nk}$ in Ref.~\cite{Braun:2014wpa}.} In particular, the previously observed approximate equality $\varphi_{10}^N\approx \varphi_{11}^N$ does not hold after taking the continuum limit. As reported in Ref.~\cite{Braun:2014wpa} for the nucleon and in Ref.~\cite{Bali:2015ykx} for hyperons, lattice simulations and light-cone sum rule calculations (cf., e.g., Ref.~\cite{Anikin:2013aka}) yield estimates of the first moments of leading-twist DAs that are significantly smaller than values obtained from traditional SVZ sum rules and models derived therefrom, cf. Refs.~\cite{Chernyak:1983ej,Chernyak:1987nu} (see, e.g., the comparison to the COZ model in Table~\ref{tab_cozparison}). Our final results confirm these findings, even though we observe a general upward trend for the first moments in the continuum extrapolation (cf.\ the right panels of Figure~\ref{fig_extrapolate_a}).\par%
It is notable that the \SU3 breaking in octet baryon DAs turns out to be very large. Some shape parameters even assume opposite signs for different baryons at the physical point. The effect on the leading-twist normalization constants can be as large as~$80\%$, for instance \mbox{$(f_T^\Xi - f^N)/f^N\approx 0.78$}, and is much stronger than estimated in QCD sum rule calculations~\cite{Chernyak:1987nu} where only a $\lesssim 10\%$ \SU3 breaking was found. For the shape parameters we find these effects to be even more pronounced such that, as a consequence, \SU3 breaking in hard exclusive reactions that are sensitive to the deviations of the DAs from their asymptotic form can be even further enhanced.\par
\begin{figure*}[p]%
\definecolor{mygreen}{rgb}{0,0.666667,0}%
\newlength{\barycentricwidth}\setlength{\barycentricwidth}{0.4523\textwidth}%
\includegraphics[width=\barycentricwidth]{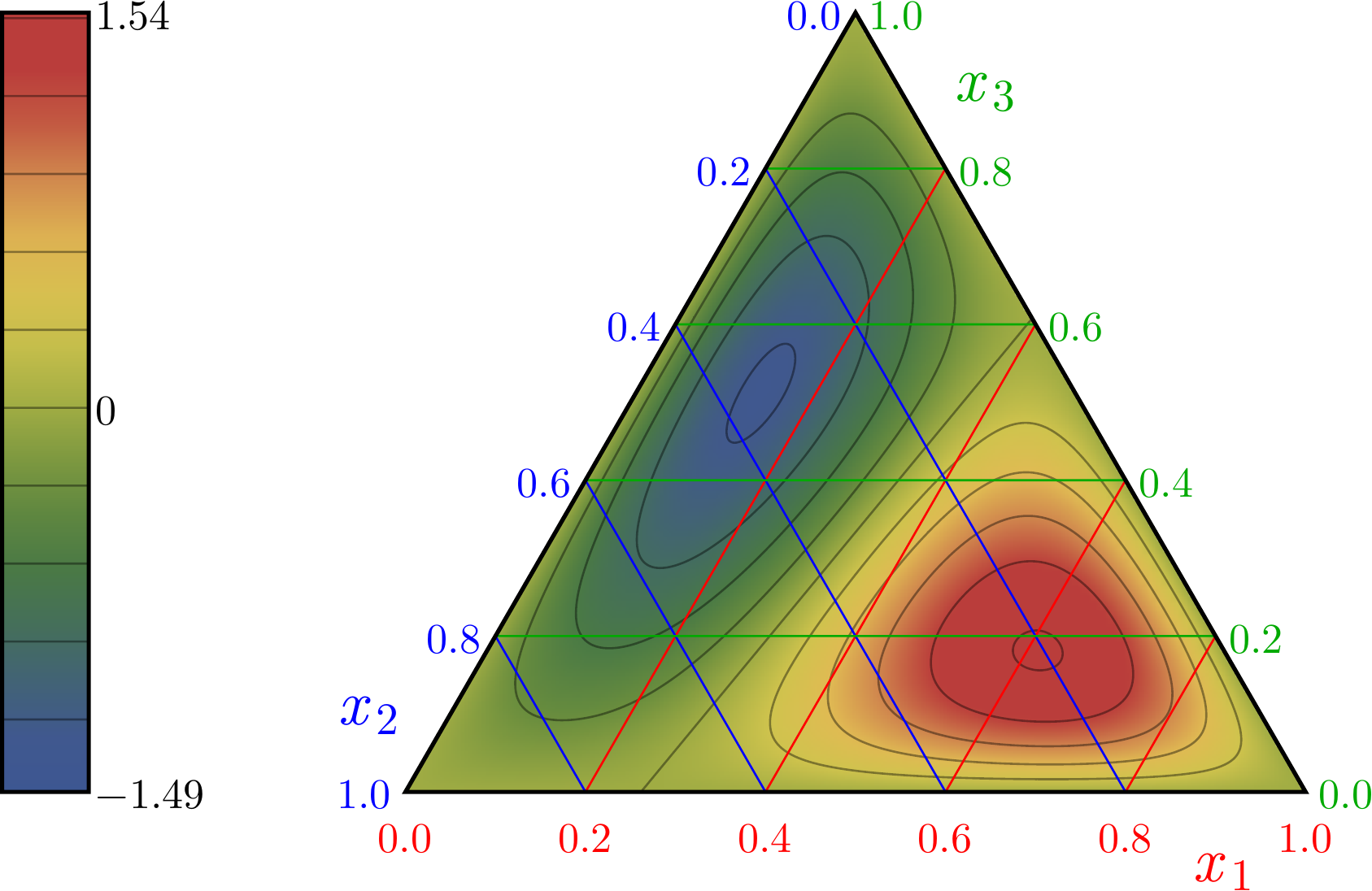}%
\llap{\raisebox{0.6419\barycentricwidth-\height}{\rlap{\clap{\boxed{\tfrac{[\VmA]^{\mathrlap{N}}}{f^N}-\phi^{\text{as}}}}}\hspace{0.6838\barycentricwidth}}}%
\llap{\raisebox{0.6419\barycentricwidth-\height}{\fock{\color{red}u^\gooduparrow \color{blue}u^\gooddownarrow \color{mygreen}d^\gooduparrow}}}\hfill%
\includegraphics[width=\barycentricwidth]{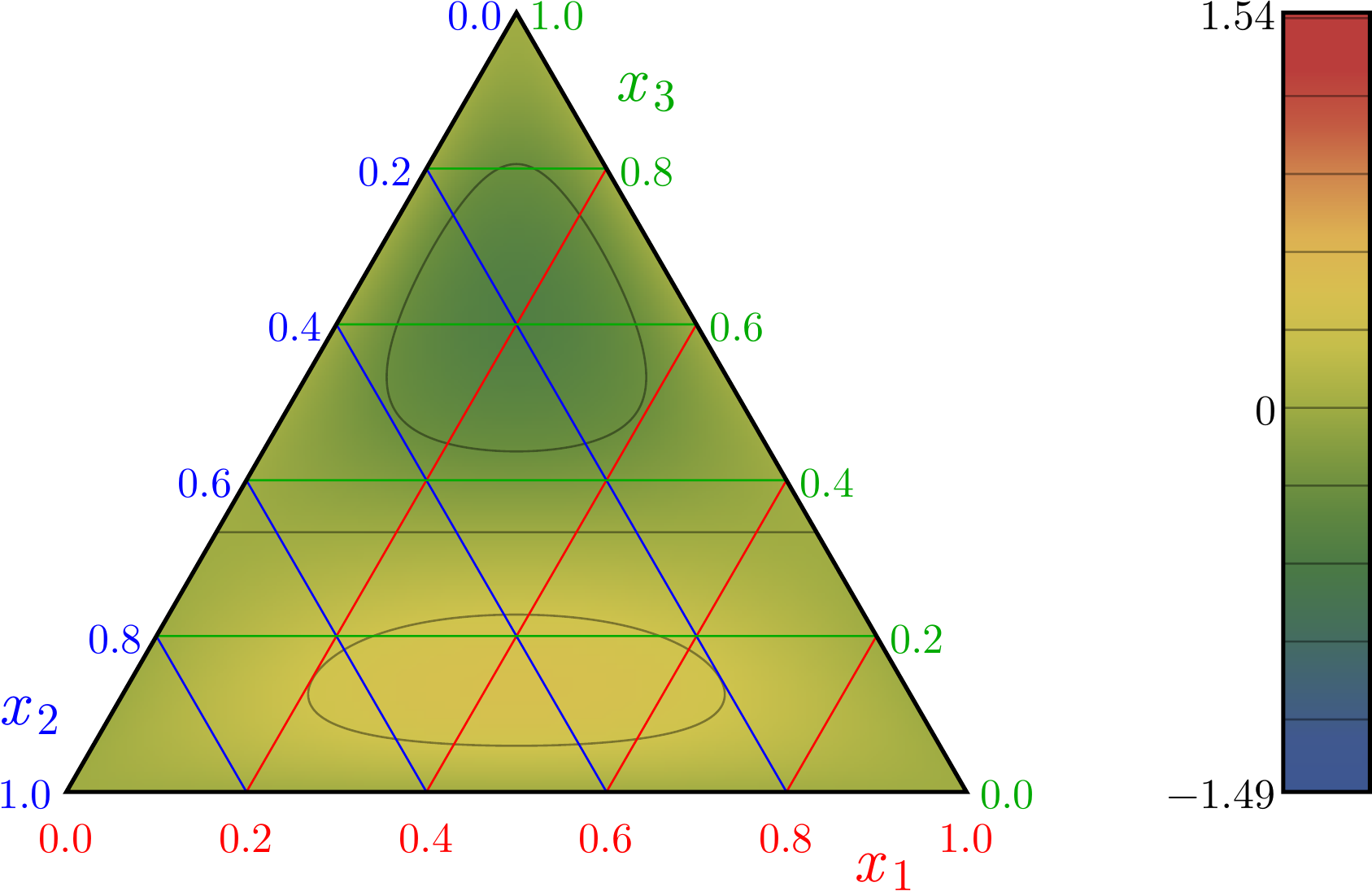}%
\llap{\raisebox{0.6419\barycentricwidth-\height}{\rlap{\fock{\color{red}u^\gooduparrow \color{blue}u^\gooduparrow \color{mygreen}d^\gooddownarrow}}\hspace{\barycentricwidth}}}%
\llap{\raisebox{0.6419\barycentricwidth-\height}{\clap{\boxed{\tfrac{T^N}{f^N}-\phi^{\text{as}}}}\hspace{0.3162\barycentricwidth}}}%
\\[\baselineskip]%
\includegraphics[width=\barycentricwidth]{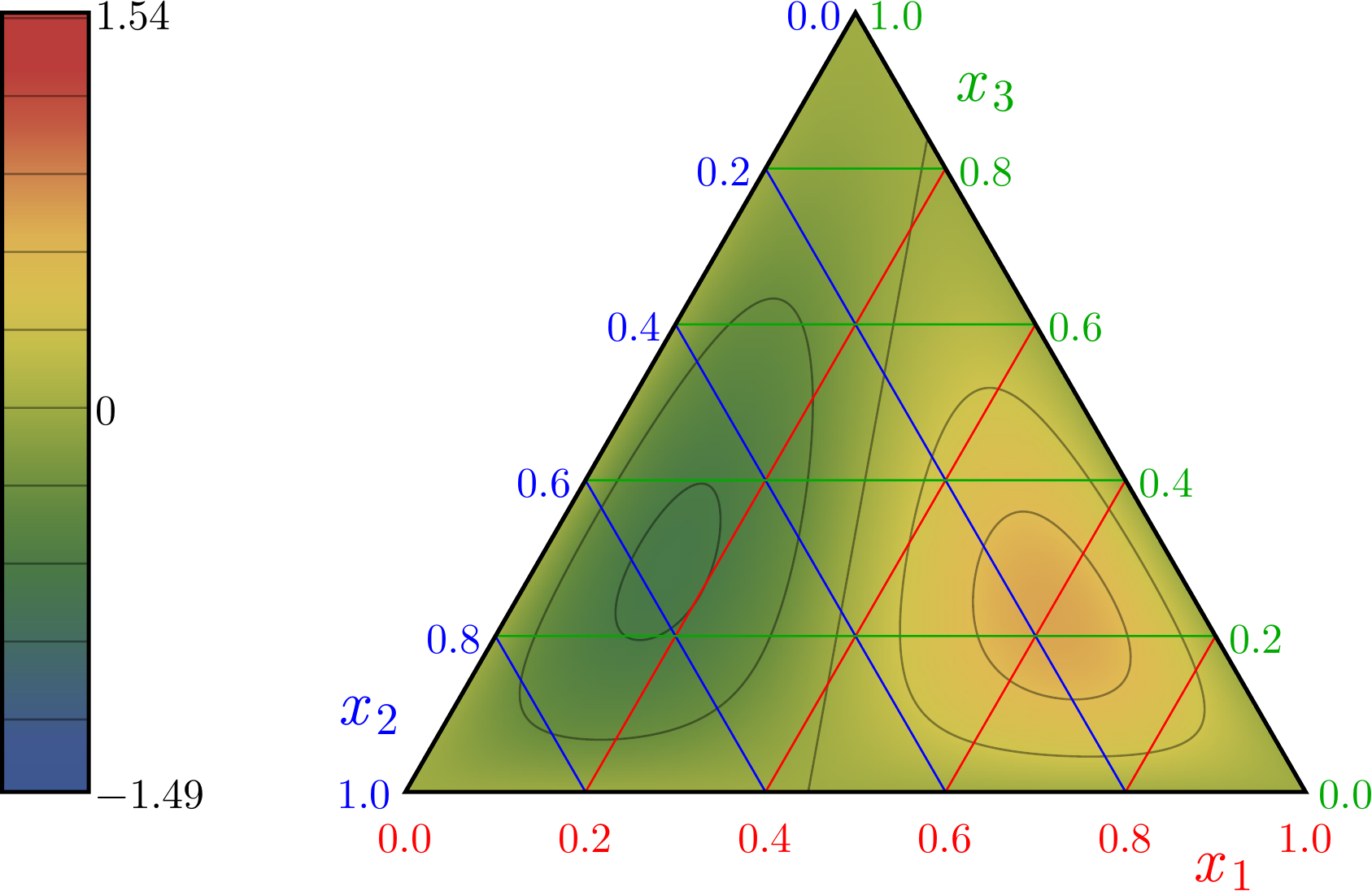}%
\llap{\raisebox{0.6419\barycentricwidth-\height}{\rlap{\clap{\boxed{\tfrac{[\VmA]^{\mathrlap{\Sigma}}}{f^\Sigma}-\phi^{\text{as}}}}}\hspace{0.6838\barycentricwidth}}}%
\llap{\raisebox{0.6419\barycentricwidth-\height}{\fock{\color{red}d^\gooduparrow \color{blue}d^\gooddownarrow \color{mygreen}s^\gooduparrow}}}\hfill%
\includegraphics[width=\barycentricwidth]{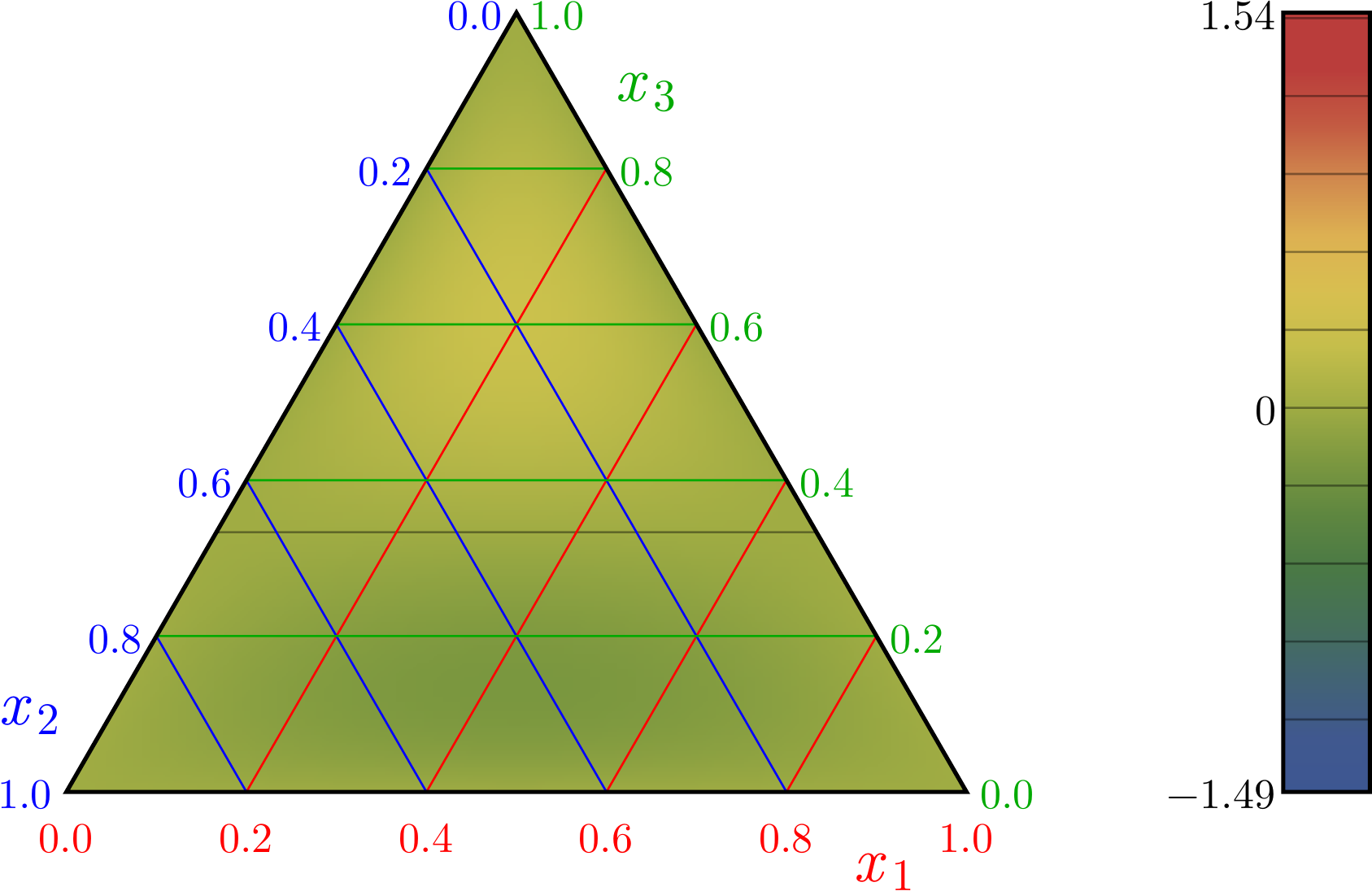}%
\llap{\raisebox{0.6419\barycentricwidth-\height}{\rlap{\fock{\color{red}d^\gooduparrow \color{blue}d^\gooduparrow \color{mygreen}s^\gooddownarrow}}\hspace{\barycentricwidth}}}%
\llap{\raisebox{0.6419\barycentricwidth-\height}{\clap{\boxed{\tfrac{T^\Sigma}{f_T^\Sigma}-\phi^{\text{as}}}}\hspace{0.3162\barycentricwidth}}}%
\\[\baselineskip]%
\includegraphics[width=\barycentricwidth]{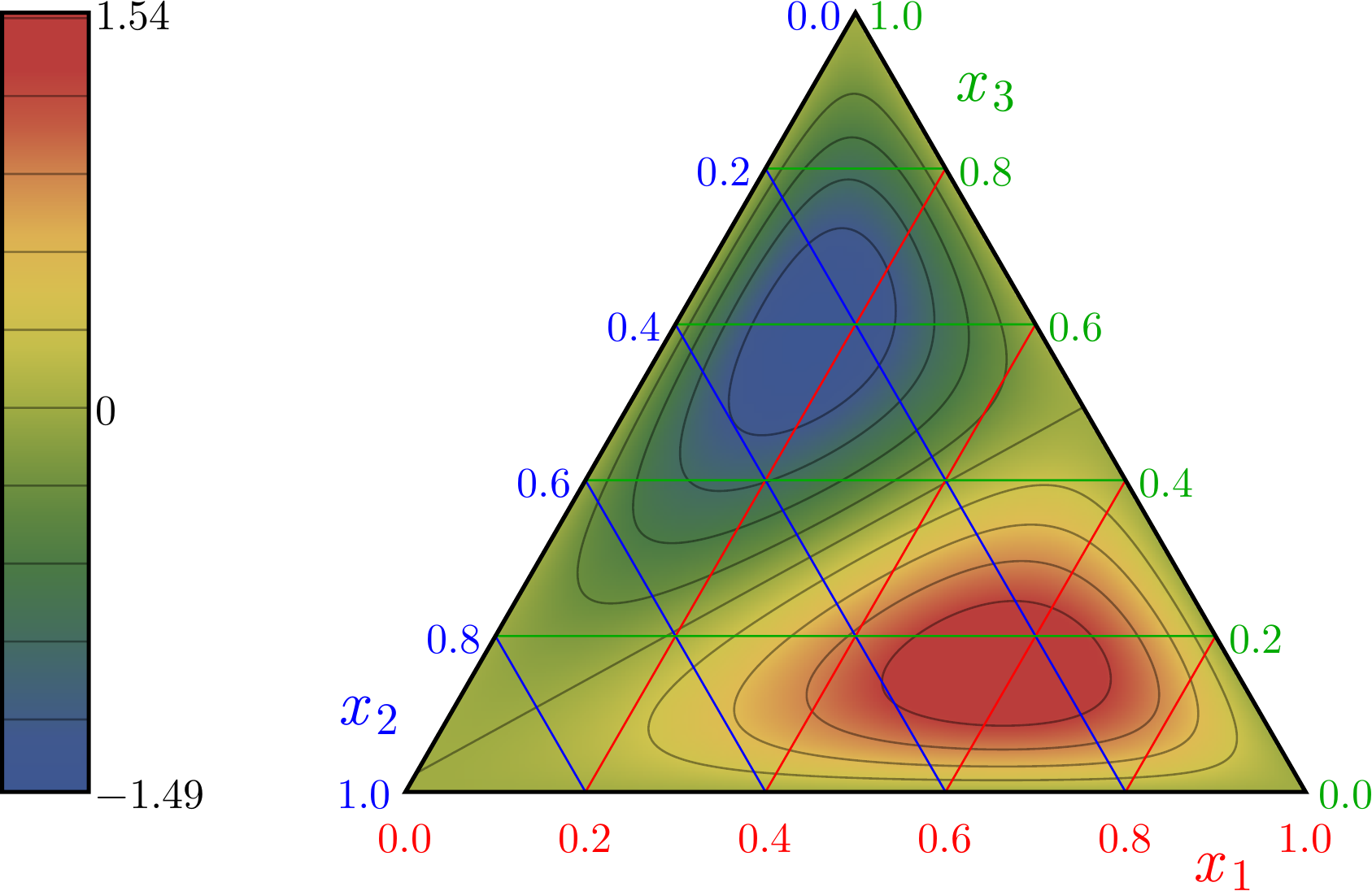}%
\llap{\raisebox{0.6419\barycentricwidth-\height}{\rlap{\clap{\boxed{\tfrac{[\VmA]^{\mathrlap{\Xi}}}{f^\Xi}-\phi^{\text{as}}}}}\hspace{0.6838\barycentricwidth}}}%
\llap{\raisebox{0.6419\barycentricwidth-\height}{\fock{\color{red}s^\gooduparrow \color{blue}s^\gooddownarrow \color{mygreen}u^\gooduparrow}}}\hfill%
\includegraphics[width=\barycentricwidth]{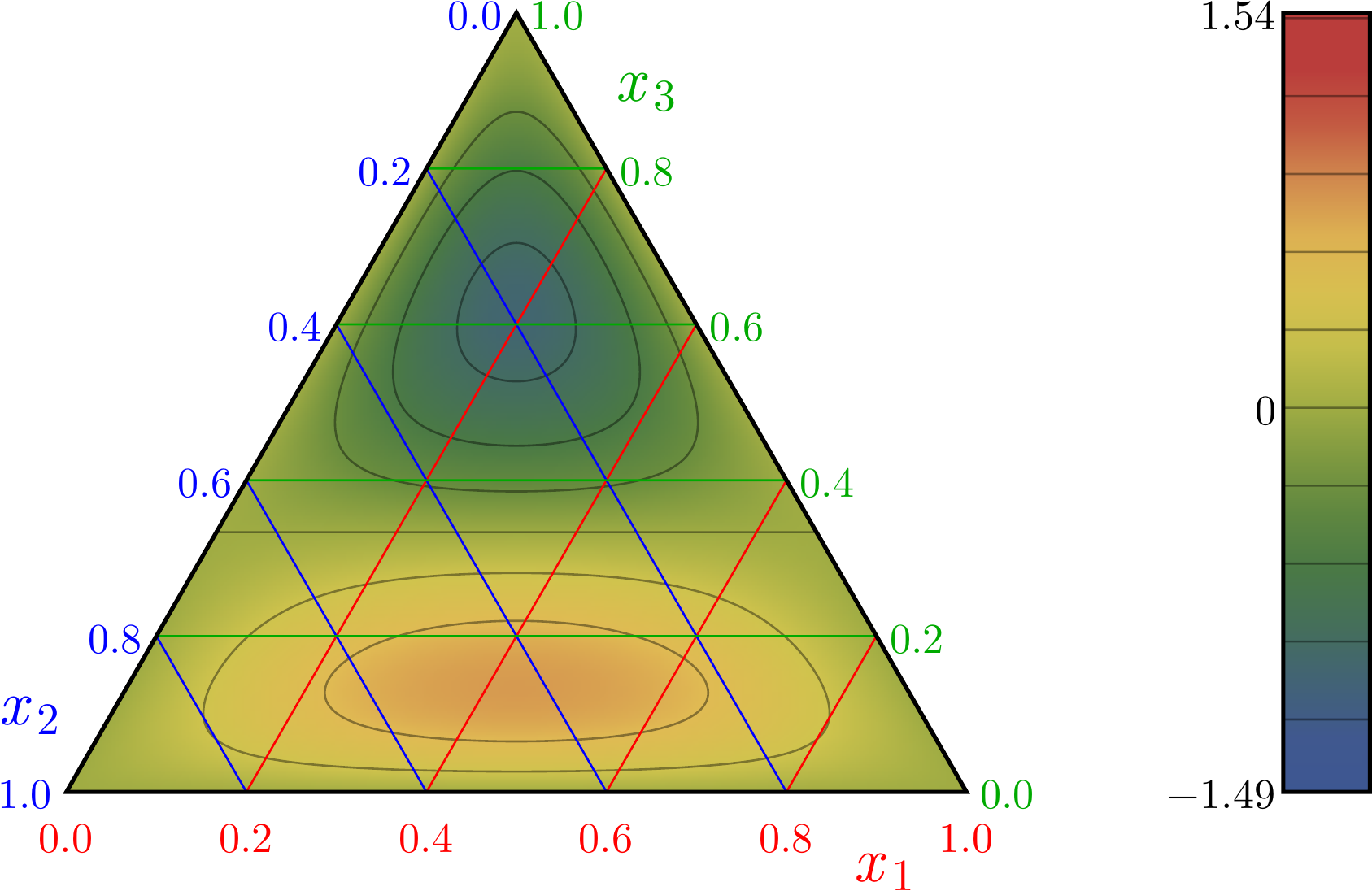}%
\llap{\raisebox{0.6419\barycentricwidth-\height}{\rlap{\fock{\color{red}s^\gooduparrow \color{blue}s^\gooduparrow \color{mygreen}u^\gooddownarrow}}\hspace{\barycentricwidth}}}%
\llap{\raisebox{0.6419\barycentricwidth-\height}{\clap{\boxed{\tfrac{T^\Xi}{f_T^\Xi}-\phi^{\text{as}}}}\hspace{0.3162\barycentricwidth}}}%
\\[\baselineskip]%
\includegraphics[width=\barycentricwidth]{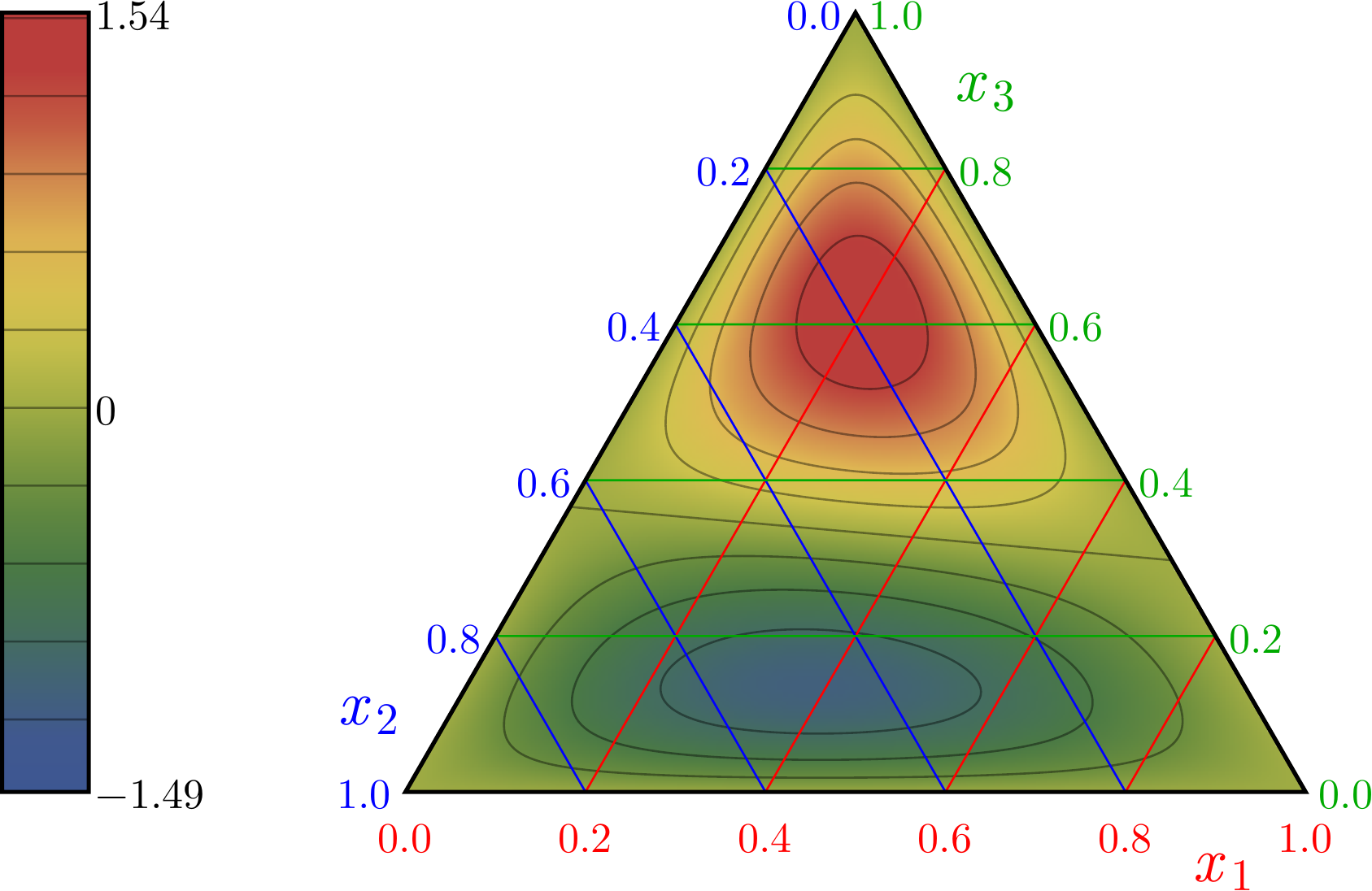}%
\llap{\raisebox{0.6419\barycentricwidth-\height}{\rlap{\clap{\boxed{\tfrac{[\VmA]^{\mathrlap{\Lambda}}}{{\scriptscriptstyle\sqrt{\frac32}}f^\Lambda}-\phi^{\text{as}}}}}\hspace{0.6838\barycentricwidth}}}%
\llap{\raisebox{0.6419\barycentricwidth-\height}{\fock{\color{red}u^\gooduparrow \color{blue}d^\gooddownarrow \color{mygreen}s^\gooduparrow}}}\hfill%
\includegraphics[width=\barycentricwidth]{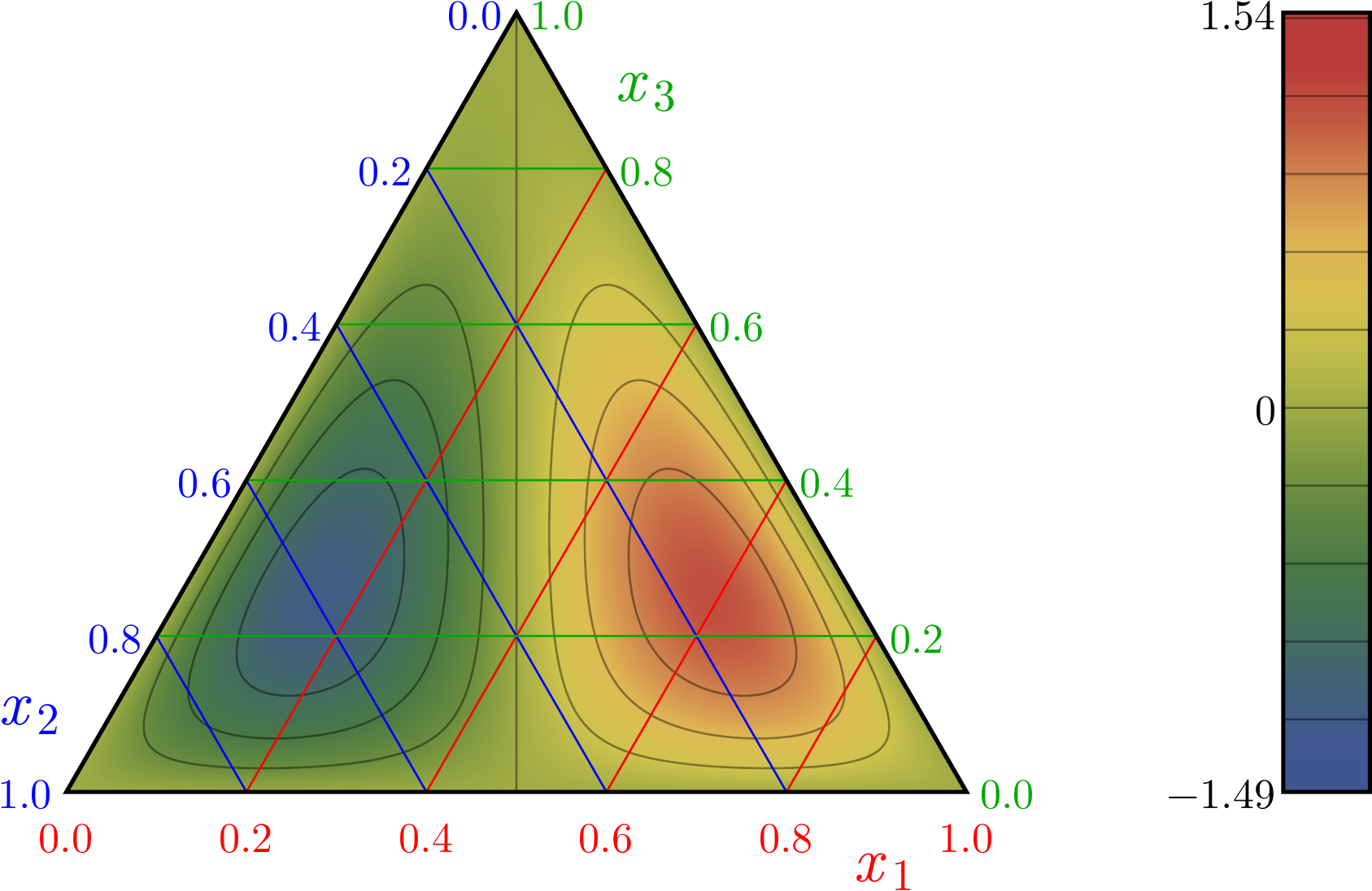}%
\llap{\raisebox{0.6419\barycentricwidth-\height}{\rlap{\fock{\color{red}u^\gooduparrow \color{blue}d^\gooduparrow \color{mygreen}s^\gooddownarrow}}\hspace{\barycentricwidth}}}%
\llap{\raisebox{0.6419\barycentricwidth-\height}{\clap{\boxed{\tfrac{T^\Lambda}{{\scriptscriptstyle\sqrt{\frac16}}f^\Lambda}}}\hspace{0.3162\barycentricwidth}}}%
\caption{\label{fig_barycentric}Barycentric plots ($x_1+x_2+x_3=1$) showing the deviations of the DAs~$[\VmA]^B$ (left column) and~$T^B$ (right column) at $\mu=\unit{2}{\giga\electronvolt}$ from the asymptotic shape $\phi^{\text{as}}\equiv120x_1x_2x_3$. ($T^\Lambda$~vanishes in the asymptotic~limit.) In this representation the coordinates~$x_i$ directly correspond to quarks of definite flavor and helicity.}%
\vspace{-\prevdepth}
\end{figure*}

%
To visualize distribution amplitudes one can make so-called barycentric plots~\cite{Mobius:1827zz}, in which the support of the DAs ($0\leq x_1,x_2,x_3\leq1$ with the additional constraint $x_1+x_2+x_3=1$) is mapped to an equilateral triangle. We shall do so for the standard DAs, $[\VmA]^B$ and~$T^B$ (see Section~\ref{sect_leadingtwist} for the relevant definitions), as these are the most convenient representations for phenomenological applications and also come with a straightforward physical interpretation: The two DAs directly correspond to the two Fock states~$f^\gooduparrow g^\gooddownarrow h^\gooduparrow$ and~$f^\gooduparrow g^\gooduparrow h^\gooddownarrow$, respectively (cf.\ Ref.~\cite{Bali:2015ykx}), and the three variables~$x_i$ are interpreted as light-cone momentum fractions of the three valence quarks. In Figure~\ref{fig_barycentric}, such plots are realized as combined density and contour plots. They are overlaid with red, blue, and green lines, which are lines of constant~$x_1$, $x_2$, and~$x_3$, respectively. For each DA the relevant overall normalization factor has been divided out and the asymptotic part has been subtracted, so that the deviation from the asymptotic shape is immediately visible and their relative strengths can be compared across different DAs. Note that, due to the symmetry properties of the DAs,%
\begin{align}%
 V^{\BnoL}(x_{213}) &= + V^{B}(x_{123}) \,, & V^{\Lambda}(x_{213}) &= - V^{\Lambda}(x_{123}) \,, \notag\\
 A^{\BnoL}(x_{213}) &= - A^{B}(x_{123}) \,, & A^{\Lambda}(x_{213}) &= + A^{\Lambda}(x_{123}) \,, \notag\\
 T^{\BnoL}(x_{213}) &= + T^{B}(x_{123}) \,, & T^{\Lambda}(x_{213}) &= - T^{\Lambda}(x_{123}) \,,
\end{align}%
the amplitudes $T^{\BnoL}$ ($T^\Lambda$) depicted in the right column are (anti-)symmetric under the interchange of $x_1$ and $x_2$.\par%
Considering the nucleon we observe, in agreement with earlier lattice studies~\cite{Braun:2014wpa,Bali:2015ykx} and Dyson--Schwinger calculations~\cite{Mezrag:2017znp}, that the ``leading'' $u^\gooduparrow$ quark, which has the same helicity as the nucleon, carries a larger momentum fraction. Historically, this statement has been the main finding of the QCD sum rule approach~\cite{Chernyak:1984bm,Chernyak:1987nu}. In the $u^\gooduparrow u^\gooduparrow d^\gooddownarrow$ nucleon state, which is described by $T^N$, the peak of the distribution is shifted towards the two $u$ quarks in a symmetric manner. $T^N$, however, is not an independent DA. Taking into account the isospin relation, the spin-flavor structure of the nucleon light-cone wave function can be represented, schematically, as $[\VmA]^N u^\gooduparrow (u^\gooddownarrow d^\gooduparrow - d^\gooddownarrow u^\gooduparrow)$. In this picture our result for $[\VmA]^N$ corresponds to a shift of the momentum distribution towards the $u^\gooduparrow$ quark, which carries the nucleon helicity, and there is some deviation from the approximate symmetry under $x_2\leftrightarrow x_3$. This symmetry could be interpreted as a scalar ``diquark'' structure for the remaining valence quarks, which is assumed in many models.\par%
For the isospin-nonsinglet baryons one can identify two competing patterns: First, the strange quarks carry an increased fraction of the momentum. Second, in the $\lvert\uparrow\downarrow\uparrow\rangle$ state the first quark has a larger momentum fraction than the second, while in the $\lvert\uparrow\uparrow\downarrow\rangle$ state the first two quarks have to behave identically. For the $\Lambda$~baryon the $u^\gooduparrow d^\gooddownarrow s^\gooduparrow$ spin orientation shows a similar behavior, as the maximum of the distribution is shifted towards the $s$~quark, while the plot for $u^\gooduparrow d^\gooduparrow s^\gooddownarrow$ reflects the antisymmetry of $T^\Lambda$ under exchange of~$x_1$ and~$x_2$. This DA is a special case, since it does not have a leading asymptotic part. However, we find large contributions of subleading conformal spin, contrary to our previous study at a finite lattice spacing~\cite{Bali:2015ykx}, where these contributions were found to be zero within error. In retrospect, this smallness was accidental and is related to the zero-crossing in the continuum extrapolation of the parameter $\pi_{10}^\Lambda$ at $a\approx\unit{0.086}{\femto\meter}$, cf.\ the brown line in the bottom right panel of Figure~\ref{fig_extrapolate_a}.\par%
To make these statements quantitative, we consider normalized first moments of $[\VmA]^B$ and~$T^{\BnoL}$,%
\begin{align}%
 \langle x_i \rangle^B         &= \frac{1}{f^B}\int \![dx]\, x_i\, [\VmA]^B\,,\notag\\
 \langle x_i \rangle^{\BnoL}_T &= \frac{1}{f_T^B} \int \![dx]\, x_i \, T^B\,,\label{eq_not_momentum_fractions}
\end{align}%
\def\mystrut{\rule[-4pt]{0sp}{13pt}}%
\begin{table}[tb]%
\centering%
\caption{Continuum results for the normalized first moments of the DAs $[\VmA]^B$ and~$T^{\BnoL}$ in the $\MSbar$~scheme at a scale $\mu=\unit{2}{\giga\electronvolt}$, see Eqs.~\eqref{eq_not_momentum_fractions}. All uncertainties from our calculation have been added in quadrature.\label{tab_not_momentum_fractions}}%
\begin{widetable}{\columnwidth}{c r@{\hspace{.5em}}l r@{\hspace{.5em}}l r@{\hspace{.5em}}l r@{\hspace{.5em}}l}
\toprule
$B$ & \multicolumn{2}{c}{$N$} & \multicolumn{2}{c}{$\Sigma$} & \multicolumn{2}{c}{$\Xi$} & \multicolumn{2}{c}{$\Lambda$}\\
\midrule
\mystrut$\langle x_1 \rangle^B$   & $u^\gooduparrow$   & $0.396_{-6}^{+7}$ & $d^\gooduparrow$   & $0.363_{-7}^{+4}$ & $s^\gooduparrow$   & $0.390_{-4}^{+4}$ & $u^\gooduparrow$   & $0.308_{-3}^{+3}$\\
\mystrut$\langle x_2 \rangle^B$   & $u^\gooddownarrow$ & $0.311_{-5}^{+5}$ & $d^\gooddownarrow$ & $0.309_{-5}^{+5}$ & $s^\gooddownarrow$ & $0.335_{-2}^{+2}$ & $d^\gooddownarrow$ & $0.300_{-7}^{+7}$\\
\mystrut$\langle x_3 \rangle^B$   & $d^\gooduparrow$   & $0.293_{-6}^{+5}$ & $s^\gooduparrow$   & $0.329_{-3}^{+6}$ & $u^\gooduparrow$   & $0.275_{-5}^{+5}$ & $s^\gooduparrow$   & $0.392_{-5}^{+5}$\\
\midrule
\mystrut$\langle x_1 \rangle^B_T$ & $u^\gooduparrow$   & $0.344_{-2}^{+2}$ & $d^\gooduparrow$   & $0.327_{-2}^{+2}$ & $s^\gooduparrow$   & $0.354_{-5}^{+5}$ & \multicolumn{2}{c}{---}\\
\mystrut$\langle x_2 \rangle^B_T$ & $u^\gooduparrow$   & $0.344_{-2}^{+2}$ & $d^\gooduparrow$   & $0.327_{-2}^{+2}$ & $s^\gooduparrow$   & $0.354_{-5}^{+5}$ & \multicolumn{2}{c}{---}\\
\mystrut$\langle x_3 \rangle^B_T$ & $d^\gooddownarrow$ & $0.311_{-5}^{+5}$ & $s^\gooddownarrow$ & $0.345_{-3}^{+3}$ & $u^\gooddownarrow$ & $0.291_{-9}^{+9}$ & \multicolumn{2}{c}{---}\\
\bottomrule
\end{widetable}%
\end{table}%
see also Eqs.~$(6.3)$ in Ref.~\cite{Bali:2015ykx}. These are sometimes referred to as momentum fractions in the literature and interpreted as the portions of the hadron's total momentum carried by the individual valence quarks. This notion is somewhat imprecise since the averaging is done with a DA instead of a squared wave function; furthermore, the interpretation as momentum fractions breaks down completely in the case of~$T^\Lambda$, which has no asymptotic part. That aside, these objects are nevertheless interesting because they provide a simple quantitative measure for the relative deviations of a DA from the asymptotic case $\langle x_1 \rangle^\text{as}=\langle x_2 \rangle^\text{as}=\langle x_3 \rangle^\text{as}=1/3$. The numerical results are summarized in Table~\ref{tab_not_momentum_fractions} and they clearly agree with the qualitative picture suggested by the above discussion of Figure~\ref{fig_barycentric}.\par%
\section{Summary\label{sect_summary}}
This study provides the first results for baryon octet light-cone distribution amplitudes from lattice QCD with control over all systematic uncertainties; they are renormalized in the $\MSbar$ scheme at $\unit{2}{\giga\electronvolt}$ and all relevant limits (infinite volume, continuum, and extrapolation to physical masses) have been taken with due diligence. As expected, finite volume effects turned out to be harmless for our ensembles, but the discretization effects have proven to be of great importance. While we find them to be already large for the wave function normalization constants (up to $\sim 20\%$ from our coarsest lattice to the continuum), they can, in the case of shape parameters, change the entire character and interpretation of the distribution. In the latter case, even zero-crossings can occur despite the fact that we only consider relatively small lattice spacings with $a<\unit{0.1}{\femto\meter}$. In comparison to our previous study at a finite lattice spacing~\cite{Bali:2015ykx}, we find even more enhanced \SU3 breaking of up to $\sim 80\%$ (at the physical point) in the phenomenologically important leading-twist normalization constants. For the shape parameters the symmetry breaking patterns show a strong lattice spacing dependence.\par%
In view of the fact that there is an intricate interplay between mass and discretization effects, resolving both dependences simultaneously is pivotal. Therefore, it is crucial to analyze a wide range of lattice spacings while covering the whole $m_\ell$-$m_s$-plane using multiple trajectories. To this end, we have employed $40$~ensembles including various lattice spacings down to $a=\unit{0.039}{\femto\meter}$, large volumes, and physical masses. Based on this vast data pool we have performed a combined fit including an estimation of all relevant uncertainties. The final numbers to be used in phenomenological applications are summarized in Table~\ref{tab_results}. These results will be useful for the study of nucleon electromagnetic form factors and baryon photoproduction using CEBAF\,@\,\unit{12}{\giga\electronvolt} at Jefferson Lab~\cite{Aznauryan:2009da,Dudek:2012vr} and weak decays of heavy baryons at LHCb, e.g., $\Lambda_b\to p\bar{\nu}_\ell\ell^-$, $\Lambda_b\to\Lambda\ell^+\ell^-$, etc.~\cite{Lu:2009cm,Aaltonen:2011qs,Aaij:2017ewm}.\par%
\begin{acknowledgement}%
The authors would like to express their gratitude towards Benjamin Gl{\"a}{\ss}le, Peter Georg, and Jakob Simeth for technical support. This work was supported by the Deut\-sche For\-schungs\-ge\-mein\-schaft (collaborative research centre SFB/TRR\nobreakdash-55), the Stu\-di\-en\-stif\-tung des deut\-schen Vol\-kes, the Narodowe Centrum Nauki (grant no.~UMO-2016/21/B/ST2/01492) and the European Union's Horizon 2020 Research and Innovation programme under the Marie Sk\l{}odowska-Curie grant agreement no.~813942 (ITN EuroPLEx).\par%
We used a modified version of the {\sc Chroma}~\cite{Edwards:2004sx} software package along with the {\sc Lib\-Hadron\-Analysis} library and improved inverters~\cite{Nobile:2010zz,Luscher:2012av,Frommer:2013fsa,Heybrock:2015kpy}. The configurations were generated as part of the CLS effort~\cite{Bruno:2014jqa,Bali:2016umi} using {\sc openQCD} (\url{https://luscher.web.cern.ch/luscher/openQCD/})~\cite{Luscher:2012av}. We thank all our CLS colleagues for the joint generation of the gauge ensembles. Additional $m_\ell=m_s$ ensembles were generated with {\sc openQCD} by members of the Mainz group on the Wilson and Clover HPC Clusters of IKP Mainz as well as by RQCD on the QPACE computer using the BQCD code~\cite{Nakamura:2010qh}.\par%
The computation of observables was carried out on the QPACE~2 and QPACE~3 systems of the SFB/TRR\nobreakdash-55, on the Regensburg QPACE~B machine, the Regensburg HPC-cluster ATHENE~2, and at various supercomputer centers. In particular, the authors gratefully acknowledge computing time granted by the John von Neumann Institute for Computing (NIC), provided on the Booster partition of the supercomputer JURECA~\cite{jureca} at J\"ulich Supercomputing Centre (JSC, \url{http://www.fz-juelich.de/ias/jsc/}), computer time granted by the Interdisciplinary Centre for Mathematical and Computational Modelling~(ICM) of the University of Warsaw, provided on Okeanos (grant nos.\ GA67-12, GA69-20), and computer time granted by the PLGRID consortium, provided on the Prometheus machine hosted by Cyfronet Krakow (grant hadronspectrum).\par%
Regarding the generation of recent gauge ensembles, the authors gratefully acknowledge the Gauss Centre for Supercomputing (\kern-.6ptGCS\kern-.3pt) for providing computing time for GCS Large-Scale Projects on the GCS share of the two supercomputers JUQUEEN~\cite{juqueen} and JUWELS~\cite{juwels} at JSC as well as on SuperMUC at Leibniz Supercomputing Centre (LRZ, \url{https://www.lrz.de}). GCS is the alliance of the three national supercomputing centres HLRS (Universit\"at Stuttgart), JSC (For\-schungs\-zen\-trum J\"ulich), and LRZ (Bayerische Akademie der Wissenschaften), funded by the German Federal Ministry of Education and Research (BMBF) and the German State Ministries for Research of Baden-W{\"u}rttemberg (MWK), Bayern (StMWFK) and Nordrhein-Westfalen (MIWF).\par%
\end{acknowledgement}%
\appendix
\def\myrule{\rule[-2pt]{0sp}{12pt}}%
\begin{table*}[t]%
\caption{\label{table_coeff_g}Coefficients for use in Eqs.~\eqref{eq_g}.}%
\begin{widetable}{\textwidth}{lrrrrrrrr}%
\toprule%
$\phi$ & $\bar{g}^1_\phi$ & $\delta g^{1\pi}_\phi$ & $\delta g^{1K}_\phi$ & $\delta g^{1\eta}_\phi$ & $\bar{g}^2_\phi$ & $\delta g^{2\pi}_\phi$ & $\delta g^{2K}_\phi$ & $\delta g^{2\eta}_\phi$\\
\midrule%
\myrule$f^N$, $\varphi_{11}^N$ & $-76$  & $-57$ & $-18$ & $-1$ & $8(5D+6F)$ & $30(D+F)$ & $12(D+F)$ & $-2(D-3F)$\\
\myrule$f^\Sigma$, $\varphi_{11}^\Sigma$ & $-76$  & $-12$ & $-60$ & $-4$ & $8(5D+6F)$ & $24D$ & $24(D+2F)$ & $-8D$\\
\myrule$f_T^\Sigma$, $\pi_{11}^\Sigma$ & $-76$  & $-24$ & $-36$ & $-16$ & $8(5D+6F)$ & $48F$ & $24D$ & $16D$\\
\myrule$f^\Xi$, $\varphi_{11}^\Xi$ & $-76$  & $-9$ & $-66$ & $-1$ & $8(5D+6F)$ & $-18(D-F)$ & $12(5D+3F)$ & $-2(D+3F)$\\
\myrule$f_T^\Xi$, $\pi_{11}^\Xi$ & $-76$  & $-9$ & $-42$ & $-25$ & $8(5D+6F)$ & $18(D-F)$ & $12(D+3F)$ & $10(D+3F)$\\
\myrule$f^\Lambda$, $\varphi_{11}^\Lambda$ & $-76$  & $-36$ & $-36$ & $-4$ & $8(5D+6F)$ & $24D$ & $8(D+6F)$ & $8D$\\
\myrule$\lambda_1^N$, $\varphi_{10}^N$ & $-28$  & $-9$ & $-18$ & $-1$ & $-40D$ & $-18(D+F)$ & $-4(5D-3F)$ & $-2(D-3F)$\\
\myrule$\lambda_1^\Sigma$, $\varphi_{10}^\Sigma$ & $-28$  & $-12$ & $-12$ & $-4$ & $-40D$ & $-8D$ & $-24D$ & $-8D$\\
\myrule$\lambda_1^\Xi$, $\varphi_{10}^\Xi$ & $-28$  & $-9$ & $-18$ & $-1$ & $-40D$ & $-18(D-F)$ & $-4(5D+3F)$ & $-2(D+3F)$\\
\myrule$\lambda_1^\Lambda$, $\varphi_{10}^\Lambda$ & $-28$  & $-36$ & $12$ & $-4$ & $-40D$ & $-72D$ & $24D$ & $8D$\\
\myrule$\lambda_T^\Lambda$, $\pi_{10}^\Lambda$ & $-28$  & $0$ & $-12$ & $-16$ & $-40D$ & $0$ & $-24D$ & $-16D$\\
\myrule$\lambda_2^N$ & $-36$  & $-9$ & $-18$ & $-9$ & $-72F$ & $-18(D+F)$ & $12(D-3F)$ & $6(D-3F)$\\
\myrule$\lambda_2^\Sigma$ & $-36$  & $-24$ & $-12$ & $0$ & $-72F$ & $-48F$ & $-24F$ & $0$\\
\myrule$\lambda_2^\Xi$ & $-36$  & $-9$ & $-18$ & $-9$ & $-72F$ & $18(D-F)$ & $-12(D+3F)$ & $-6(D+3F)$\\
\myrule$\lambda_2^\Lambda$ & $-36$  & $0$ & $-36$ & $0$ & $-72F$ & $0$ & $-72F$ & $0$\\
\bottomrule
\end{widetable}%
\end{table*}%
\begin{table}[t]%
\caption{\label{table_coeff_sigma}Coefficients for use in Eqs.~\eqref{eq_deltaselfenergyprime}.  $\bar{\sigma}=\tfrac43(5D^2+9F^2)$.}%
\begin{widetable}{\columnwidth}{lrrr}%
\toprule
$B$ & $\delta\sigma^\pi_\B$ & $\delta\sigma^K_\B$ & $\delta\sigma^\eta_\B$\\
\midrule
\myrule$N$       & $(D+F)^2$            & $\tfrac56D^2-DF+\tfrac32F^2$  & $\tfrac13(D-3F)^2$\\
\myrule$\Sigma$  & $\tfrac49(D^2+6F^2)$ & $D^2+F^2$                     & $\tfrac43D^2$\\
\myrule$\Xi$     & $(D-F)^2$            & $\tfrac56D^2+DF+\tfrac32F^2$  & $\tfrac13(D+3F)^2$\\
\myrule$\Lambda$ & $\tfrac43D^2$        & $\tfrac13(D^2+9F^2)$          & $\tfrac43D^2$\\
\bottomrule
\end{widetable}%
\end{table}%
\section{Extrapolation formulae\label{sect_beauty}}%
In the following we will give a detailed description of the prefactors~$g_\phi$ in the master fit formula~\eqref{eq_master_fit_formula}. These contain the contributions from one-loop diagrams in BChPT, see Ref.~\cite{Wein:2015oqa} for details (cf.\ also Ref.~\cite{Wein:2011ix}). We write down all formulae in a factorized form and disentangle the flavor symmetric and flavor symmetry breaking terms:%
\begin{align}%
g_\phi &= \sqrt{\smash[b]{Z_B}}\bigl(1+\bar{g}_\phi+\delta g_\phi\bigr) \,.
\end{align}%
The baryon-dependent (but DA-independent) $Z$-factor is written as%
\begin{align}%
\sqrt{\smash[b]{Z_B}} &= 1+\tfrac12 \bar{\Sigma}^\prime+\tfrac12\delta\?\Sigma^\prime_B
\end{align}%
in terms of the self-energy contributions%
\begin{align}%
\bar{\Sigma}^\prime &= \bar{\sigma} H_3(\bar{m})\,, \label{eq_deltaselfenergyprime} \\
\delta\?\Sigma^\prime_B &= 3\,\delta\sigma^\pi_\B\,\delta\?H_3(m_\pi) + 4\,\delta\sigma^K_\B\delta\?H_3(m_K) + \delta\sigma^\eta_\B\,\delta\?H_3(m_\eta)\,, \notag
\end{align}%
whereas the remaining loop contributions, which are both, baryon- and DA-dependent, are given by\footnote{As input values for the various low-energy constants we use $F_0=\unit{87}{\mega\electronvolt}$, $D=0.623$, $F=0.441$, and $m_b=\unit{880}{\mega\electronvolt}$ from Ref.~\cite{Ledwig:2014rfa}.}%
\begin{align}%
\bar{g}_\phi &= \frac{1}{24F_0^2}\bigl(\bar{g}^1_\phi H_1(\bar{m})+\bar{g}^2_\phi H_2(\bar{m})\bigr)\,, \label{eq_g}\\
\delta g_\phi &= \frac{1}{24F_0^2} \smashoperator[r]{\sum_{M=\pi,K,\eta}}\bigl(\delta g_\phi^{1M}\delta\?H_1(m_M)+\delta g_\phi^{2M}\delta\?H_2(m_M)\bigr) \,. \notag
\end{align}%
The coefficients $\bar{\sigma}$, $\delta \sigma_\B^M$, $\bar{g}^i_\phi$, and $\delta g_\phi^{iM}$ are given in Tables~\ref{table_coeff_g} and~\ref{table_coeff_sigma}. They satisfy the equations%
\begin{align}
\bar{\sigma} &= 2\,\delta\sigma^M_{\!N} + 3\,\delta\sigma^M_{\!\Sigma} + 2\,\delta\sigma^M_{\!\Xi} + \delta\sigma^M_{\!\Lambda} &&\forall M\,,\notag\\
\bar{\sigma} &= 3\,\delta\sigma^\pi_\B + 4\,\delta\sigma^K_\B + \delta\sigma^\eta_\B &&\forall B\,, \notag\\
\bar{g}_\phi^i &= \delta g^{i\pi}_\phi + \delta g^{iK}_\phi + \delta g^{i\eta}_\phi &&\forall\phi,i\,.
\end{align}
Using IR regularization~\cite{Becher:1999he}, the loop-integrals $H_i(m)$ and $\delta\?H_i(m)=H_i(m)-H_i(\bar{m})$ are given by%
\def\sqrtstrut{\rule[-5.5pt]{0sp}{20pt}}%
\def\sqrtstruttwo{\rule[-5pt]{0sp}{15pt}}%
\begin{align}%
H_1(m) &= 2m^2\biggl(L_d+\frac{1}{32\pi^2}\ln\frac{m^2}{\mu_\chi^2}\biggr) - \delta\?I_{1,0}(m) \,, \notag\\
H_2(m) &= \frac{m^4}{m_b^2}\biggl(L_d+\frac{1}{32\pi^2}\ln\frac{m^2}{\mu_\chi^2}\biggr) - \frac{m^4}{32\pi^2m_b^2} \notag\\ 
&\hphantom{{}={}}{+}\frac{m^3}{8\pi^3m_b}\sqrt{\sqrtstrut\smash{1-\frac{m^2}{4m_b^2}}}\arccos\frac{-m}{2m_b} + m^2\delta\?I_{1,1}(m) \,, \notag\displaybreak[0]\\
H_3(m) &= -\frac{3m^2}{2F_0^2}\biggl(1-\frac{2m^2}{3m_b^2}\biggr)\biggl(L_d+\frac{1}{32\pi^2}\ln\frac{m^2}{\mu_\chi^2}\biggr) \notag\\ 
&\hphantom{{}={}}{-}\frac{m^2}{32\pi^2F_0^2} + \frac{3m^3}{32\pi^2F_0^2m_b}{\frac{1-\frac{m^2}{3m_b^2}}{\sqrt{\sqrtstruttwo\smash{1-\frac{m^2}{4m_b^2}}}}}\arccos\frac{-m}{2m_b}\notag \\
&\hphantom{{}={}}{+}\smash[t]{\frac{1}{4F_0^2}}\bigl(\begin{aligned}[t]&\delta\?I_{1,0}(m)+2m^2\delta\?I_{1,1}(m)\\&{-}2m^2(2m_b^2-m^2)\delta\?I_{1,2}(m)\bigr)\,.\taghere\end{aligned}
\end{align}%
As shown in Ref.~\cite{Wein:2015oqa}, the divergent terms~$L_d$ can be absorbed into appropriate counter-terms, and thus can be simply set to zero in the formulae above. The scale~$\mu_\chi$ is set to the typical hadronic value~$\unit{1}{\giga\electronvolt}$. The~$\delta\?I_{j,k}$ take into account the leading finite volume effects:%
\def\sqrtstrut{\rule[-1.5pt]{0sp}{10pt}}%
\begin{flalign}%
\delta\?I_{1,0}(m) &= \frac{-1}{4\pi^2}\smashoperator[l]{\sum_{\mathbf{n}\neq\mathbf{0}}} \frac{m}{\lvert\mathbf{n}\rvert L} K_1\bigl(m\lvert\mathbf{n}\rvert L\bigr) \,,\notag\\
\delta\?I_{1,1}(m) &= \frac{-1}{8\pi^2}\int_0^\infty\mkern-18mudu\smashoperator{\sum_{\mathbf{n}\neq\mathbf{0}}} K_0\bigl(w\lvert\mathbf{n}\rvert L\bigr) \cos\bigl(u\mathbf{p}\cdot\mathbf{n}L\bigr)\,,\notag\\
\delta\?I_{1,2}(m) &= \smash[b]{\frac{-1}{8\pi^2}\int_0^\infty\mkern-18mudu\smashoperator{\sum_{\mathbf{n}\neq\mathbf{0}}} \frac{u\lvert\mathbf{n}\rvert L}{2w} K_1\bigl(w\lvert\mathbf{n}\rvert L\bigr)  \cos\bigl(u\mathbf{p}\cdot\mathbf{n}L}\bigr) \,,
\end{flalign}%
where $K_i$ denote modified Bessel functions of the second kind, $w=\sqrt{\sqrtstrut\smash{u^2m_b^2+(1-u)m^2}}$, and $\mathbf{n}$ is a three-vector of integers.\par%
\section{Ensemble details and additional plots\label{sect_plots}}%
\begin{table}[t]%
\centering
\caption{Lattice spacings~$a$, corresponding to the five different inverse couplings~$\beta$ used in this study. These have been obtained by determining the Wilson flow time at the \SU3 symmetric point in lattice units~$t_0^*/a^2$ and equating $t_0^*$ with the result $\mu_{\mathrm{ref}}^*=(8t_0^*)^{-1/2}\approx\unit{478}{\mega\electronvolt}$ of Ref.~\cite{Bruno:2017gxd}.\label{table_spacings}}%
\begin{widetable}{\columnwidth}{lccccc}%
\toprule
$\beta$ & $3.40$ & $3.46$ & $3.55$ & $3.70$ & $3.85$\\
\midrule
$a\,[\femto\meter]$ & $0.086$ & $0.076$ & $0.064$ & $0.050$ & $0.039$\\
\bottomrule
\end{widetable}%
\end{table}%
Our lattice spacings are listed in Table~\ref{table_spacings}. Table~\ref{table_ensembles_renorm} lists the ensembles used for the determination of the renormalization factors. A full list of the CLS and RQCD ensembles used in the main analysis is given along with their properties in Table~\ref{table_ensembles}. In Figures~\mbox{\ref{fig_f}--\ref{fig_l2}} we show the global fits for all measured quantities. In contrast to the figures in the main text, these are expanded to show both, individual lattice spacings and quark mass trajectories.\par%
\begin{table}[tbp]%
\centering%
\caption{List of ensembles with $m_u=m_d=m_s$ used for the determination of the renormalization factors.\label{table_ensembles_renorm}}%
\begin{widetable}{\columnwidth}{rccrclcc}%
\toprule
\multicolumn{1}{c}{Ens.} & $\beta$ & $N_s$ & \multicolumn{1}{c}{$N_t$} & bc & \multicolumn{1}{c}{$\kappa_\ell=\kappa_s$} & $m_\pi\,[\mega\electronvolt]$ & $m_\pi L$\\
\midrule
rqcd017 & $3.40$ & $32$ & $ 32$ & p & $0.136865         $ & $235$ & $3.3$\\
rqcd021 & $3.40$ & $32$ & $ 32$ & p & $0.136813         $ & $338$ & $4.7$\\
rqcd016 & $3.40$ & $32$ & $ 32$ & p & $0.13675962       $ & $425$ & $5.9$\\
rqcd019 & $3.40$ & $32$ & $ 32$ & p & $0.1366           $ & $604$ & $8.4$\\
\midrule
X450    & $3.46$ & $48$ & $ 64$ & p & $0.136994         $ & $264$ & $4.9$\\
rqcd030 & $3.46$ & $32$ & $ 64$ & p & $0.1369587        $ & $320$ & $3.9$\\
B450    & $3.46$ & $32$ & $ 64$ & p & $0.13689          $ & $418$ & $5.2$\\
rqcd029 & $3.46$ & $32$ & $ 64$ & p & $0.1366           $ & $708$ & $8.7$\\
\midrule
X251    & $3.55$ & $48$ & $ 64$ & p & $0.1371           $ & $268$ & $4.2$\\
X250    & $3.55$ & $48$ & $ 64$ & p & $0.13705          $ & $348$ & $5.4$\\
rqcd025 & $3.55$ & $32$ & $ 64$ & p & $0.137            $ & $411$ & $4.3$\\
B250    & $3.55$ & $32$ & $ 64$ & p & $0.1367           $ & $708$ & $7.4$\\
\midrule
N300    & $3.70$ & $48$ & $128$ & o & $0.137            $ & $421$ & $5.1$\\
N303    & $3.70$ & $48$ & $128$ & o & $0.1368           $ & $641$ & $7.8$\\
\midrule
J500    & $3.85$ & $64$ & $192$ & o & $0.136852         $ & $411$ & $5.2$\\
N500    & $3.85$ & $48$ & $128$ & o & $0.13672514       $ & $599$ & $5.7$\\
\bottomrule
\end{widetable}%
\end{table}%
\begin{table*}[p]%
\centering%
\caption{\label{table_ensembles}List of the ensembles used in this work, labeled by their identifier and sorted by the inverse coupling $\beta$ and the pion masses. We specify the geometries $N_s^3\times N_t$ as well as the boundary condition in time (periodic~(p) or open~(o)). The light and strange hopping parameters used in the simulation are given by $\kappa_\ell$ and $\kappa_s$, respectively, and the resulting approximate meson masses $m_\pi$ and $m_K$ have been obtained from suitable two-point functions. $\#\text{conf.}$ gives the number of configurations analyzed. Each of the ensembles has been tuned such that it lies close to at least one of the following trajectories in the quark mass plane: $m_u+m_d+m_s\approx\text{phys.}$ (\trm), $m_s\approx\text{phys.}$ (\msc), or $m_u=m_d=m_s$ (\sym). These correspond to the green, red, and blue lines in Figure~\ref{figure_ensembles}. An in-depth description of the ensemble generation can be found in Ref.~\cite{Bruno:2014jqa}.}%
\begin{widetable}{\textwidth}{rccrcllcccrc}%
\toprule
\multicolumn{1}{c}{Ens.} & $\beta$ & $N_s$ & \multicolumn{1}{c}{$N_t$} & bc & \multicolumn{1}{c}{$\kappa_\ell$} & \multicolumn{1}{c}{$\kappa_s$} & $m_\pi\,[\mega\electronvolt]$ & $m_K\,[\mega\electronvolt]$ & $m_\pi L$ & \multicolumn{1}{c}{\#conf.} & traj.\\
\midrule
D150    & $3.40$ & $64$ & $128$ & p & $0.137088         $ & $0.13610755       $ & $126$ & $479$ & $3.5$ & $ 374$ & \trm, \msc\\
D101    & $3.40$ & $64$ & $128$ & o & $0.13703          $ & $0.136222041      $ & $217$ & $473$ & $6.0$ & $ 609$ & \trm\\
C101    & $3.40$ & $48$ & $ 96$ & o & $0.13703          $ & $0.136222041      $ & $220$ & $473$ & $4.6$ & $1596$ & \trm\\
C102    & $3.40$ & $48$ & $ 96$ & o & $0.13705084580022 $ & $0.13612906255557 $ & $222$ & $501$ & $4.6$ & $1500$ & \msc\\
rqcd017 & $3.40$ & $32$ & $ 32$ & p & $0.136865         $ & $0.136865         $ & $235$ & $235$ & $3.3$ & $1849$ & \sym\\
H106    & $3.40$ & $32$ & $ 96$ & o & $0.137015570024   $ & $0.136148704478   $ & $272$ & $517$ & $3.8$ & $1553$ & \msc\\
U101    & $3.40$ & $24$ & $128$ & o & $0.13697          $ & $0.13634079       $ & $273$ & $461$ & $2.8$ & $ 795$ & \trm\\
N101    & $3.40$ & $48$ & $128$ & o & $0.13697          $ & $0.13634079       $ & $279$ & $463$ & $5.8$ & $1456$ & \trm\\
H105    & $3.40$ & $32$ & $ 96$ & o & $0.13697          $ & $0.13634079       $ & $280$ & $465$ & $3.9$ & $2833$ & \trm\\
rqcd021 & $3.40$ & $32$ & $ 32$ & p & $0.136813         $ & $0.136813         $ & $338$ & $338$ & $4.7$ & $1541$ & \sym\\
H102    & $3.40$ & $32$ & $ 96$ & o & $0.136865         $ & $0.136549339      $ & $354$ & $440$ & $4.9$ & $1997$ & \trm\\
U102    & $3.40$ & $24$ & $128$ & o & $0.136865         $ & $0.136549339      $ & $357$ & $443$ & $3.7$ & $2117$ & \trm\\
H107    & $3.40$ & $32$ & $ 96$ & o & $0.13694566590798 $ & $0.136203165143476$ & $366$ & $546$ & $5.1$ & $1564$ & \msc\\
U103    & $3.40$ & $24$ & $128$ & o & $0.13675962       $ & $0.13675962       $ & $417$ & $417$ & $4.4$ & $2942$ & \trm, \sym\\
H101    & $3.40$ & $32$ & $ 96$ & o & $0.13675962       $ & $0.13675962       $ & $420$ & $420$ & $5.9$ & $2000$ & \trm, \sym\\
\midrule
X450    & $3.46$ & $48$ & $ 64$ & p & $0.136994         $ & $0.136994         $ & $264$ & $264$ & $4.9$ & $ 400$ & \sym\\
N450    & $3.46$ & $48$ & $128$ & p & $0.1370986        $ & $0.136352601      $ & $285$ & $524$ & $5.3$ & $ 680$ & \msc\\
N401    & $3.46$ & $48$ & $128$ & o & $0.1370616        $ & $0.1365480771     $ & $286$ & $462$ & $5.3$ & $1100$ & \trm\\
rqcd030 & $3.46$ & $32$ & $ 64$ & p & $0.1369587        $ & $0.1369587        $ & $320$ & $320$ & $3.9$ & $1224$ & \sym\\
B452    & $3.46$ & $32$ & $ 64$ & p & $0.1370455        $ & $0.136378044      $ & $350$ & $545$ & $4.3$ & $1943$ & \msc\\
S400    & $3.46$ & $32$ & $128$ & o & $0.136984         $ & $0.136702387      $ & $352$ & $443$ & $4.3$ & $1742$ & \trm\\
B450    & $3.46$ & $32$ & $ 64$ & p & $0.13689          $ & $0.13689          $ & $418$ & $418$ & $5.2$ & $1612$ & \trm, \sym\\
\midrule
D201    & $3.55$ & $64$ & $128$ & o & $0.1372067        $ & $0.136546844      $ & $199$ & $501$ & $4.1$ & $1078$ & \msc\\
D200    & $3.55$ & $64$ & $128$ & o & $0.1372           $ & $0.136601748      $ & $201$ & $481$ & $4.2$ & $1000$ & \trm\\
X251    & $3.55$ & $48$ & $ 64$ & p & $0.1371           $ & $0.1371           $ & $268$ & $268$ & $4.2$ & $ 346$ & \sym\\
N200    & $3.55$ & $48$ & $128$ & o & $0.13714          $ & $0.13672086       $ & $284$ & $463$ & $4.4$ & $1712$ & \trm\\
N201    & $3.55$ & $48$ & $128$ & o & $0.13715968       $ & $0.136561319      $ & $285$ & $523$ & $4.5$ & $1500$ & \msc\\
S201    & $3.55$ & $32$ & $128$ & o & $0.13714          $ & $0.13672086       $ & $290$ & $468$ & $3.0$ & $2093$ & \trm\\
N203    & $3.55$ & $48$ & $128$ & o & $0.13708          $ & $0.136840284      $ & $346$ & $442$ & $5.4$ & $1543$ & \trm\\
X250    & $3.55$ & $48$ & $ 64$ & p & $0.13705          $ & $0.13705          $ & $348$ & $348$ & $5.4$ & $ 345$ & \sym\\
N204    & $3.55$ & $48$ & $128$ & o & $0.137112         $ & $0.136575049      $ & $351$ & $545$ & $5.5$ & $ 950$ & \msc\\
H200    & $3.55$ & $32$ & $ 96$ & o & $0.137            $ & $0.137            $ & $411$ & $411$ & $4.3$ & $2000$ & \trm, \sym\\
N202    & $3.55$ & $48$ & $128$ & o & $0.137            $ & $0.137            $ & $412$ & $412$ & $6.4$ & $ 884$ & \trm, \sym\\
\midrule
J303    & $3.70$ & $64$ & $192$ & o & $0.137123         $ & $0.1367546608     $ & $258$ & $475$ & $4.2$ & $ 631$ & \trm\\
J304    & $3.70$ & $64$ & $192$ & o & $0.13713          $ & $0.1366569203     $ & $260$ & $523$ & $4.2$ & $1406$ & \msc\\
N302    & $3.70$ & $48$ & $128$ & o & $0.137064         $ & $0.1368721791358  $ & $345$ & $452$ & $4.2$ & $1383$ & \trm\\
N304    & $3.70$ & $48$ & $128$ & o & $0.137079325093654$ & $0.136665430105663$ & $352$ & $554$ & $4.3$ & $1463$ & \msc\\
N300    & $3.70$ & $48$ & $128$ & o & $0.137            $ & $0.137            $ & $421$ & $421$ & $5.1$ & $2027$ & \trm, \sym\\
\midrule
J501    & $3.85$ & $64$ & $192$ & o & $0.1369032        $ & $0.136749715      $ & $334$ & $445$ & $4.2$ & $1506$ & \trm\\
J500    & $3.85$ & $64$ & $192$ & o & $0.136852         $ & $0.136852         $ & $411$ & $411$ & $5.2$ & $ 842$ & \trm, \sym\\
\bottomrule
\end{widetable}%
\end{table*}%
\clearpage
\DeclareRobustCommand{\captiontail}{plotted in the infinite volume limit as a function of~$m_\pi^2$ on the three different trajectories as well as for five different lattice spacings and the continuum limit (for illustrative purposes the points shown have been obtained by translating the data along the fitted function). The dotted vertical lines mark the physical mass point.}%
\def\figureheight{.925\textheight}
\begin{figure*}[p]%
\centering
\includegraphics[height=\figureheight]{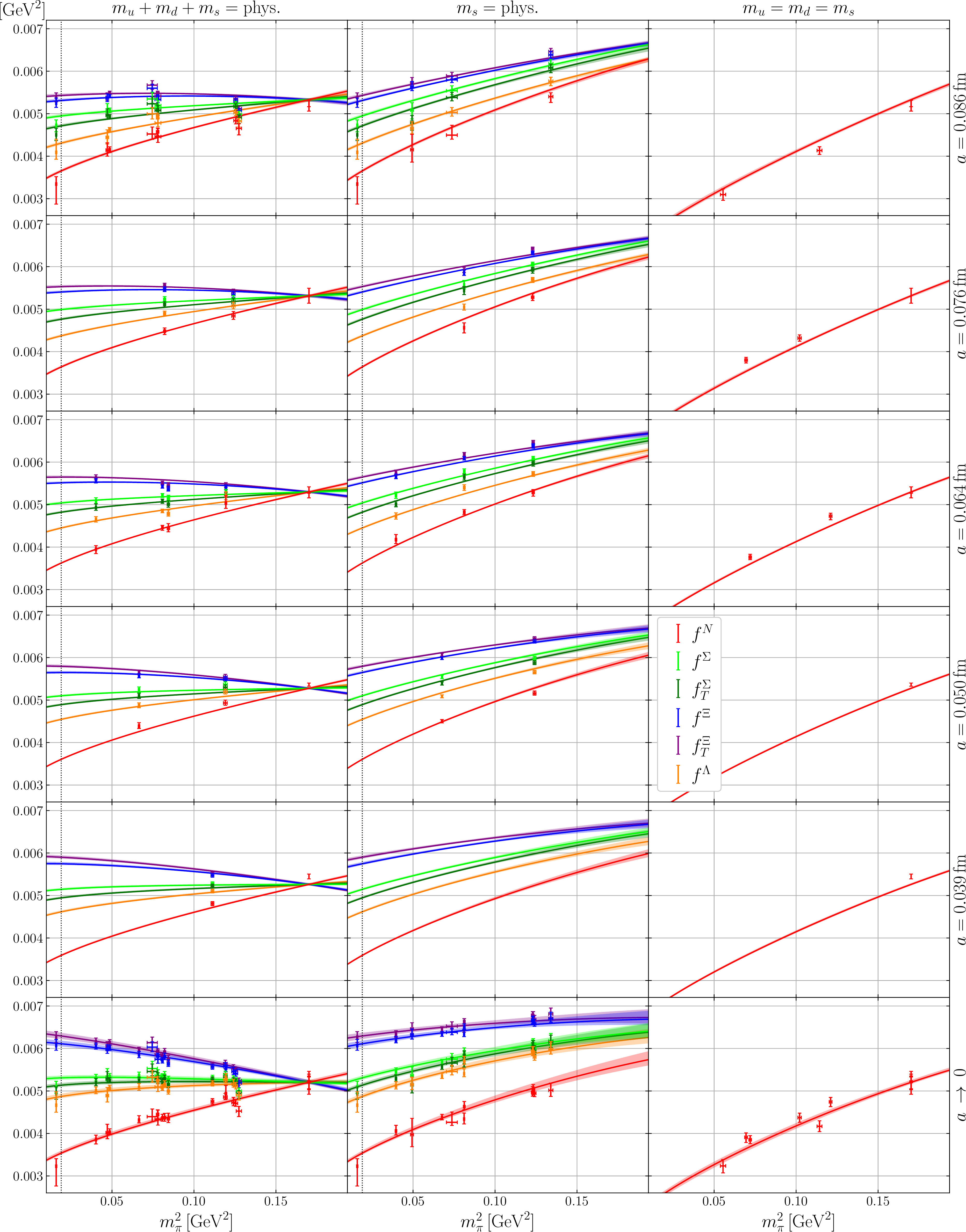}%
\caption{\label{fig_f}Global fit (using $13$~free parameters) for the leading-twist normalization constants~$f^B$ and~$f_T^{\BnoL}$, \captiontail}%
\end{figure*}%
\begin{figure*}[p]%
\centering
\includegraphics[height=\figureheight]{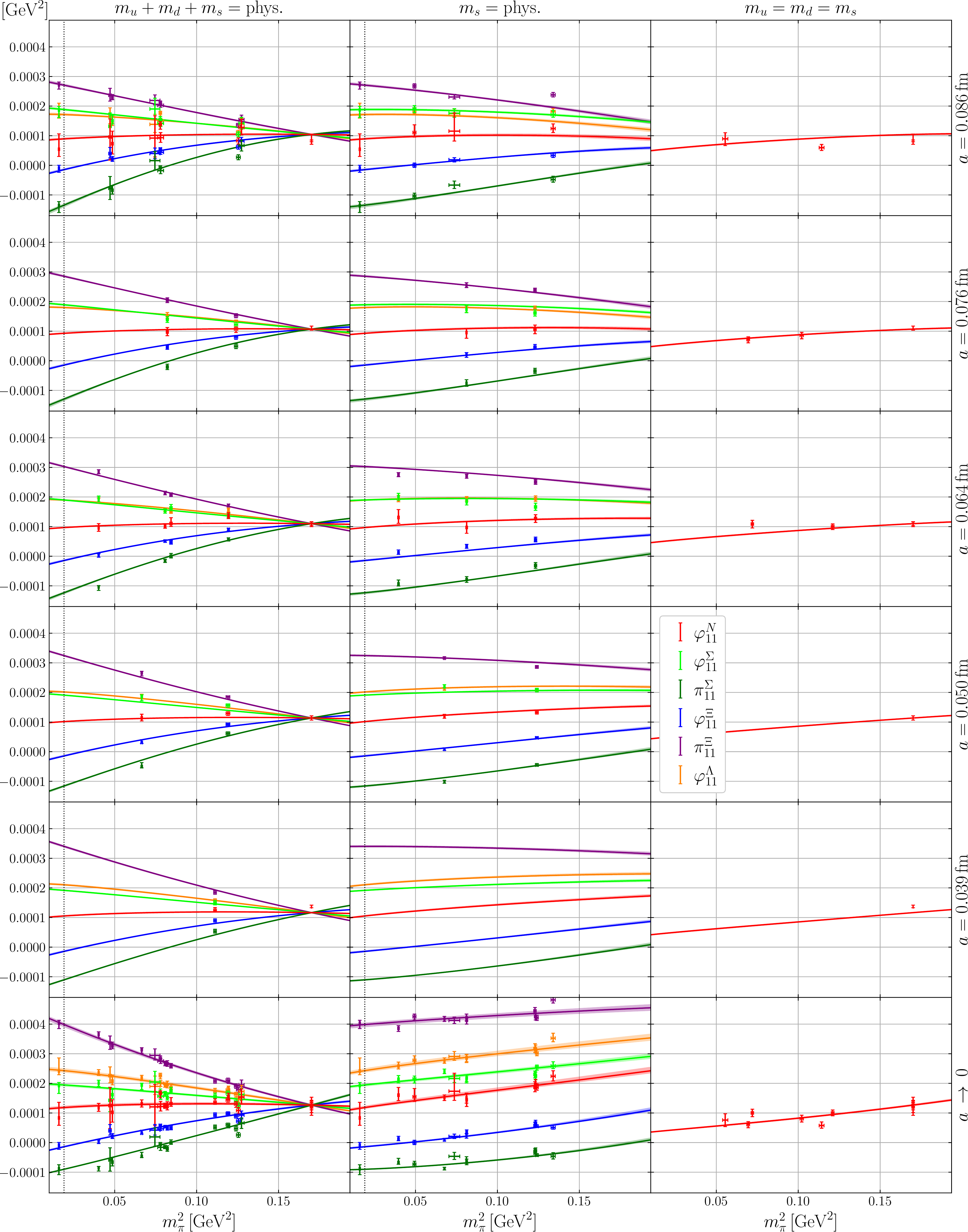}%
\caption{Global fit (using $13$~free parameters) for the leading-twist first moments~$\varphi_{11}^B$ and~$\pi_{11}^{\BnoL}$, \captiontail}%
\end{figure*}%
\begin{figure*}[p]%
\centering
\includegraphics[height=\figureheight]{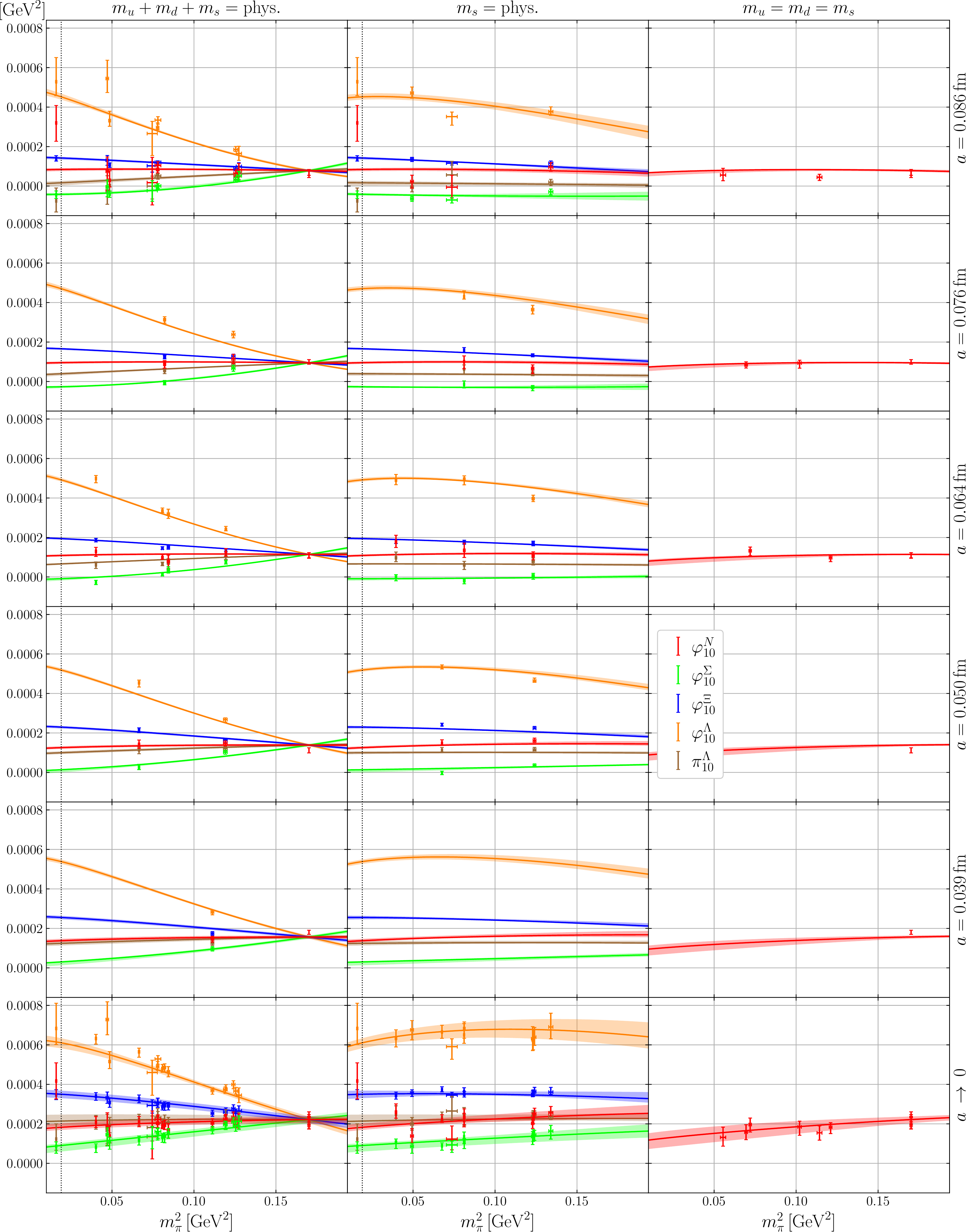}%
\caption{Global fit (using $12$~free parameters) for the leading-twist first moments~$\varphi_{10}^B$ and~$\pi_{10}^\Lambda$, \captiontail}%
\end{figure*}%
\begin{figure*}[p]%
\centering
\includegraphics[height=\figureheight]{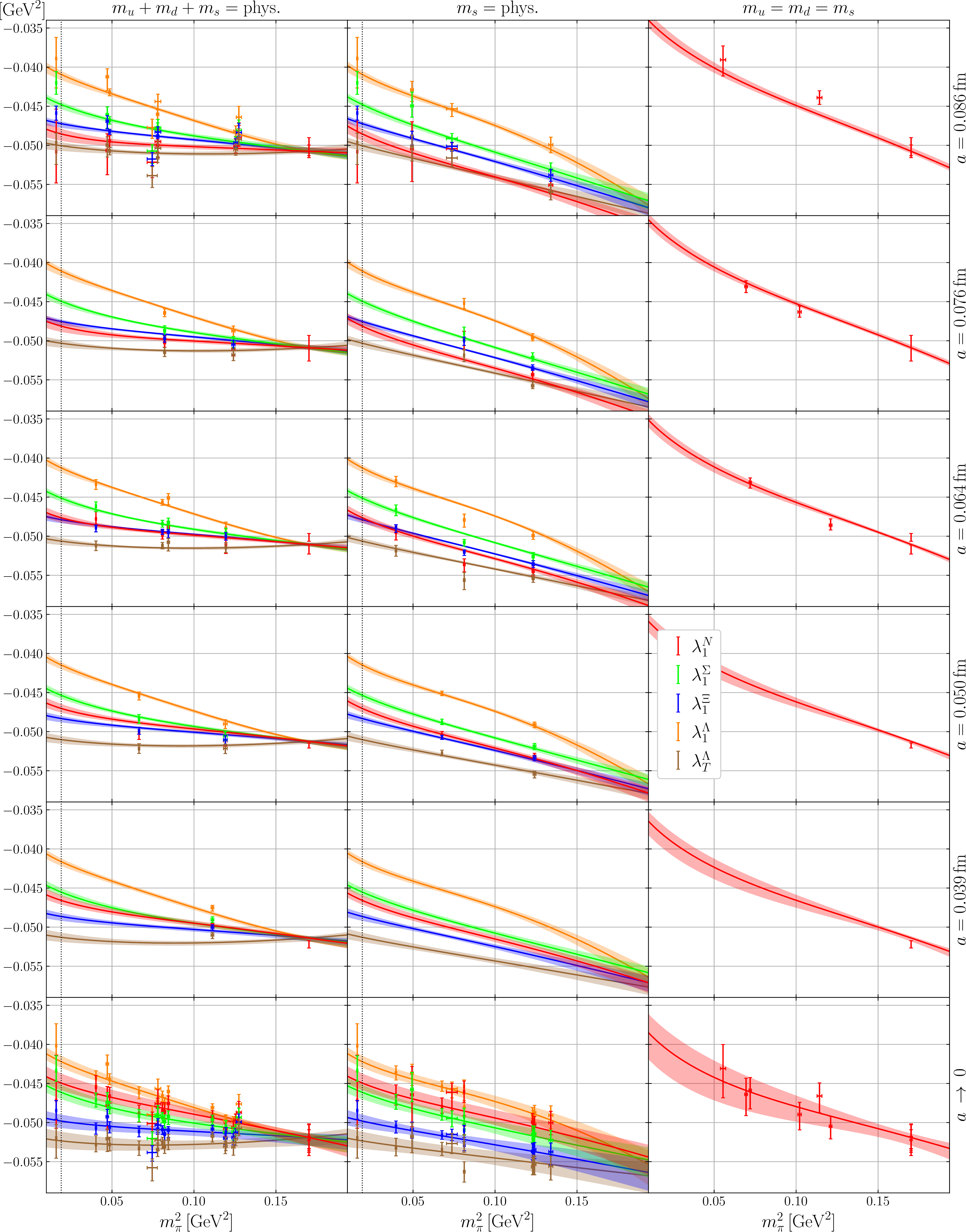}%
\caption{Global fit (using $23$~free parameters) for the higher-twist normalization constants~$\lambda_1^B$ and~$\lambda_T^\Lambda$, \captiontail}%
\end{figure*}%
\begin{figure*}[p]%
\centering
\includegraphics[height=\figureheight]{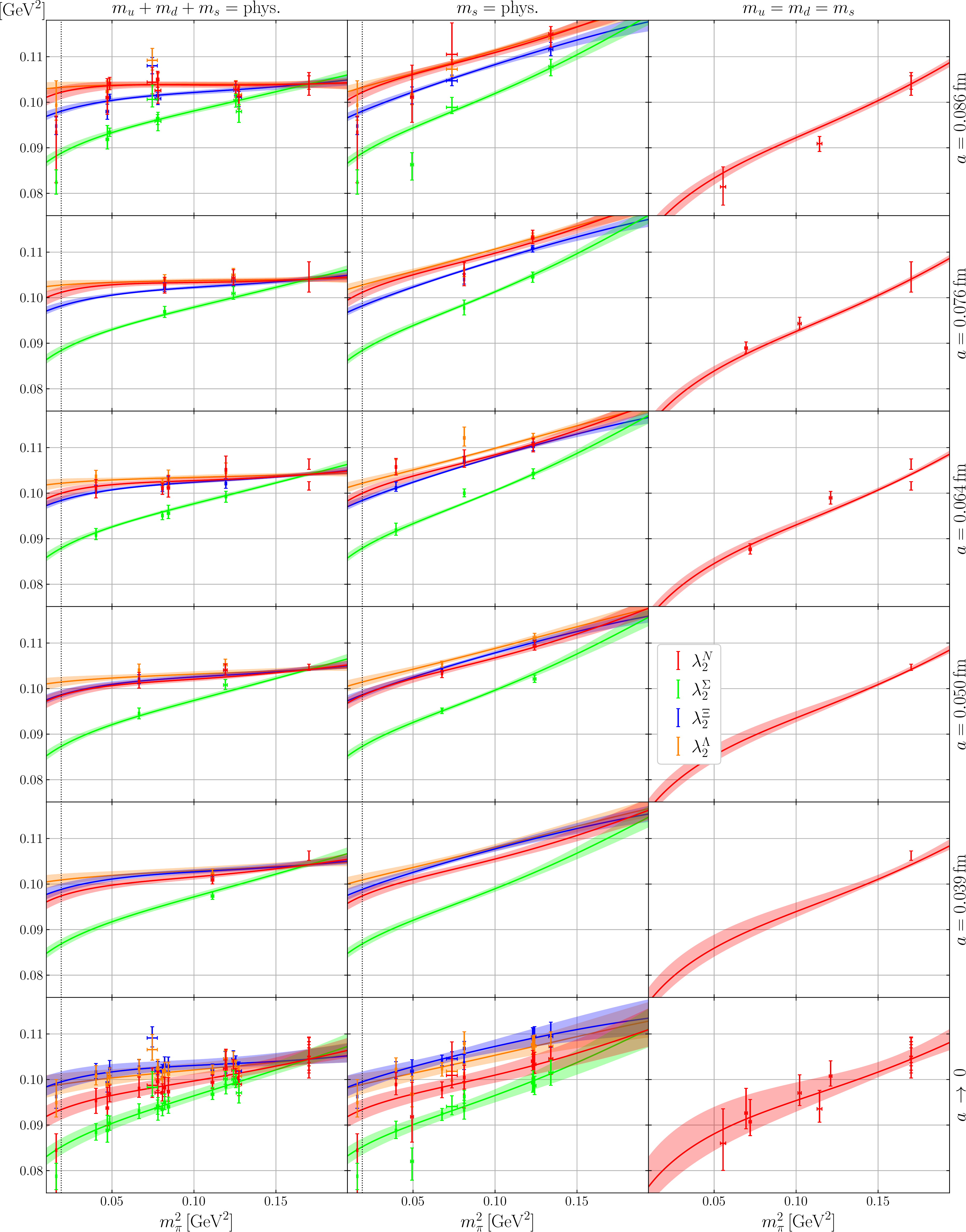}%
\caption{\label{fig_l2}Global fit (using $19$~free parameters) for the higher-twist normalization constants~$\lambda_2^B$, \captiontail}%
\end{figure*}%
\clearpage%
\raggedbottom
\raggedright

%

\begin{thebibliography}{10}
\providecommand{\url}[1]{{#1}}
\providecommand{\urlprefix}{URL }
\expandafter\ifx\csname urlstyle\endcsname\relax
  \providecommand{\doi}[1]{DOI \discretionary{}{}{}#1}\else
  \providecommand{\doi}{DOI \discretionary{}{}{}\begingroup
  \urlstyle{rm}\Url}\fi

\bibitem{Lepage:1979zb}
G.P. Lepage, S.J. Brodsky,
  \href{http://dx.doi.org/10.1016/0370-2693(79)90554-9}{Phys. Lett.
  \textbf{87B}, 359 (1979)}

\bibitem{Lepage:1980fj}
G.P. Lepage, S.J. Brodsky,
  \href{http://dx.doi.org/10.1103/PhysRevD.22.2157}{Phys. Rev. \textbf{D22},
  2157 (1980)}

\bibitem{Efremov:1979qk}
A.V. Efremov, A.V. Radyushkin,
  \href{http://dx.doi.org/10.1016/0370-2693(80)90869-2}{Phys. Lett.
  \textbf{94B}, 245 (1980)}

\bibitem{Braun:1988qv}
V.M. Braun, I.E. Filyanov, \href{http://dx.doi.org/10.1007/BF01548594}{Z. Phys.
  \textbf{C44}, 157 (1989)}

\bibitem{Balitsky:1989ry}
I.I. Balitsky, V.M. Braun, A.V. Kolesnichenko,
  \href{http://dx.doi.org/10.1016/0550-3213(89)90570-1}{Nucl. Phys.
  \textbf{B312}, 509 (1989)}

\bibitem{Chernyak:1990ag}
V.L. Chernyak, I.R. Zhitnitsky,
  \href{http://dx.doi.org/10.1016/0550-3213(90)90612-H}{Nucl. Phys.
  \textbf{B345}, 137 (1990)}

\bibitem{Gockeler:2008xv}
M.~G{\"o}ckeler, et~al.,
  \href{http://dx.doi.org/10.1103/PhysRevLett.101.112002}{Phys. Rev. Lett.
  \textbf{101}, 112002 (2008)}

\bibitem{Braun:2008ur}
V.M. Braun, et~al., \href{http://dx.doi.org/10.1103/PhysRevD.79.034504}{Phys.
  Rev. \textbf{D79}, 034504 (2009)}

\bibitem{Braun:2009jy}
V.M. Braun, et~al.,
  \href{http://dx.doi.org/10.1103/PhysRevLett.103.072001}{Phys. Rev. Lett.
  \textbf{103}, 072001 (2009)}

\bibitem{Schiel:2011av}
R.W. Schiel, G.S. Bali, V.M. Braun, S.~Collins, M.~G{\"o}ckeler, C.~Hagen,
  R.~Horsley, Y.~Nakamura, D.~Pleiter, P.E.L. Rakow, A.~Sch{\"a}fer,
  G.~Schierholz, H.~St{\"u}ben, P.~Wein, J.M. Zanotti,
  \href{http://dx.doi.org/10.22323/1.139.0175}{PoS \textbf{Lattice 2011}, 175
  (2012)}

\bibitem{Braun:2012zza}
V.M. Braun, M.~G{\"o}ckeler, R.~Horsley, Y.~Nakamura, D.~Pleiter, P.E.L. Rakow,
  A.~Sch{\"a}fer, R.~Schiel, G.~Schierholz, H.~St{\"u}ben, P.~Wein, J.M.
  Zanotti, \href{http://dx.doi.org/10.1063/1.3701193}{AIP Conf. Proc.
  \textbf{1432}, 80 (2012)}

\bibitem{Braun:2014wpa}
V.M. Braun, S.~Collins, B.~Gl{\"a}{\ss}le, M.~G{\"o}ckeler, A.~Sch{\"a}fer,
  R.W. Schiel, W.~S{\"o}ldner, A.~Sternbeck, P.~Wein,
  \href{http://dx.doi.org/10.1103/PhysRevD.89.094511}{Phys. Rev. \textbf{D89},
  094511 (2014)}

\bibitem{Bruno:2014jqa}
M.~Bruno, et~al., \href{http://dx.doi.org/10.1007/JHEP02(2015)043}{JHEP
  \textbf{02}, 043 (2015)}

\bibitem{Bali:2015ykx}
G.S. Bali, V.M. Braun, M.~G{\"o}ckeler, M.~Gruber, F.~Hutzler, A.~Sch{\"a}fer,
  R.W. Schiel, J.~Simeth, W.~S{\"o}ldner, A.~Sternbeck, P.~Wein,
  \href{http://dx.doi.org/10.1007/JHEP02(2016)070}{JHEP \textbf{02}, 070
  (2016)}

\bibitem{Bali:2019dqc}
G.S. Bali, V.M. Braun, S.~B{\"u}rger, M.~G{\"o}ckeler, M.~Gruber, F.~Hutzler,
  P.~Korcyl, A.~Sch{\"a}fer, A.~Sternbeck, P.~Wein,   (2019).
\newblock \href{https://arxiv.org/abs/1903.08038} {\texttt{arXiv:1903.08038
  [hep-lat]}}

\bibitem{Wein:2016ozo}
P.~Wein, {Chiral perturbation theory for generalized parton distributions and
  baryon distribution amplitudes}.
\newblock Ph.D. thesis, Universit{\"a}t Regensburg (Germany) (2016).
\newblock \href{http://nbn-resolving.org/urn:nbn:de:bvb:355-epub-347744}
  {\texttt{urn:nbn:de:bvb:355-epub-347744}}

\bibitem{Gruber:2017ozo}
M.~Gruber, {Renormalization of three-quark operators for baryon distribution
  amplitudes}.
\newblock Ph.D. thesis, Universit{\"a}t Regensburg (Germany) (2017).
\newblock \href{http://nbn-resolving.org/urn:nbn:de:bvb:355-epub-364421}
  {\texttt{urn:nbn:de:bvb:355-epub-364421}}

\bibitem{Hutzler:2018ozo}
F.~Hutzler, {Hadron Distribution Amplitudes from Lattice QCD Simulations}.
\newblock Ph.D. thesis, Universit{\"a}t Regensburg (Germany) (2018).
\newblock \href{http://nbn-resolving.org/urn:nbn:de:bvb:355-epub-374743}
  {\texttt{urn:nbn:de:bvb:355-epub-374743}}

\bibitem{Chernyak:1983ej}
V.L. Chernyak, A.R. Zhitnitsky,
  \href{http://dx.doi.org/10.1016/0370-1573(84)90126-1}{Phys. Rept.
  \textbf{112}, 173 (1984)}

\bibitem{Kraenkl:2011qb}
S.~Kr{\"a}nkl, A.~Manashov,
  \href{http://dx.doi.org/10.1016/j.physletb.2011.08.028}{Phys. Lett.
  \textbf{B703}, 519 (2011)}

\bibitem{Braun:2000kw}
V.~Braun, R.J. Fries, N.~Mahnke, E.~Stein,
  \href{http://dx.doi.org/10.1016/S0550-3213(00)00516-2}{Nucl. Phys.
  \textbf{B589}, 381 (2000)}.
\newblock [Erratum:\ Nucl.\ Phys.\ \textbf{B607}, 433 (2001)]

\bibitem{Anikin:2013aka}
I.V. Anikin, V.M. Braun, N.~Offen,
  \href{http://dx.doi.org/10.1103/PhysRevD.88.114021}{Phys. Rev. \textbf{D88},
  114021 (2013)}

\bibitem{Braun:2008ia}
V.M. Braun, A.N. Manashov, J.~Rohrwild,
  \href{http://dx.doi.org/10.1016/j.nuclphysb.2008.08.012}{Nucl. Phys.
  \textbf{B807}, 89 (2009)}

\bibitem{Wein:2015oqa}
P.~Wein, A.~Sch{\"a}fer, \href{http://dx.doi.org/10.1007/JHEP05(2015)073}{JHEP
  \textbf{05}, 073 (2015)}

\bibitem{Claudson:1981gh}
M.~Claudson, M.B. Wise, L.J. Hall,
  \href{http://dx.doi.org/10.1016/0550-3213(82)90401-1}{Nucl. Phys.
  \textbf{B195}, 297 (1982)}

\bibitem{Nakamura:2010qh}
Y.~Nakamura, H.~St{\"u}ben, \href{http://dx.doi.org/10.22323/1.105.0040}{PoS
  \textbf{Lattice 2010}, 040 (2011)}

\bibitem{Luscher:2012av}
M.~L{\"u}scher, S.~Schaefer,
  \href{http://dx.doi.org/10.1016/j.cpc.2012.10.003}{Comput. Phys. Commun.
  \textbf{184}, 519 (2013)}

\bibitem{Gusken:1989qx}
S.~G{\"u}sken, \href{http://dx.doi.org/10.1016/0920-5632(90)90273-W}{Nucl.
  Phys. B (Proc. Suppl.) \textbf{17}, 361 (1990)}

\bibitem{Falcioni:1984ei}
M.~Falcioni, M.L. Paciello, G.~Parisi, B.~Taglienti,
  \href{http://dx.doi.org/10.1016/0550-3213(85)90280-9}{Nucl. Phys.
  \textbf{B251}, 624 (1985)}

\bibitem{Luscher:2011kk}
M.~L{\"u}scher, S.~Schaefer,
  \href{http://dx.doi.org/10.1007/JHEP07(2011)036}{JHEP \textbf{07}, 036
  (2011)}

\bibitem{Gracey:2012gx}
J.A. Gracey, \href{http://dx.doi.org/10.1007/JHEP09(2012)052}{JHEP \textbf{09},
  052 (2012)}

\bibitem{Martinelli:1994ty}
G.~Martinelli, C.~Pittori, C.T. Sachrajda, M.~Testa, A.~Vladikas,
  \href{http://dx.doi.org/10.1016/0550-3213(95)00126-D}{Nucl. Phys.
  \textbf{B445}, 81 (1995)}

\bibitem{Sturm:2009kb}
C.~Sturm, Y.~Aoki, N.H. Christ, T.~Izubuchi, C.T.C. Sachrajda, A.~Soni,
  \href{http://dx.doi.org/10.1103/PhysRevD.80.014501}{Phys. Rev. \textbf{D80},
  014501 (2009)}

\bibitem{Gockeler:2008we}
M.~G{\"o}ckeler, et~al.,
  \href{http://dx.doi.org/10.1016/j.nuclphysb.2008.12.015}{Nucl. Phys.
  \textbf{B812}, 205 (2009)}

\bibitem{Kaltenbrunner:2008zz}
T.~Kaltenbrunner, {Renormalization of Three-Quark Operators for the Nucleon
  Distribution Amplitude}.
\newblock Ph.D. thesis, Universit{\"a}t Regensburg (Germany) (2008).
\newblock \href{http://nbn-resolving.de/urn:nbn:de:bvb:355-opus-11093}
  {\texttt{urn:nbn:de:bvb:355-opus-11093}}

\bibitem{Braun:2016wnx}
V.M. Braun, P.C. Bruns, S.~Collins, J.A. Gracey, M.~Gruber, M.~G{\"o}ckeler,
  F.~Hutzler, P.~P{\'e}rez-Rubio, A.~Sch{\"a}fer, W.~S{\"o}ldner, A.~Sternbeck,
  P.~Wein, \href{http://dx.doi.org/10.1007/JHEP04(2017)082}{JHEP \textbf{04},
  082 (2017)}

\bibitem{Bruno:2017gxd}
M.~Bruno, et~al., \href{http://dx.doi.org/10.1103/PhysRevLett.119.102001}{Phys.
  Rev. Lett. \textbf{119}, 102001 (2017)}

\bibitem{Chernyak:1987nu}
V.L. Chernyak, A.A. Ogloblin, I.R. Zhitnitsky,
  \href{http://dx.doi.org/10.1007/BF01557663}{Z. Phys. \textbf{C42}, 569
  (1989)}

\bibitem{Mobius:1827zz}
A.F. M{\"o}bius, \emph{{Der barycentrische Calcul}} (Verlag von Johann
  Ambrosius Barth (Leipzig, Germany), 1827).
\newblock \href{http://nbn-resolving.org/urn:nbn:de:bvb:12-bsb10082429-2}
  {\texttt{urn:nbn:de:bvb:12-bsb10082429-2}}

\bibitem{Mezrag:2017znp}
C.~Mezrag, J.~Segovia, L.~Chang, C.D. Roberts,
  \href{http://dx.doi.org/10.1016/j.physletb.2018.06.062}{Phys. Lett.
  \textbf{B783}, 263 (2018)}

\bibitem{Chernyak:1984bm}
V.L. Chernyak, I.R. Zhitnitsky,
  \href{http://dx.doi.org/10.1016/0550-3213(84)90114-7}{Nucl. Phys.
  \textbf{B246}, 52 (1984)}

\bibitem{Aznauryan:2009da}
I.~Aznauryan, et~al.,   (2009).
\newblock \href{https://arxiv.org/abs/0907.1901} {\texttt{arXiv:0907.1901
  [nucl-th]}}

\bibitem{Dudek:2012vr}
J.~Dudek, et~al., \href{http://dx.doi.org/10.1140/epja/i2012-12187-1}{Eur.
  Phys. J. \textbf{A48}, 187 (2012)}

\bibitem{Lu:2009cm}
C.D. L{\"u}, Y.M. Wang, H.~Zou, A.~Ali, G.~Kramer,
  \href{http://dx.doi.org/10.1103/PhysRevD.80.034011}{Phys. Rev. \textbf{D80},
  034011 (2009)}

\bibitem{Aaltonen:2011qs}
T.~Aaltonen, et~al.,
  \href{http://dx.doi.org/10.1103/PhysRevLett.107.201802}{Phys. Rev. Lett.
  \textbf{107}, 201802 (2011)}

\bibitem{Aaij:2017ewm}
R.~Aaij, et~al., \href{http://dx.doi.org/10.1007/JHEP04(2017)029}{JHEP
  \textbf{04}, 029 (2017)}

\bibitem{Edwards:2004sx}
R.G. Edwards, B.~Jo{\'o},
  \href{http://dx.doi.org/10.1016/j.nuclphysbps.2004.11.254}{Nucl. Phys. B
  (Proc. Suppl.) \textbf{140}, 832 (2005)}

\bibitem{Nobile:2010zz}
A.~Nobile, \href{http://dx.doi.org/10.22323/1.105.0034}{PoS \textbf{Lattice
  2010}, 034 (2011)}

\bibitem{Frommer:2013fsa}
A.~Frommer, K.~Kahl, S.~Krieg, B.~Leder, M.~{Rott\-mann},
  \href{http://dx.doi.org/10.1137/130919507}{SIAM J. Sci. Comput. \textbf{36},
  A1581 (2014)}

\bibitem{Heybrock:2015kpy}
S.~Heybrock, M.~Rottmann, P.~Georg, T.~Wettig,
  \href{http://dx.doi.org/10.22323/1.251.0036}{PoS \textbf{LATTICE 2015}, 036
  (2016)}

\bibitem{Bali:2016umi}
G.S. Bali, E.E. Scholz, J.~Simeth, W.~S{\"o}ldner,
  \href{http://dx.doi.org/10.1103/PhysRevD.94.074501}{Phys. Rev. \textbf{D94},
  074501 (2016)}

\bibitem{jureca}
{J{\"u}lich Supercomputing Centre},
  \href{http://dx.doi.org/10.17815/jlsrf-4-121-1}{JLSRF \textbf{4}, A132
  (2018)}

\bibitem{juqueen}
{J{\"u}lich Supercomputing Centre},
  \href{http://dx.doi.org/10.17815/jlsrf-1-18}{JLSRF \textbf{1}, A1 (2015)}

\bibitem{juwels}
{J{\"u}lich Supercomputing Centre},
  \href{http://dx.doi.org/10.17815/jlsrf-5-171}{JLSRF \textbf{5}, A135 (2019)}

\bibitem{Wein:2011ix}
P.~Wein, P.C. Bruns, T.R. Hemmert, A.~Sch{\"a}fer,
  \href{http://dx.doi.org/10.1140/epja/i2011-11149-5}{Eur. Phys. J.
  \textbf{A47}, 149 (2011)}

\bibitem{Ledwig:2014rfa}
T.~Ledwig, J.~Martin~Camalich, L.S. Geng, M.J. Vicente~Vacas,
  \href{http://dx.doi.org/10.1103/PhysRevD.90.054502}{Phys. Rev. \textbf{D90},
  054502 (2014)}

\bibitem{Becher:1999he}
T.~Becher, H.~Leutwyler, \href{http://dx.doi.org/10.1007/PL00021673}{Eur. Phys.
  J. \textbf{C9}, 643 (1999)}

\end{thebibliography}
\end{document}